\documentclass[notitlepage,reprint,onecolumn, amsmath,amssymb, aps,prd, nofootinbib]{revtex4-1} 

\usepackage{graphicx}
\usepackage{axodraw2}
\usepackage{bm}
\usepackage{color}
\usepackage[utf8]{inputenc} 
\usepackage[T1]{fontenc}

\usepackage{booktabs} 
\usepackage{array} 
\usepackage{verbatim} 
\usepackage{slashed}
\usepackage{comment}
\usepackage[compat=1.1.0]{tikz-feynman}
\usepackage{hyperref}
\usepackage{caption}
\usepackage{subcaption}

\numberwithin{equation}{section}

\definecolor{DGreen}{rgb}{0.08,0.5,0.15}

\definecolor{gray}{rgb}{0.5,0.5,0.5}
\definecolor{violet}{rgb}{0.5,0,0.5}

\newcommand{\bit}{\begin{itemize}}
\newcommand{\eit}{\end{itemize}}
\newcommand{\ben}{\begin{enumerate}}
\newcommand{\een}{\end{enumerate}}
\newcommand{\beq}{\begin{equation}}
\newcommand{\eeq}{\end{equation}}
\newcommand{\bea}{\begin{eqnarray}}
\newcommand{\eea}{\end{eqnarray}}
\newcommand{ \lsim}{\mathrel{\vcenter
     {\hbox{$<$}\nointerlineskip\hbox{$\sim$}}}}
\newcommand{ \gsim}{\mathrel{\vcenter
     {\hbox{$>$}\nointerlineskip\hbox{$\sim$}}}}
\newcommand{\gappeq}{\mathrel{\rlap {\raise.5ex\hbox{$>$}}
{\lower.5ex\hbox{$\sim$}}}}
\newcommand{\lappeq}{\mathrel{\rlap{\raise.5ex\hbox{$<$}}
{\lower.5ex\hbox{$\sim$}}}}

\newcommand{\Zslash}{ \, Z  \! \! \! \! / ~ }

\newcommand{\LNP}{\Lambda_{NP}}

\newcommand{\tl}{\tau \to l}

\newcommand{\me}{\mu \to e}

\newcommand{\muc}{\mu A \to \! eA }
\newcommand{\mucL}{\mu A \to \! e_LA }

\newcommand{\teg}{\tau \to e \gamma}
\newcommand{\meg}{\mu \to e \gamma}
\newcommand{\tmg}{\tau \to \mu \gamma}

\newcommand{\tlg}{\tau \to l \gamma}

\newcommand{\meee}{\mu \to e \bar{e} e}
\newcommand{\teee}{\tau \to e \bar{e} e}
\newcommand{\tmmm}{\tau \to \mu \bar{\mu} \mu}
\newcommand{\tlll}{\tau \to 3 l }

\def\a{\alpha}
\def\b{\beta}
\def\g{\gamma}
\def\d{\delta}

\def\m{\mu}

\def\r{\rho}
\def\s{\sigma}

\begin{document}

\title{ Constraining New Physics models from $\mu \to e $ observables in bottom-up EFT \\}

\author{Marco Ardu}
\email{E-mail address: marco.ardu@ific.uv.es}
\affiliation{ Departament de Fısica Teorica, Universitat de Valencia, Dr. Moliner 50,
	E-46100 Burjassot \& IFIC, Universitat de Valencia \& CSIC, E-46071, Paterna, Spain}

\author{Sacha Davidson}
\email{E-mail address: s.davidson@lupm.in2p3.fr}
\affiliation{LUPM, CNRS,
Université Montpellier
Place Eugene Bataillon, F-34095 Montpellier, Cedex 5, France
}
\author{Stéphane Lavignac}
\email{E-mail address: stephane.lavignac@ipht.fr}
\affiliation{Institut de Physique Th{\'e}orique, Universit{\'e} Paris Saclay,
CNRS, CEA, F-91191 Gif-sur-Yvette, France
}

\begin{abstract}
\noindent
{Upcoming experiments will improve the sensitivity to $\mu\to e$ processes by several orders of magnitude,  and could  observe lepton flavour-changing contact interactions for the first time. In this paper, we investigate what could be learned about New Physics from the measurements of these $\me$ observables, using a bottom-up {effective field theory (EFT)} approach and focusing on three popular models with new particles around the TeV scale (the type II seesaw, the inverse seesaw and a scalar leptoquark). We showed in a previous publication that $\mu\to e$ observables have the ability to rule out these models because none can fill the whole experimentally accessible parameter space. In this work we give more details on our  {EFT} formalism and present more complete results. We discuss the impact of some observables complementary to $\mu\to e$ transitions (such as the neutrino mass scale and ordering, and LFV $\tau$ decays) and draw attention to the interesting appearance of Jarlskog-like invariants in our expressions for the low-energy Wilson coefficients.  }
\end{abstract}

\maketitle


\section{Introduction }
\label{sec:intro}

The observed neutrino mass matrix  $[m_\nu]$ \cite{ParticleDataGroup:2022pth} requires New Physics beyond the Standard Model (SM) which  is  lepton flavour-changing.   Observing    lepton flavour-changing  processes other than neutrino oscillations, would therefore  give   complementary information  on New Physics in   the lepton sector.

Flavour-changing contact interactions among charged leptons (which we refer to as LFV), have not yet been observed,
but  upcoming experiments  aim to improve the sensitivity  to a few processes in the   $\mu\to e$ sector  by  orders of magnitude, and to probe a  wide palet of $\tau\to l$ processes with lesser sensitivity.
  This is summarised  in Table~\ref{tab:bds}; it  suggests that LFV could be discovered in  $\me$, while $\tl$ is more promising for distinguishing among models. For a review of $\me$ LFV, see {\it e.g.} \cite{KO}. 
  
  The aim of this project is  to explore  what can  be learned  about  New Physics in the lepton sector  from    observations of $\meg,\meee$ and/or $\muc$.
   For instance,  it would be ideal if  the data could indicate  properties of the  New Physics model, such as whether new particles interact with lepton doublets  or singlets or both,  whether LFV occurs amoung SM particles  at loop or tree level,    or whether LFV is related to $[m_\nu]$,  baryogenesis or New Physics in the quark flavour sector.

  In order to quantify what $\me$ data could tell us about models, 
  we study  these questions in a bottom-up EFT approach,  assuming
  $\LNP\gsim$ TeV. 
  We  translate the data from the experimental scale  to $\LNP$ using EFT,
  then match  three ``representative'' models  to the  experimentally
  allowed {Wilson}  coefficient space, and explore which differences among the models can be identified by the data.

  Model predictions for LFV have  been widely studied; in particular, there is a large literature  devoted to calculating LFV rates in neutrino mass models (for a review, see {\it e.g} \cite{Calibbi:2017uvl,Ardu:2022sbt};  or for example 
  \cite{Hisano:1995cp,Antusch:2006vw,Rossi:2002zb,Altmannshofer:2009ne,Cirigliano:2004mv,Omura:2015nja,Arganda:2014dta,Deppisch:2004fa,Agashe:2006iy,Blanke:2007db,Aliev:2007qw,Cai:2017jrq,Escribano:2021uhf,Lopez-Ibanez:2021yzu,Magg:1980ut,Lazarides:1980nt,Schechter:1980gr,Mohapatra:1980yp,Chun:2003ej,Kakizaki:2003jk,Akeroyd:2009nu,Dinh:2012bp,Ferreira:2019qpf,Calibbi:2022wko,Wyler:1982dd,Mohapatra:1986aw,Mohapatra:1986bd,Ilakovac:1994kj,Tommasini:1995ii,Ibarra:2011xn,Alonso:2012ji,Abada:2015oba,Coy:2018bxr,Crivellin:2022cve,Zhang:2021jdf})  and other  Standard Model extensions such as leptoquarks 
 \cite{Buchmuller:1986zs,Dorsner:2016wpm,Bauer:2015knc,Sakaki:2013bfa,Cai:2017wry,Angelescu:2018tyl,Lee:2021jdr,Shi:2019gxi,Desai:2023jxh,Bordone:2020lnb}.
 We hope that our bottom-up EFT approach could give a complementary perspective on the well-studied relations between models and observables.
  Our study differs  from top-down analyses in that firstly, we suppose that upcoming $\me$ experiments will measure 12 Wilson coefficients, and not just three rates.
  So we are using a more optimistic/futuristic parametrisation of the  LFV observables, using as input  everything they  could tell us.
  This allowed us  to   show in a previous publication~\cite{Ardu:2023yyw},  that the three models we consider could be ruled out by upcoming data.
  Secondly, model  studies  frequently scan  over the model parameter space;
  this allows to  estimate    correlations among LFV  observables, but the results depend on  the  choice of measure on  model  parameter space. We circumvent  the issue  of measure and the need to scan    by parametrising the models in terms of ``Jarlskog-like'' invariants~\cite{Jarlskog:1985ht}, which contribute to the observables.
Also, the operator coefficients are allowed to be complex, which is consistent with  the hints of leptonic CP violation in  neutrino observations \cite{T2K:2019bcf}.

This manuscript   gives more complete results  than  presented in \cite{Ardu:2023yyw} --- in particular  highlighting the appearance of model ``invariants'' in the coefficients at the experimental scale, and   explores the impact of  some complementary  observables on  the model  predictions for $\me$. 
 Section  \ref{sec:notn} reviews  $\me$   observables and  the EFT formalism implemented here, then  Section  \ref{sec:models} summarises   the three TeV-scale  models that we consider, which are
 the type II seesaw\cite{Magg:1980ut,Lazarides:1980nt,Schechter:1980gr,Mohapatra:1980yp}, an inverse seesaw~\cite{Wyler:1982dd,Mohapatra:1986aw,Mohapatra:1986bd}, and a scalar leptoquark\cite{Buchmuller:1986zs,Dorsner:2016wpm,Bauer:2015knc,Sakaki:2013bfa,Cai:2017wry,Angelescu:2018tyl,Lee:2021jdr,Shi:2019gxi,Desai:2023jxh,Bordone:2020lnb} which can fit the $R_D$ anomaly~\cite{BaBar:2012obs,Belle:2015qfa,Belle:2019rba,LHCb:2023zxo,LHCb:2023cjr}. 
 The matching of the models to the EFT is relegated to Appendix \ref{app:matching}. 
 Section \ref{sec:obs+invar}  gives   the twelve  observable Wilson coefficients at the experimental scale (including the coefficients  for $\muc$ on heavy targets, missing in \cite{Ardu:2023yyw}), expressed in terms of ``invariant'' combinations of  model and SM parameters.  Section \ref{sec:pheno}
 explores the interplay of $\me$ flavour change with some complementary observables in the models we consider.
Then we discuss  what we learned about bottom-up reconstruction  in Section  \ref{sec:disc}, and  conclude in Section  \ref{sec:sum}.


\section{ Observables and Notation }
\label{sec:notn}

In a bottom-up perspective, one starts from data and how to parametrise it, which is reviewed in Section \ref{ssec:obs}. Section \ref{ssec:EFT} reviews our EFT formalism; it  allows to obtain  expressions for  the observable  Wilson coefficients in terms of operator coefficients at the weak scale, which are given in Appendix  \ref{ssec:CLNP}.

\subsection{Observables}
\label{ssec:obs}

Some models  considered here  generate a Majorana mass matrix for the three light neutrinos. At energies below the weak scale (taken $\simeq m_W$), it can be  included in the Lagrangian  as \cite{ParticleDataGroup:2022pth}
  \beq
     \delta {\cal L}_{<m_W}= - \overline{\nu_{\a}}\frac{[m_\nu]^{\a\b}}{2}\nu^c_\b
+h.c. 
      \label{Lmnu}
  \eeq
  where  greek indices  indicate the charged lepton mass eigenstate basis,
square brackets indicate a matrix, 
  and   $[m_\nu]_{\a\b} = U_{\a i} m_i U_{\b i}$ for $ m_i$ real and positive. 
  The leptonic mixing matrix   $U$ is   parametrised as in \cite{ParticleDataGroup:2022pth}  by three mixing angles, one ``Dirac'' phase and two ``Majorana'' phases. In our  study, we approximate the Dirac phase $\delta \approx 3\pi/2$~\cite{T2K:2023smv},  and take the Majorana phases as free parameters.  The Majorana phases  and the lightest neutrino mass $m_{min}$ affect  the elements of $[m_\nu]$, but not the flavour-changing elements of
  \beq
      [m_\nu m^\dagger _\nu]_{\a\b} = [U D_mD_m U^\dagger]_{\a\b} - m_j^2 [U  U^\dagger]_{\a\b}
      \label{mmdag}
\eeq
for $\a \neq \b \in\{e,\mu,\tau\}$  and $D_m = {\rm Diag}\, (m_1,m_2, m_3)$, because the Majorana phases cancel and  the $|m_i^2 -m_j^2|$ are determined in neutrino oscillations. 
  
 For LFV, we focus on  $\me$ interactions that are otherwise flavour diagonal
  (so we do not consider  weak meson decays such as $K\to \mu^\pm e^\mp$).
A feature of  this sector is that there are  very restrictive bounds on a handful of processes,  and the experimental sensitivity is planned to improve by several orders of magnitude in  the next few years (see Table~\ref{tab:bds}).  This differs from the  $\tau\to l$ $(l\in\{ \mu,e\})$ sector, where
  a large variety of  LFV $\tau$ decays currently have BRs of order $ 10^{-8}$, and
  Belle II aims for sensitivities ${\cal O} (10^{-9}\to 10^{-10})$.
  So from an EFT perspective, the $\tl$ sector allows to independently measure almost every Wilson coefficient  up to a New Physics scale $\gsim 10 $ TeV (at tree level)\cite{Banerjee:2022xuw}; whereas in the $\me$ sector, fewer coefficients are probed at greater accuracy, motivating   the use of Renormalization Group Equations (RGEs) in the $\me$ sector.

\renewcommand{\arraystretch}{1.3}
\begin{table}[th]
\begin{center}
\begin{tabular}{|l|l|l|}
\hline
Process & Current bound on BR   & Upcoming Sensitivity    \\
\hline
$\mu\to e \gamma $ & $ < 3.1 \times 10^{-13}$ \cite{MEG:2016leq}
		   & $  {\sim 10^{-14}}$ \cite{MEG II:2018kmf} \\
$\mu\to \bar{e}ee $ & $ < 1.0  \times 10^{-12}$ \cite{SINDRUM:1987nra}
			& $ 10^{-14} \to 10^{-16}$ \cite{Blondel:2013ia} \\
$\mu Ti \to e Ti$ &  $< 6.1 \times 10^{-13}$ \cite{Wintz:1998rp}
&$\sim 10^{-16}$ \cite{COMET:2009qeh,Mu2e:2014fns}  \\
$\mu Au \to e Au$ &  $< 7 \times 10^{-13}$ \cite{SINDRUMII:2006dvw}
&(?~$\to  10^{-18})$ \cite{Kuno:2005mm,CGroup:2022tli}) \\
\hline
$\tau\to l\gamma$ & $<3.3\times 10^{-8}$ \cite{tau1} & $  \lsim 10^{-8} $\cite{belle2t3l} \\
$\tau\to 3l$ & $< {\rm few}\times 10^{-8}$ \cite{tau2}
			& few$ \times 10^{-10} $\cite{belle2t3l} \\
			...&...&...\\			\hline
		\end{tabular}	\end{center} 
\caption{Some $\mu \to e$  processes,  with  current experimental bounds, and the estimated reach  of upcoming  (and future)  experiments. For
  $\mu\to e \gamma$,  MEG II~\cite{MEG II:2018kmf}  aims to reach $BR \sim 6\times 10^{-14}$.  The Mu3e experiment \cite{Blondel:2013ia} aims for $BR(\meee) \sim 10^{-14}$  with the current beam and data-taking starting in 2025, then $\sim 10^{-16}$ with a beam upgrade~\cite{Aiba:2021bxe}. The COMET~\cite{COMET:2009qeh} and Mu2e~\cite{Mu2e:2014fns} experiments will search for $\mu\to e$ conversion on light nuclei where the $\mu$-$A$ bound state is long-lived; a different experimental approach, {\it eg} the PRISM/PRIME~\cite{Kuno:2005mm} or ENIGMA~\cite{CGroup:2022tli} proposals would be required to improve  also the sensitivity to  heavy  nuclei with shorter lifetimes. Some $\tl$ processes are included for comparaison, with Belle II expectations  for a luminosity of  50 ab$^{-1}$~\cite{belle2t3l}.
   \label{tab:bds}} 
\end{table}
 \renewcommand{\arraystretch}{1}

The three processes we consider are $\meg$, $\meee$ and $\muc$. In the latter,  a $\mu^-$ is captured by a nucleus,  where it can transform into an electron via various interactions; we restrict to those which are  coherent across the nucleus, or ``spin-independent''\footnote{We neglect $\mu e\g\g$ operators  which can contribute to $\muc$~\cite{Davidson:2020ord}.} (some details are given in Appendix \ref{app:ops}).  
  At the experimental scale,  these three  processes   can be parametrised  by the dimensionless operator coefficients $\{C\}$ of the following Lagrangian \cite{KO,KKO,Davidson:2020hkf,Davidson:2022nnl}:
\bea
\d {\cal L}_{<m_W} &=& \frac{1}{v^2}\sum_{X\in\{L,R\}}{\Big (}
  C^{e\mu}_{D,X} (m_\mu \overline{e} \sigma^{\a\b}P_{X} \mu) F_{\a\b}+
  C^{e\mu ee}_{S,XX} (\overline{e} P_X \mu ) (\overline{e} P_X e )
  + C^{e\mu ee}_{V,LX} (\overline{e} \gamma^\a P_L \mu ) (\overline{e} \gamma_\a P_X e )
\nonumber\\
&&
  + C^{e\mu ee}_{V,RX} (\overline{e} \gamma^\a P_R \mu ) (\overline{e} \gamma_\a P_X e ) + C^{e\mu}_{Al,X} {\cal O}_{Al,X} + C^{e\mu}_{Au\perp,X} {\cal O}_{Au \perp,X} {\Big )} + h.c.
\label{Lag1}
\eea
where  ${\cal O}_{Al} $ is  the combination of operators  contributing to 
SI $\mu\to e$  conversion on light targets\footnote{These operators are refered to as  ${\cal O}_{Alight} $ and  ${\cal O}_{Aheavy} $ in \cite{Davidson:2022nnl,Ardu:2023yyw}.}  such as Titanium (used by SINDRUM~\cite{SINDRUMII:2006dvw,Wintz:1998rp})  or Aluminium(to be used by COMET~\cite{COMET:2009qeh} and Mu2e~\cite{Mu2e:2014fns}), and $ {\cal O}_{Au\perp}$ is an independent combination probed by heavy targets 
such as Gold (used by SINDRUM~\cite{SINDRUMII:2006dvw}).
At the experimental scale, these operators would  describe $\mu\to e$ interactions with nucleons, which can be  matched to operators involving quarks at a scale  $\sim$ 2 GeV.  We use the quark basis here, because it is more convenient  for  comparing to models.   At  2 GeV, the
operators are\footnote{Notice that the quark currents are not chiral, which introduces  factors of 2  as given in Eq. (\ref{factor2}).}
\bea
    {\cal O}_{Al,X} &\simeq&
    (\overline{e}  P_X \mu) {\Big (} 0.692  (\overline{u}  u)
      +   0.699 (\overline{d}  d) + 0.0341 (\overline{s}  s)
      +  0.00440 (\overline{c}  c)+  0.00128  (\overline{b}  b)      {\Big )}
\nonumber\\
   && +
    (\overline{e} \g^\a P_X \mu) {\Big (}
 0.125 (\overline{u} \g_\a  u) +  0.128 (\overline{d} \g_\a  d)
       {\Big )}
\label{OAlight}\\
  {\cal O}_{Au,X}&\simeq &\cos \theta_A  {\cal O}_{Al,X} + \sin \theta_A{\cal O}_{Au\perp,X}
\label{OAheavy}\\
      {\cal O}_{Au\perp,X}&\simeq &
      -(\overline{e}  P_X \mu) {\Big (}
        0.2 (\overline{u}  u)  +0.1 (\overline{d}  d)+
        0.008(\overline{s}  s)+  0.001(\overline{c}  c)+    0.0003(\overline{b}  b) 
 {\Big )} \nonumber\\&& +
    (\overline{e} \g^\a P_X \mu) {\Big (}
 0.56 (\overline{u} \g_\a  u) +  0.8 (\overline{d} \g_\a  d)
       {\Big )}
\label{OAheavyperp}
\eea
where $\theta_A$ is the misalignement angle between Gold and Aluminium, 
and  Appendix~\ref{app:ops} briefly reviews these results.

  The Branching Ratios (BRs) 
  can  be written as
\bea
BR(\meg)&=& 384 \pi^2 (|C^{e\mu}_{D L}|^2 + |C^{e\mu}_{D R}|^2) 
\label{BRmeg}\\
BR(\meee) 
& =&  \frac{|C^{e\mu ee}_{S,LL}|^2+ |C^{e\mu ee}_{S,RR}|^2}{8}
+2 |C^{e\mu ee}_{V,RR}  + 4eC^{e\mu}_{D, L}|^2
 +2| C^{e\mu ee}_{V,LL}  + 4eC^{e\mu}_{D ,R}|^2  \label{BRmeee}\\
&&+ (64 \ln\frac{m_\mu}{m_e} -136) 
(|eC^{e\mu}_{D ,R}|^2 +|eC^{e\mu}_{D ,L}|^2) + |C^{e\mu ee}_{V,RL}  + 4eC^{e\mu}_{D, L}|^2
 + |C^{e\mu ee}_{V,LR}  + 4eC^{e\mu}_{D, R} |^2 
 \nonumber\\
BR_{SI}(\mu Al \to e Al)&=& B_{Al} ( |d_{Al} C^{e\mu}_{D R}+ C_{Al,L}|^2 + |d_{Al} C^{e\mu}_{D L} + C_{Al, R}|^2)
\label{BRmucAl}\\
BR_{SI}(\mu Au \to e Au)&=& B_{Au} (|d_{Au}C^{e\mu}_{D R}+ C_{Au,L}|^2 + |d_{Au} C^{e\mu}_{D L} + C_{Au , R}|^2) 
\label{BRmucAu}
\eea
  where  only the Spin Independent contribution to $\muc$ is  included based on the results of \cite{KKO} (we  do not use the recent results of \cite{Haxton:2022piv}),  the $B_A$ are target-nucleus dependent  constants discussed in Appendix
  \ref{app:ops}, and $d_A\equiv I_{A,D}/(4|\vec{u}_A|)$ are given after Eqn (\ref{dA}). 
In all cases, the outgoing electrons are approximated as chiral because they
are relativistic.  The  final  states
containing electrons of different chirality  therefore do not interfere,
giving  independent constraints  on the coefficients
generating those final states. The experimental  limits  on  $\meg$ and $\meee$ therefore constrain eight coefficients --- the two dipoles and {six} four-lepton coefficients--- to be  in the vicinity of zero. The correlation matrix and resulting bounds are  discussed in \cite{Davidson:2018kud} and  we repeat the bounds here for completeness:
 \bea
|C_{D,X}|^2   \leq
\frac{B_{\meg} }{ 384\pi^2 +  e^2(64\ln\frac{m_\mu}{m_e}- 136) \frac{B_{\meg}}{ B_{\meee}} } \nonumber \\
|C_{V,XX}|^2   \leq \frac{ B_{\meee}}{ 2}\left(
1 +  \frac{32e^2B_{\meg}}{ 205e^2B_{\meg} +384\pi^2 B_{\meee}}\right)\nonumber \\
|C_{V,XY}|^2   \leq  B_{\meee}
\left( 1 +  \frac{16e^2B_{\meg}}{ 205e^2B_{\meg} +384\pi^2 B_{\meee}}\right)
\eea
where  $B_{process}$  is the
experimental upper bound on the Branching Ratio for the process,
and  $e^2(64\ln\frac{m_\mu}{m_e}- 136)$ is written as $ 205e^2$. {Although the discussion  of
\cite{Davidson:2018kud} implicitly supposed the coefficients were real, the same bounds apply for complex coefficients, because the real and imaginary components of the coefficients are constrained  to lie  inside the same ellipse around the origin.}

{
The dipole coefficient  is not constrained by $\mu A \to e A$ because
it contributes in interference, but the bound on $|C_{Al,X}|^2 $  (and
 in principle $|C_{Au\perp,X}|^2 $)   is
affected in the expected way by the independent bounds on the dipole:  
\bea
|C_{Al,X}|^2   \leq
\frac{B_{\mu Al \to e Al}}{B_{Al}} +  \frac{d_{Al}^2B_{\meg} }{ 384\pi^2} ~~.\nonumber
\eea
In this work, we do not  consider future improvements in the experimental reach for $\muc$ on heavy targets, so the  current  MEG bound on the dipole coefficient ensures that it is negligible  in $\mu Au \to e Au$. 
The current bound on Gold therefore  implies
\bea
|C_{Au\perp,X}|^2 \leq \frac{{B}_{\mu Au \to e Au} }{B_{Au} \sin^2\theta_A}+
\frac{\cos^2\theta_A {B}_{\mu Al \to e Al}}{B_{Al} \sin^2\theta_A}
\eea
where $\theta_A$ is the misalignement angle between Gold and Aluminium and given after Eq. (\ref{thetaAlAu}).

The current bounds on the twelve  observable coefficients are given  in Table~\ref{tab:2}, as well as the bounds that could be set, if upcoming experiments do not observe $\me$ processes. The $\muc$  bounds given  here differ from  our previous paper \cite{Ardu:2023yyw} because   ${\cal O}_{Au\perp,X}$  is here defined  in  terms of operators with quarks, so  differs from the nucleon definition of  \cite{Ardu:2023yyw}. The matching of  quarks  with nucleons is discussed in Appendix \ref{app:ops}.
}

\begin{table}[ht]
\begin{center}
\begin{tabular}{|l|l|l|l|}
\hline
 & current bound & upcoming & process\\
 \hline
$|C_{D,X}|$ &$9.0\times 10^{-9}$ & $ \sim 10^{-9}$ &$\meee $ \\
$|C_{V,XX}|$ &$7.0\times 10^{-7}$ & $\sim 10^{-8}$ &$\meee $ \\
$|C_{V,XY}|$ &$ 1.0 \times 10^{-6}$ & $\sim 10^{-8}$&$\meee $ \\
 $|C_{S,XX}|$ &$2.8\times 10^{-6}$ & $ \sim 10^{-8}$  &$\meee $\\
 $C_{Al,X}$ & $4.3\times 10^{-9}$ & $7\times 10^{-11}$&$\mu {\rm Ti}\to e   {\rm Ti}$\\
$C_{Au\perp,X}$ & $7.4\times 10^{-8}$ & ${7.4\times 10^{-8}}$&$\mu {\rm Au}\to e   {\rm Au}$\\
\hline
\end{tabular}
\caption{ Current and possible future   bounds on the  observable  coefficients, with $X,Y \in \{L,R\}, X\neq Y$. { The current bound on $C_{Al,X}$ is from
    $\mu {\rm Ti}\to e   {\rm Ti}$ \cite{Wintz:1998rp}, and the future  bound assumes  BR( $\mu{\rm Al}\to e   {\rm Al}) \leq 10^{-16}$.
The current and future bounds on $C_{Au\perp,X}$ are identical because we do not include the  reach of the proposed PRISM/PRIME and ENIGMA experiments among our  upcoming results.
\label{tab:2}}}
\end{center}
\end{table}

In this theoretical  study,  we  optimistically
consider   that these   twelve  coefficients can be measured with  the same reach that they can be constrained,  given in Table~\ref{tab:2}.   Polarising the muon could allow to distinguish  between  coefficients of operators with an L vs an  R projector in the $\bar{e}-\mu$ bilinear in $\meg$ and $\meee$\cite{KO} as well as in $\muc$ \cite{Kuno:1987dp}.  And in the case of $\meee$,
the  angular distributions of  the three-particle final state
 allow to determine the  magnitude of  various coefficients
 and some phases,   as discussed in \cite{Okadameee}. Vector operators ${\cal O}_{VXX}$ and scalar operators ${\cal O}_{SYY}$ (for $Y \neq X$)  induce the same angular distributions, but could in principle be distinguished by measuring  final state helicities~\cite{Okadameee}. Our models do not  generate the scalar operators within the  reach of upcoming experiments (see Subsection~\ref{ssec:tLFV}).

\subsection{bottom-up EFT}
\label{ssec:EFT}

For an introduction to EFT, see for instance \cite{Georgi,burashouches,LesHouches}. 
Our EFT consists of  an operator basis and  Renormalisation Group Equations (RGEs) to  look after scale evolution. The previous section  presented an operator basis  describing the observables at low energy;  a more complete set of operators is required in translating the observables  from the experimental to the New Physics scale, because   the  RGEs mix operators.

Our New Physics scale is at $\LNP \approx$ TeV,  so we need 
a    QED$\times$QCD invariant basis of  operators below the weak scale,
included  in the Lagrangian as 
\bea
   \delta {\cal L}_{<m_W}&=& - \overline{\nu_{\a}}\frac{[m_\nu]^{\a\b}}{2}\nu^c_\b
+
 \sum_{n \geq 2} \sum_{A,\zeta}\frac{C_A^{\zeta} \mathcal{O}_A^{(4+n)\zeta}}{v^n} + h.c. 
 \label{LEFT}
 \eea
 where $v = 174$ GeV,  the superscript $\zeta$ gives  the flavour indices, and  the   operator subscripts indicate the Lorentz structure and particle content. So for instance
 $$
 \mathcal{O}_{V,XY}^{e\mu uu} = (\overline{e} \g^\a P_X \mu)  (\overline{u} \g_\a P_Y u) ~~~,~~~X,Y \in\{L,R\}. 
 $$
We write the SM Yukawa matrices as $[Y_e], [Y_u]$ and $ [Y_d]$, and the Yukawa eigenvalue of fermion $f$ as $y_f$.

 As discussed in Appendix \ref{assec:quelEFT},  we did not find a simple and reliable recipe  to  express the coefficients of the Lagrangian (\ref{LEFT})  in terms of model parameters,    because a TeV is not so far from the weak scale.
 The SMEFT operator basis~\cite{BW,polonais} and RGEs~\cite{JMT} are appropriate above
 the weak scale, but the  EFT expansions (in operator dimension and loop$\times$logarithm) are not converging fast.  In principle, we could match the model directly to  the Lagragian~(\ref{LEFT}), but that requires to  calculate  many loop diagrams.
Finally, we  opt to  present the matching results   in   the  basis  of Eq.~(\ref{LEFT}).
 So we will not be using the SMEFT basis, but  for comparing to the literature, we implement it as
\bea
  \delta {\cal L}_{SMEFT}&=&
   - \frac{C_5^{\a \b}}{2v} (\overline{\ell_{\a}} \varepsilon H^*)(\ell^c_{\b} \varepsilon H^*) +  
 \sum_{A,\zeta}\frac{C_A^{\zeta} \mathcal{O}_A^{\zeta}}{v^2} + h.c. 
 \label{LSMEFT}
 \eea
where $[m_\nu]^{\a\b} = C_5^{\a \b}v $.

The low energy EFT includes all  LFV operators of dimension six, and some relevant operators of dimension seven (see \cite{Davidson:2020hkf} for a list). 
In this low-energy EFT,   we  ensure that  operators appear only once 
 by requiring  $\zeta = e \mu...$. 
Operators of  dimension five and six 
($n= 1,2$) are included in SMEFT,  
 but  without the $+h.c.$ for hermitian operators.
In SMEFT, we follow the convention
that each  flavour index  in $\zeta$   runs over  all three  generations
(so some operators are repeated in the SMEFT Lagrangian, causing some factor of 2 or 4 differences  between coefficients   in SMEFT vs the low-energy EFT).

  The  doublet and singlet leptons  are  in the charged lepton mass eigenstates $\{e,\mu,\tau\}$\cite{Ardu:2021koz},  which can differ from the  diagonal-Yukawa basis as discussed in Appendix \ref{assec:EH}. The singlet quarks are   labelled by their flavour, and the  quark doublets are in the $u$-type mass basis, with  generation indices that  run  $1\to 3$.

  In  quantum field theories, the coefficients of renormalizable and non-renormalizable operators evolve with scale.  For non-renormalisable operators,  this evolution of the operator coefficients lined up in a row vector $\vec{C}$ can be described
  as 
  \bea
  \mu\frac{d}{d\mu} \vec{C} \equiv  \frac{d}{d t} \vec{C}
 =  \vec{C} [\Gamma] + \vec{C}[\vec{X}] \vec{C}^\dagger + ... 
  \label{RGE}
  \eea
  where $t =\log \mu$, the effects of renormalisable interactions on the non-renormalisable operators are described by the matrix $[\Gamma]$, and $[\vec{X}]$ is a three-index tensor that schematically represents the effect of non-renormalisable interactions on the evolution of other non-renormalisable interactions  (for instance, the mixing of  a pair of four-fermion operators into another four-fermion operator via a fish diagram).
  
The RGEs we implement automatically  for the QED$\times$ QCD-invariant EFT are an improved leading log  approximation to $[\Gamma]$ \cite{Crivellin:2017rmk}, where  most of the anomalous dimension are at  one-loop, augmented by  the  two-loop vector to dipole mixing (because this mixing vanishes at one loop and is ${\cal O} (10^{-3})$ \cite{Ciuchini:1993fk}). We also consider  the mixing of two dimension six operators into a dimension eight operator described by $[\vec{X}]$, and  when relevant, include these contributions by hand in  matching  the model to EFT at $\LNP$. Since  the dimension eight operators have two additional Higgs legs,  they  are particularily relevant  when the Higgs has ${\cal O} (1)$ couplings to loop particles, such as the top quark in the leptoquark model.   These dimension eight contributions are  discussed more fully in Section \ref{ssec:tLFV} and  Appendix \ref{app:matching}.
  
Solving the RGEs  allows to translate  operator coefficients   from the experimental scale to the New Physics scale --- where the coefficients  can be calculated in  a model. We aim to resum the QCD running, and include the electroweak
 effects in perturbation theory (${\cal O}(\a \log)$ and occasionally ${\cal O}(\a^2 \log^2)$). If   the matrix $[\Gamma]$ were   scale-independent, then the solution would be  $\sim \exp \Gamma t$; however, $[\Gamma]$ decribes  SM loop corrections, and  SM parameters  run in various ways  with scale. For    $[\Gamma]$  scale-dependent,  the RGEs can be solved by scale-ordering,  analogous to the  familiar time-ordering that allows to solve a similiar  equation for the time-translation operator in quantum field theory.
We  include the  scale-dependence  of $\a_s$ (at one loop), and the associated running of quark Yukawas, but neglect  the running  due to other SM couplings (including the top Yukawa).
So the effect of QCD is to renormalise some coefficients and rescale some electroweak  anomalous dimensions, such that
the  first terms in the perturbative solution of the  RGE (\ref{RGE}) are
  \bea
 C_J(m )&\simeq &  C_J(\Lambda ) \eta_m^{a_J}  -
 C_K(\Lambda  ) \eta_m^{a_K}    \widetilde{\Gamma}_{KJ} \ln \frac{\Lambda }{m}
\eea
where $\widetilde{\Gamma}_{KJ}  = f_{KJ} \Gamma_{KJ} (m)$  \cite{Bellucci:1981bs} with  $\Gamma$  the electroweak anomalous dimension matrix, no sum on $J,K$ and : 
\bea
f_{KJ}  =
\frac {( 1 - \eta_m^{a_J-a_K -a_I+1 }) }{(1 +a_J - a_K-a_I)(1-\eta_m)} ~~,~~~\eta_m = \frac{\a_s(m(m))}{\a_s(\LNP)} ~~,~~~\eta = \frac{\a_s(2{\rm GeV})}{\a_s(\LNP)} 
\label{soln1tG}
\eea
where $a_{J} = -\gamma^s_{J}/2\b_0$ for $\a_s\gamma^s_{J}/4\pi$  the QCD anomalous dimension of the  coefficients $C_{J}$ ($a_T=-4/23$, $a_S= a_D = 12/23$ with 5 flavours),  $\b_0$ is the 1-loop QCD  beta function coefficient, 
and  the parameters in the electroweak anomalous dimension
cause it to run as $\Gamma_{KJ}(m) = \eta^{a_I} \Gamma_{KJ}(\Lambda)$. 
Finally, once the RGEs have been solved,  a coefficient at the experimental scale (for instance $C_{D,X} (m_\mu)$), can be written   at $\LNP$ as  a weighted  sum of the coefficients of operators that can contribute  via loops to  $\meg$.

{It can be shown~\cite{Davidson:2020hkf} that almost every operator involving 3 or 4 legs
that induces a $\me$ interaction (but no other flavour change) contributes to  the amplitude for $\meg$, $\meee$ and/or $\muc$ suppressed at most by a factor of order $10^{-3}$.}
  This suggests that the RGEs are relevant to include, because they  ensure that  a small handful of processes are sensitive to almost any  $\me$  operator.
 Despite that almost all operators can contribute, there remain only 12 constraints. Rather than dealing with a large correlation matrix corresponding to the usual operator basis, we  use a scale-dependent basis that corresponds to the  twelve experimentally probed directions. This was proposed in \cite{Davidson:2020hkf}, and the recipe we follow  is  outlined  in \cite{Davidson:2022nnl}.

    The RGEs can be solved to express an operator coefficient
    at the experimental scale, {\it eg} the dipole coefficient,  in terms of operator coefficients at the NP scale:
$$C_D(m_\mu)   \simeq \vec{C} (\LNP)  \cdot \vec{e}_D ~~~.  $$
    The directions in coefficient space corresponding to the twelve vectors $\{e_O\}$ then   form a basis for the  observable subspace at $\LNP$.  The  elements of most of these vectors at  or just above the electroweak scale are given
    in  \cite{Davidson:2020hkf} (the  coefficient combinations probed by $(\mu Al \to e Al)$ and  $(\mu Au \to e Au)$ are given  for completeness in Appendix \ref{ssec:CLNP}).


\section{ The models}
\label{sec:models}

The  three  TeV-scale New Physics models that we consider are  the type II and inverse seesaw models,  and  a scalar leptoquark.
These models are  selected for their   diverse\footnote{Ref.~\cite{Kanemura:2015cca}   makes an interesting classification of   Majorana  neutrino mass models into three sets,  based  on the flavour structures that are multiplied in the definition of $[m_\nu]$. However, both seesaw models enter the same set.} lepton-flavour-changing predictions:  LFV is controlled by the neutrino mass matrix in the type II seesaw,  is  independent of the neutrino mass matrix in the inverse seesaw, and  the leptoquark  can mediate $\muc$ at tree level, as well as addressing  anomalies in the quark flavour sector.

The type II seesaw model  \cite{
  Mohapatra:1980yp,Magg:1980ut,Schechter:1980gr,
  Lazarides:1980nt}  is an economical neutrino mass model,
where    the SM particle content is extended
  with a colour-singlet, SU(2) triplet
   scalar $\Delta$, of  hypercharge
   $Y=+1$ (in the normalization where the lepton doublets have $Y=-1/2$).
   The  SM  Lagrangian at the mass scale of the triplet  is augmented by
\bea
\d{\cal L}_\Delta&=& (D_\rho \Delta^I)^\dagger D^\rho \Delta^I  -M_\Delta^2|\Delta|^2
+\frac12\left(f_{\alpha\beta}\,\overline{\ell^c_\alpha} (i\tau_2)\tau_I \ell_\beta \Delta^I
+M_\Delta \lambda_H\, H^T (i\tau_2) \tau_I H \Delta^{*I}+{\rm h.c.}\right)
\nonumber          \\
&&     + \lambda_3 (H^\dagger H) (\Delta^{I*} \Delta^{I})+\lambda_4{\rm Tr}(\Delta^{I*}\tau_{I}\tau_{J}\tau_{K}\Delta^{K})(H^\dagger \tau_J H) + \dots ~~,
\label{LII}
\eea
where 
$\ell$ are the left-handed SU(2) doublets,
 $M_\Delta$ is  the  triplet  mass which we take $\sim$ TeV,
$f$ is a symmetric complex $3\times3$ matrix  proportional to the light neutrino mass matrix and whose indices $
\a,\b$ run   over $\{e,\mu,\tau\}$,
$\{\tau_I\}$ are the Pauli matrices,
and the $\lambda$'s are real dimensionless couplings. 
A feature of this model, that is shared with some  other neutrino mass models, is that the SU(2) singlet leptons do not  acquire new interactions, so  LFV is expected to    involve the doublet leptons.
 The phenomenology of the type II seesaw has been widely studied at colliders ~\cite{Akeroyd:2005gt,FileviezPerez:2008jbu,Melfo:2011nx, Freitas:2014fda,Ghosh:2017pxl,Antusch:2018svb} and for low-energy LFV \cite{Konetschny:1977bn,Chun:2003ej,Ma:2000xh,Kakizaki:2003jk,Akeroyd:2009nu,Dinh:2012bp,Ferreira:2019qpf,Calibbi:2022wko,Mandal:2022zmy}.

In Effective Field Theory, the neutrino mass matrix  generated in the type II seesaw model  can be obtained by matching, at $M_\Delta$, the  tree-level diagram {in the left panel} of Figure \ref{fig:t2match}  onto the   dimension five neutrino mass operator appearing in Eq. (\ref{LSMEFT}).  This gives a  neutrino mass  matrix
\bea  
 [m_\nu]^{\a\b}\simeq 
\frac{ [f]^{\a\b *} \lambda_H v^2}{M_\Delta}
 \label{mnutypeII}
 \eea
 which can also be obtained in the model by minimising the potential for the Higgs and the electrically neutral component of  {the triplet = $(\Delta^1 + i\Delta ^2)/\sqrt{2}\,$, which  obtains a {\it vev}  $ -\lambda_H v^2/(\sqrt{2}M_\Delta)$.
 This model is reputed to be predictive for LFV, because the lepton  flavour-changing  couplings $f^{\a\b}$ are proportional to the light neutrino mass matrix. However, LFV is not suppressed by the small neutrino mass scale, because  lepton number change involves  $[f]^{\a\b *} \lambda_H$,    so for a sufficiently small Higgs to triplet coupling, $\lambda_H \sim 10^{-12}$, it is possible to have $M_\Delta\sim$ TeV and
 $f^{\a\b}$ of ${\cal O}(1)$.

The second model we consider is the inverse (type I) seesaw~\cite{Wyler:1982dd,Mohapatra:1986aw,Mohapatra:1986bd}, which, like the type II seesaw,   naturally generates  small  Majorana neutrino masses from  new particles  that can be  at the TeV-scale with ${\cal O}(1)$  LFV couplings.  
In this model,   $n$   gauge singlet Dirac  fermions, $\Psi _a^T = (S_a, N_a)$, are added to the SM  particle content, with  approximately lepton number conserving interactions. Lepton number changing interactions can be included  via Yukawa couplings  and/or Majorana masses(see \cite{Fernandez-Martinez:2022gsu} for a ``basis-independent'' discussion of the options);  we choose to  allow  small Majorana masses for  $S_a$ and write the Lagrangian as
\begin{equation}
	\delta\mathcal{L}_{NS}=i\overline{N}\slashed{\partial} N+i\overline{S}\slashed{\partial} S-\left(Y^{\alpha a}_\nu(\overline{\ell}_\alpha\tilde{H} N_a)+M_{a}\overline{S}_a N_a+\frac{1}{2}\mu_{ab}\overline{S_a} S^c_b+{\rm h.c}\right),
\label{Linverse}
\end{equation}
where  $a,b$ run from 1..$n$, $N_a$  and  $S_a$ are  respectively right- and left-handed and in the  eigenbases of $M$,  $Y_\nu$ is a complex $3\times n$ dimensionless matrix,  we take  $M_a$ of ${\cal O}$(TeV), and $\mu$ is an $n\times n$ Majorana mass matrix with $\mu_{ab}\ll M_c$.  For vanishing $\mu$,  lepton number is conserved, and  the $N_a$ combine with the $S_a$ into  Dirac singlet neutrinos, which can have lepton flavour changing interactions $Y_\nu$ with the SM doublets. Like the type II seesaw, this model is expected to induce LFV among doublet leptons, but unlike the type II seesaw, it  has  several  non-degenerate  heavy new particles, and will induce low energy LFV processes via loop diagrams because  the flavour-changing $Y_\nu$ couples $\ell$ to two heavy particles.

The inverse seesaw models  induce LFV \cite{Ilakovac:1994kj,Tommasini:1995ii,Ibarra:2011xn,Alonso:2012ji,Abada:2015oba,Coy:2018bxr,Crivellin:2022cve,Zhang:2021jdf},   they can lead to non-unitarity of the lepton mixing matrix ~\cite{Antusch:2014woa, Fernandez-Martinez:2016lgt, Chrzaszcz:2019inj, Blennow:2023mqx}, and the singlets  could be discovered at colliders for suitable mass ranges ~\cite{delAguila:2007qnc, Atre:2009rg, Deppisch:2015qwa, Banerjee:2015gca, Antusch:2016vyf, Das:2018usr, Mekala:2022cmm}.
With  small   $\mu_{ab}$, the  leading contribution to the active neutrino mass matrix  is
\bea
    [m_\nu]^{\a\b} \simeq  [Y_\nu M^{-1} \mu M^{-1} Y_\nu^T]^{\a\b} v^2~~~.
    \label{mnuinverse}
\eea
So  the flavour-changing $Y_\nu$  can be  ${\cal O}$(1) because  small $\mu_{ab}$ gives small $m_\nu$, and the flavour change is expected   to be  independent of the active  neutrino mass matrix as can be seen for $n$ = 3  by solving Eq. (\ref{mnuinverse}) for $\mu_{ab}$.

Finally, our third model includes  a leptoquark --- for a review of this class of coloured and charged bosons, see {\it e.g.} \cite{Dorsner:2016wpm}---
 which is chosen to  fit the anomalies  in $R_{D^*}$  and/or $R_D$~\cite{BaBar:2012obs,Belle:2015qfa,Belle:2019rba,LHCb:2023zxo,LHCb:2023cjr}. It is
{an SU(2)-singlet scalar}
 denoted  $S_1$  in the notation of \cite{Buchmuller:1986zs} -- not to be confused with the singlet fermions  $\{S_a\}$ of Eq.~(\ref{Linverse}) --
 and has   interactions
\bea
{\cal L}_S & = & 
(D_\r S_1)^\dagger D^\r S_1 - m_{LQ}^2 S_1^\dagger S_1  + 
(- \lambda^{\a j}_{L} \overline{\ell}_\a  i \tau_2 q_j^c
+ \lambda^{\a j}_{R} \overline{e}_\a u_j^c ) S_1 
+  (\lambda^{\a j*}_{L} \overline{q^c}_j  i \tau_2  \ell_\a +
\lambda^{ \a j*}_{R} \overline{u^c}_j e_\a) S_1^{\dagger} \nonumber\\
&&+ \lambda_4 H^\dagger H S_1^\dagger S_1 + ...
\label{LLQ}
\eea
where
the  generation indices are $\a \in\{ e,\mu,\tau \}$
and $j\in \{u,c,t\}$, and the
sign of the doublet  contraction
is taken to give $+ \lambda^{\alpha j}_L \overline{e_L} (u_L)^c S_1$.
The leptoquark mass is  $m_{LQ} \approx $ TeV, consistent with the
CMS and ATLAS  searches   \cite{CMSLQ,ATLASLQ}  which exclude  leptoquarks
{with sub-TeV masses}
 that are  pair-produced  via  strong interactions,  and  decay   to specific final states. 
Some leptoquark-Higgs interactions are included  in Eq. (\ref{LLQ}) because they appear in  the matching  results of Appendix \ref{ssec:LQmatchsum}, but  their contributions to LFV observables are negligible assuming perturbative couplings.
This Lagrangian  does not  lead to  neutrino masses (as mentioned in Appendix \ref{ssec:LQmatchsum}), but  features $\muc$ at tree level and  LFV interactions for singlet  and doublet leptons (so  unlike the seesaw models, it induces scalar and tensor operators).

The  low-energy  phenomenology of leptoquarks   (see {\it eg}   \cite{Bauer:2015knc,Sakaki:2013bfa,Cai:2017wry,Angelescu:2018tyl,Lee:2021jdr,Shi:2019gxi,Desai:2023jxh,Bordone:2020lnb}) attracted attention in recent years  due to various B-physics anomalies.   In particular, the excesses in the ratios~\cite{BaBar:2012obs,Belle:2015qfa,Belle:2019rba,LHCb:2023zxo,LHCb:2023cjr}
\bea
R_X  = \frac{BR(B\to X_c \bar{\tau}\nu)}{BR(B\to X_c \bar{\ell}\nu)}
\eea
where  $X_c = D, D^*...$, could indicate a new charged-current four-fermion interaction involving $b$ and $c$ quarks, a $\tau$ and a  (tau) neutrino. The $S_1$ leptoquark can generate this interaction  with various Lorentz structures, while preserving lepton flavour (if $S_1$ only  interacts with $\tau$'s and $\nu_\tau$'s, one could attribute $\tau$ flavour to the leptoquark.) So fitting $R_D$ and/or $R_{D^*}$ involves  leptoquark  coupling constants that  differ from  the $\lambda_X^{e q}$, $\lambda_X^{\mu q}$ that are relevant for  $\me$  processes,
and in this manuscript,   we neglect the $\lambda_X^{\tau q}$  couplings  and possible   correlations  among $\me$ and $\tl$ processes that could arise in this model\footnote{It is clear that if the leptoquark has couplings $\lambda_X^{e q}$, $\lambda_X^{\mu q}$ and $\lambda_X^{\tau q}$, then it can mediate $\tau\to e$, $\tau\to \mu$ and  $\mu\to e$ processes. Furthermore,  if $|\lambda_X^{\tau q}|\sim 0.2$ to fit  the B-physics anomalies, then  experimental bounds $B_{(\tl)}$  on $\tl$ processes could constrain  $\lambda_X^{l q}< B_{(\tau \to l)}/\lambda_X^{\tau q} $. In all the cases we checked, we find $$  \frac{B_{(\tau \to e)}}{\lambda_X^{\tau q}}
 \frac{B_{(\tau \to \mu)}}{\lambda_X^{\tau q}} > B_{(\mu \to e)}
$$
that is, that $\me$ processes can probe
{smaller $\lambda_X^{l q}$ couplings}
than $\tl$ can exclude.}. The LFV predictions of  the $S_1$ leptoquark  have been discussed, for instance,  in  \cite{Crivellin:2020mjs,Bigaran:2020jil}.

\section{The observables coefficients  in terms of model parameters}
\label{sec:obs+invar}

In this section, we give the  coefficients of the observable  Lagrangian
(\ref{Lag1}) as a function  of model and SM parameters. Some of these results were already presented in \cite{Ardu:2023yyw}.    The coefficients are expressed in terms of $\mu\to e$ flavour-changing combinations of model and SM parameters, which we refer to as ``invariants''. These are a convenient stepping-stone  between the models and the observables, because they concisely identify the masses and couplings constants  that the observables depend on, and give the functional dependance. We briefly discuss the invariants in Section \ref{ssec:invar} and list the coefficients in Section \ref{ssec:CLex}.

\subsection{Comments on invariants}
\label{ssec:invar}

Invariants were  introduced \cite{Jarlskog:1985ht,Bernabeu:1986fc}
   as products of   Lagrangian coupling constants  (or matrices), in order to    have a Lagrangian-basis-independent measure of  symmetry breaking in a model. However,  it may be  unclear how these elegant constructions
   relate to observable  $S$-matrix elements  for  symmetry-violating processes, because  the mass and scale-dependence of  $S$-matrix elements can be  intricate.
   
For instance, the  original invariant constructed to measure CP violation in the quark sector of the Standard Model can  be written in terms of  Lagrangian Yukawa matrices as \cite{Jarlskog:1985ht,Bernabeu:1986fc,Branco:1999fs}
\bea
   \frac{1}{3} {\rm Tr} {\Big  \{} \left [  Y_uY_u^\dagger, Y_dY_d^\dagger \right]^3 {\Big \} }& =& 2i J
   (y_t^2 - y_c^2) (y_t^2 - y_u^2) (y_c^2 - y_u^2)
   (y_b^2 - y_s^2) (y_b^2 - y_d^2) (y_s^2 - y_d^2) ~~,~~
   \label{Jinvar}\\
{\rm with}~~   J&= &  {\rm Im}\{ V_{us}V_{cs}^* V_{cb}V_{ub}^*\}\nonumber
\eea
but it is unclear  at what scale to evaluate the quark masses (or equivalently, yukawa couplings).  This invariant   can be compared to  the parameter $\epsilon_K$  (see {\it eg}~\cite{burashouches} for a brief review) that contributes to CP-violation  in $K-\bar{K}$ mixing,  and  for which the result at  next-to or next-to-next-to leading log has been expressed in a rephasing invariant form  in  \cite{Brod:2019rzc}. Restricting to the leading log QCD corrections for simplicity,  one can write 
\bea
\epsilon_K &\propto&  J 
 {\Big(}
   {\rm Re}\{V_{td}V_{ts}^*V_{ud}^*V_{us}\}\eta^{tt} S(x_t)   + 2 |V_{ud}|^2|V_{us}|^2\eta^{ut} (S(x_c)  -S(x_c,x_t)){\Big)}
 \label{eK}
 \eea
 where $ x_Q = m_Q(m_Q)/m_W$,  the Inami-Lim functions\cite{Inami:1980fz} are
 \bea
S(x_t)& =& \frac{4x_t -11 x^2_t + x_t^3}{4(1-x_t)^2} - \frac{3x_t^2}{2(1-x_t)^3}\ln x_t \nonumber\\
 S(x_c)-S(x_c,x_t)& =& x_c\left(1-\ln \frac{x_t}{x_c}\right) + \frac{3x_cx_t}{4(1-x_t)} + \frac{3x_c x_t^2}{4(1-x_t)^2}\ln x_t ~~~, \nonumber
 \eea
and $\eta^{pq}$ are the appropriate QCD corrections for the Inami-Lim functions (see~\cite{Brod:2019rzc}).
So one sees that there is only a remote relation between  the  Lagrangian invariant of Eq. (\ref{Jinvar}), and Eq. (\ref{eK}), which allows a numerically precise prediction for the observable  $\epsilon_K$,

In the models considered here,   we obtain expressions for the observable coefficients at  $m_\mu$ (which are $\approx$ S-matrix elements)   in terms of  ``invariants'' , which  encode the  dependence of  LFV observables on running  SM and model coupling constants and masses.
 For instance, in the type II seesaw model,    photon penguin diagrams (contributing to  $\meee$) generate the coefficient $C_{V,LR}^{e\mu ee}$  proportional to  the invariant :
\bea
\frac{v^2}{M_\Delta^2}\, [m_\nu \ln \frac{[\tilde{m}_e \tilde{m}_e^\dagger]} {M_\Delta^2}\, m_\nu^\dagger]^{e \mu}\,
& =\, & \frac{v^2}{M_\Delta^2}\left( [m_\nu m_\nu^\dagger]^{e \mu}  \ln \frac{m_\mu^2} {M_\Delta^2} +       [m_\nu]^{e\tau} [m_\nu^*]^{ \mu \tau } \ln \frac {\tilde{m }_\tau^2} {m_\mu^2}+[m_\nu]^{ee} [m_\nu^*]^{ \mu e } \ln \frac {\tilde{m }_e^2}{m_\mu^2}\right) ~~~, \label{invar1}
       \eea
where
 $[\tilde{m}_e]$ is the charged lepton mass matrix,  but with its eigenvalues replaced  by  
$m_\a \to \tilde{m }_\a = \max \{m_\a,q^2\}$, which  is  the kinematic cutoff of the logarithm, with  $ 4m_e^2 \leq q^2 \lsim m_\mu^2$  the (four-momentum)$^2$ of the photon.
(So the cutoff of the $\tau$ loop is $m_\tau$,
but $\tilde{m}_e \approx m_\mu$  so the last term in the parentheses of Eq.~(\ref{invar1}) vanishes, or equivalently there is no long-distance contribution to the matrix element,  because $q^2 \sim m_\mu^2$ in most of the phase space~\cite{Marciano:1977wx}.)   
Eq. (\ref{invar1})  exemplifies our relatively simple invariants, constructed by multiplying matrices, which   encode the correct  scale evolution of  coupling constants and masses,
provided that the scale separations are large enough for EFT to be reliable\footnote{But notice the impact of phase space:  the tilde on $m_e$ is from kinematics that the Lagrangian does not know about, so the Lagrangian  invariant would include a ``large''
{$\ln \frac{m_\mu}{m_e}$}
that the coefficient does not contain because the large log is  only relevant in a small  phase space.}.

However,  our simple invariants  apply  only  for models  with  a single mass scale  for LFV;   for models  with  many heavy LFV  particles  around the scale $\LNP$,   such as  the inverse seesaw,  the operator coefficients  obtained in matching are  not linear products of matrices in flavour space (see $eg$  Eq.s \ref{CVLLinv}, \ref{CVLRinv}).
  Constructions that involve Inami-Lim functions  of mass matrix eigenvalues are no doubt also ``invariant'' under Lagrangian  basis transformations, but like Eq. (\ref{eK}), this invariance is not manifest.

  Our invariants have other   attractive features, beyond the correct scale-dependence to parametrise  $S$-matrix elements.
  As expected they measure  $\me$ flavour change in the model,  and they also  identify  the products of  model parameters relevant to observables.
  This second feature  allows to circumvent the necessity of scanning over model parameters.  For instance, in the  inverse seesaw model, $[Y_\nu]$ is a 3$\times$3  matrix of unknown complex numbers, so naively one must scan over them all, and possibly  impose some texture. However, in the case of  degenerate singlets,   there are only two  invariants ---  $[Y_\nu Y_\nu ^\dagger]_{e\mu}$  and  $[Y_\nu Y_\nu ^\dagger Y_\nu Y_\nu ^\dagger]_{e\mu}$ --- which are complex numbers of magnitude 0$\to$1. This raises the  question  whether one could reconstruct the model Lagrangian from a sufficient number of invariants?

Finally, the invariant of Eq. (\ref{invar1}) illustrates an interesting dynamical mechanism to break properties of the model. 
Recall that $ [m_\nu m_\nu^\dagger]^{e \mu}$  is a function only of   neutrino oscillation parameters--- which are measured ---  so the  second  term of the second expression exhibits  the  log-induced dependence on the  unknown Majorana phases  and neutrino mass scale. This logarithmic breaking of a relation between model-matrices is reminiscent of the ``log-GIM'' mechanism in the quark sector \cite{Gavela:1981sk}, where $\Delta F = 1$ FCNC operators can be mediated by similar penguin diagrams  at ${\cal O}(G_F \a_e\ln m_W/m_c)$, with a charm quark  in the loop.\footnote{ In the lepton sector below the weak scale, the  (quadratic) GIM  mechanism suppresses LFV amplitudes $\propto m^2_\nu/m_W^2$, so implementing log-GIM  would be very interesting ($\ln (m_\nu/m_W)\gg m^2_\nu/m_W^2$). However, this appears difficult  \cite{Hernandez-Tome:2018fbq,Blackstone:2019njl},  because there are no ``penguin diagrams'' in the SM with a massless gauge boson attached to a neutrino loop, and because the logarithm  in log-GIM is cutoff by the largest of the  mass in the loop, or  the external momentum traversing the loop, and this latter is of order charged lepton masses in LFV processes.
}

\subsection{Predictions at the experimental scale}
\label{ssec:CLex}
  
This section lists the model predictions for the observable coefficients of the Lagrangian (\ref{Lag1}),  partially presented in Ref. \cite{Ardu:2023yyw}. The model parameters  are given in the Lagrangians of Section  \ref{sec:models}.  The expressions given here occasionally differ from \cite{Ardu:2023yyw}, because we  made numerically insignificant  modifications of  the lower cutoff of some logarithms, in order to obtain more elegant invariants with the physically correct cutoffs. 

\subsubsection{$\meg$}
\label{sssec:meg}

In the   inverse   seesaw model,  the dipole coefficients are
\bea
C^{e\mu}_{D,R}&\simeq &- \frac{e}{32 \pi^2 } [Y_\nu M^{-1} (M^\dagger)^{-1}Y_\nu^\dagger]^{e\mu} v^2  {\left(1- 16 \frac{\alpha_e}{4\pi} \ln\frac{M}{m_\mu}\right)}~~,
\label{meginv}
\eea
where the  parentheses include  the  flavour-universal ${\cal O}(10\%)$ QED running,  and $M$ is the singlet mass scale. 
The couplings $Y^{\a a}_\nu$ can be of order one --- which could be  especially motivated in the $\tau$ sector---so ${\cal O}(Y^4)$  combinations can be  larger than  ${\cal O}(Y^2)$, and could
appear at dimension eight when two additional Higgs legs on the sterile neutrino line are replaced by the Higgs vev, or at two-loop when the Higgs legs are closed.
We estimate the $\mathcal{O}(Y^4)$ terms to be suppressed with respect to the coefficient in Eq.~\ref{meginv} by $\sim v^2/M^2$ or $\sim 1/(16\pi^2)$, {and we therefore expect them to not modify significantly the correlations between the $\mu\to e$ observables that the model can predict \cite{Ardu:2023yyw}.}

For the type II seesaw, 
{
\bea
C^{e\mu}_{D,R} &\simeq &  \frac{3e}{128 \pi^2}  {\Big [}\frac{[m_\nu m_\nu^\dagger]_{e\mu}}{\lambda^2_H v^2}\left( 1-16 \frac{\a_e}{4\pi}\ln\frac{M_\Delta}{m_\mu} \right) +
  \frac{ \alpha_e}{\pi }  \frac{116}{27}\left( \ln\frac{M_\Delta}{m_\mu}
 \frac{[m_\nu m_\nu^\dagger]_{e\mu}}{\lambda_H^2 v^2 }
-  \ln\frac{m_\tau}{m_\mu} \frac{[m_\nu^*]_{\mu \tau} [m_\nu]_{e\tau} }{\lambda_H^2 v^2 }\right){\Big]}
  \label{CDt2}\\
& \simeq &
  \frac{3e}{128 \pi^2}  {\Big [}\frac{[m_\nu m_\nu^\dagger]_{e\mu}}{\lambda^2_H v^2}
  {\Big (}1  + \frac{32}{27}  \frac{\alpha_e}{4\pi} \ln\frac{M_\Delta}{m_\tau} 
    {\Big )}   +   
  \frac{116 \alpha_e}{27\pi }  \ln\frac{m_\tau}{m_\mu} \sum_{\a \in e\mu} \frac{[m^*_\nu]_{\mu \a} [m_\nu]_{e\a} }{\lambda_H^2 v^2 }{\Big ]}
\label{megt2}
\eea
}
where  the second term in the  bracket in Eq. (\ref{CDt2}) arises from  the two-loop vector to dipole mixing \cite{Ciuchini:1993fk}.
In both seesaw  models, the coefficient $C^{e\mu}_{D,L}$ is suppressed by a factor $m_e/m_\mu$, and can be obtained from Eq.s (\ref{megt2},\ref{meginv})  by multiplying by $m_e/m_\mu$.

 For the leptoquark, which interacts with singlet and doublet leptons, the dipole coefficients are
\bea
 \frac{ m_{LQ}^2}{ v^2}C^{e\mu}_{D,X} (m_\mu) &\simeq& 
  \frac{ e  [\lambda_Y\lambda^\dagger_Y]^{e\mu} }{128\pi^2}
\left(1- 16 \frac{\alpha_e}{4\pi} \ln\frac{m_{LQ}}{m_\mu}\right)
+ \frac{ 2 \alpha^2_e }{9 \pi^2 e} 
\left[ \lambda_Y \ln\frac{m_{LQ}}{\tilde{m}_Q}  \lambda^\dagger_Y \right]^{e\mu}  
 \nonumber \\&&
 -\frac{ \alpha_e  }{2\pi e y_\mu} f_{TD}
 \left[ \lambda_Y Y_u \left(\eta^{a_T}_{\tilde{m}_Q}  \ln\frac{m_{LQ}}{\tilde{m}_Q} - \frac{5}{4} \right) \lambda^\dagger_X \right]^{e\mu}
 ~~~~~~~~~~~~~
 \label{megLQ}
 \eea
 where $X\neq Y\in\{L,R\}$,
 $f_{TD}$, $a_T$  and $\eta$ are related to the QCD running and defined at Eq. \ref{soln1tG}, and the   $\tilde{m}_Q$  serving as lower cutoff for the logarithms  (here and further in the manuscript) is
 $$
\tilde{m}_Q = {\rm max} \{ m_Q(m_Q), 2~{\rm GeV}\}$$ because the quarks are matched to nucleons at 2 GeV.  The first term   in Eq. (\ref{megLQ})  is the matching contribution (times  its QED running),the second term is the 2-loop  mixing of tree vector operators into the dipole, and the  third term is the  one loop  mixing of   tensor operators to dipoles. 

 {The last term  of Eq. (\ref{megLQ}) requires some discussion, because
   the  first  log-enhanced term  in the parentheses arises in the RGEs between $m_{LQ}\to m_\m$,    but the  second  ``finite'', or  not-log-enhanced  term 
   is formally of higher order in EFT.  It is included because it is of comparable magnitude  to the log-enhanced term in the case of   internal top quarks --- that is, as discussed in Appendix \ref{assec:quelEFT},  the scale ratio $m_{LQ}/m_t$ is not large, so our matching conditions at $\LNP$ are constructed to reproduce
the results of  matching  to a QCD$\times$QED invariant EFT at $m_W$. And finally, although the finite part is only required  for the top quark, a quark-flavour-summed expression  is given in order to retain the ``invariant'' formulation. (The light quark contributions are numerically negligible in this expression,  which is fortunate because their QCD running is also not correct.)}

  We do not consider dimension 8 contributions to the dipole, because  they are suppressed  $\propto v^2/M^2$, and   do not allow to circumvent the parametric suppression of the dimension six term. For instance,  in the inverse seesaw,  such terms also have  the loop and $y_\mu$ suppression applying to (\ref{meginv}).

\subsubsection{$\meee$}
\label{sssec:meee}

The decay $\meee$ can be mediated at the experimental scale by the dipole operators, and vector and scalar four-lepton  operators (see Eq. (\ref{BRmeee}).
We do not  give results for the scalar coefficients, because they are effectively vanishing: in matching, all three models induce  coefficients  that are  smaller than the upcoming experimental sensitivity,  and  SM interactions that could  transform some other LFV operator into a scalar are suppressed by lepton Yukawas, so negligible as well. Unfortunately, scalar coefficients $C_{S,XX}$ are indistinguishable from vectors $C_{V,YY}$ in the angular distribution of $\meee$, so the absence of scalar coefficients in these models  would be  challenging to test.

For the type II seesaw, the vector four-lepton coefficients arise at tree
level, {as illustrated in the middle  diagram of Figure~\ref{fig:t2match},}  and in the RGEs via QED penguin  diagrams:
\bea
C^{e\mu ee}_{V,LL}& 
\simeq &\frac{[m^*_{\nu}]_{\mu e}[m_{\nu}]_{e e}}{2 \lambda_H^2v^2} + { \frac{\alpha_e}{{ 3}\pi\lambda_H^2v^2}
{\Big[} m^\dagger_{\nu} \ln \left(\frac{M_\Delta}{\tilde{m}_\a}\right)  m_{\nu} {\Big]}_{\mu e }} 
\label{CVLLt2} \\
C^{e\mu ee}_{V,LR}& \simeq & { \frac{\alpha_e}{ 3\pi\lambda_H^2v^2}
{\Big[} m^\dagger_{\nu} \ln \left(\frac{M_\Delta}{\tilde{m}_\a}\right)  m_{\nu} {\Big]}_{\mu e }}
\label{pingt2} ~~~.
\eea
where $\a$ is the  index of the intermediate charged lepton.
The operators ${\cal O}^{e\mu ff }_{V,RR}$ and  ${\cal O}^{e\mu ff }_{V,RL}$, where the flavour-change is among singlet leptons, have  coefficients below upcoming experimental sensitivities because they are suppressed $\propto y_e y_\mu$.  This is also the case for the inverse seesaw.  

In the case of the inverse seesaw, both vector operators arise via loop diagrams,  with propagating singlets and Higgses: 
\bea
C^{e\mu ee}_{V,LL}
&\simeq &
v^2\frac{\alpha_e}{4\pi}\bigg(
-1.8[Y_\nu M_a^{-2}\left(\frac{11}{6}+\ln\left(\frac{m^2_W}{M_a^2}\right)\right)
  Y_\nu^\dagger]_{e\mu}
+ 2.7 [Y_{\nu}(Y^\dagger_\nu Y_\nu)_{ab}\frac{1}{M^2_{a}-M^2_{b}}
  \ln\left(\frac{M^2_{a}}{M^2_{b}}\right)Y_\nu^\dagger]_{e\mu}
\nonumber\\
&&
+2.5 Y^{e a}_\nu Y^{*\mu a}_\nu Y^{e b}_\nu Y^{*e b}_\nu\frac{1}{M^2_{a}-M^2_{b}}\ln\left(\frac{M^2_{a}}{M^2_{b}}\right)+\mathcal{O}\left(\frac{\alpha_e}{4\pi}\right)\bigg)\label{CVLLinv} \\
&\to& 
	\frac{v^2}{M^2}\left(3.3\times 10^{-3}(Y_\nu Y^\dagger_\nu)_{e\mu}(1+0.56 (Y_\nu Y^\dagger_\nu)_{ee})+1.55\times 10^{-3}(Y_\nu Y^\dagger_\nu Y_\nu Y^\dagger_\nu)_{e\mu}\right) \label{CVLLinvdeg}\\
C^{e\mu ee}_{V,LR}&\simeq & v^2\frac{\alpha_e}{4\pi}
\bigg( 1.5[Y_\nu M_a^{-2}
  \left( \frac{11}{6} +\ln \left(\frac{m^2_W}{M_a^2}\right) \right)
  Y_\nu^\dagger]_{e\mu}
-2.7 [Y_{\nu}(Y^\dagger_\nu Y_\nu)_{ab}
  \frac{1}{M^2_{a}-M^2_{b}} \ln \left(\frac{M^2_{a}}{M^2_{b}}\right)
  Y_\nu^\dagger]_{e\mu} \nonumber\\& &
+\mathcal{O}\left(\frac{\alpha_e}{4\pi}\right)\bigg)\label{CVLRinv}\\
&\to& 	\frac{v^2}{M^2}\left(-2.8\times 10^{-3}(Y_\nu Y^\dagger_\nu)_{e\mu}-1.6\times 10^{-3}(Y_\nu Y^\dagger_\nu Y_\nu Y^\dagger_\nu)_{e\mu}\right) \label{CVLRinvdeg}
\eea
where  the first expression  for a coefficient  is for  arbitrary  singlet
masses  $\gsim$ TeV , and after the arrow is the simplified formula when the singlets  mass$^2$ differences are less than $v^2$  \cite{Ardu:2023yyw}.
In the first expression, the first two terms arise from  $Z$ and $\gamma$ penguins  above the electroweak scale (the Higgs propagates in the loop), and the last one is from boxes.

Finally  the leptoquark  can  generate flavour-changing lepton currrents  involving either   singlet or doublet leptons. We give here the coefficient  for left-handed leptons ; the result for singlets is obtained by interchanging $L\leftrightarrow R$:
\bea
 \frac{ m_{LQ}^2}{ v^2} C_{V,LX}^{e\mu ee} (m_\mu) & \simeq &
-\frac{N_c }{64\pi^2} 
[\lambda_L \lambda_L^\dagger]^{e \mu} [\lambda_X \lambda_X^\dagger]^{ee}
     \left( 1\mp 12 \frac{\a_e}{4\pi} \ln\frac{m_{LQ}}{m_\mu} \right) 
{ +} \frac{\a_e}{3\pi} \left[ \lambda_L  \ln\frac{m_{LQ}}{m_Q}   \lambda_L^\dagger \right]^{e\mu}
\label{CVLLeeLQ}\\
&&
- g^e_X \frac{N_c }{16\pi^2} 
 \left[ \lambda_L Y_u \left(\ln \frac{m_{LQ}}{m_Q}  -\frac{5}{6} \right)Y_u^\dagger   \lambda^\dagger_L \right]^{e\mu}
 \nonumber
 \eea
where  $g^e_L= -1+2\sin^2\theta_W$, $g^e_R=2\sin ^2\theta_W$,
the first  term represents   the box  diagram at $m_{LQ}$  (and its QED running to $m_\mu$, with $-$/$+$ for $X=$/$\neq L$),
the second term is  the  log-enhanced  photon penguin  that mixes  the 
tree operators  ${\cal O}_{VLL}^{QQ}$ (for $Q\in \{u,c,t\}$)
into 4-lepton operators, 
and the last term   is  the contribution of the $Z$-penguins
without  the  negligible QED running. {Since  we only retain the part of the  $Z$-penguin  that is proportional to $y_Q^2$(see Appendix \ref{assec:ping}),  the dominant contribution  arises from the top quark,  where the ``finite'' (not log-enhanced) part of the  diagrams is included because the logarithm is not large (see the discussion  after Eq. (\ref{megLQ}) or in Appendix \ref{assec:quelEFT}).     }

\subsubsection{$\mu {\rm Al}\to e {\rm Al}$} 
\label{sssec:mucAl}

The SINDRUMII experiment  searched for $\muc$ on Titanium ($Z = 22$)  and Gold ($Z= 79$), setting the bounds listed  in Table~\ref{tab:bds}.  Upcoming experiments will start with an Aluminium target,  which  probes a similar combination of coefficients as Titanium in the analysis of \cite{KKO}. So  this section  gives expressions for the coefficient  on  Aluminium $C_{Al,X}$,
expressed in the quark operator basis, where  the   conversion ratio on Aluminium is given in Eq. (\ref{dA}).

The photon penguin diagrams in the  type II seesaw model generate a vector $\me$ operator on  $u$ and $d$ quarks,  giving 
\bea
{C}^{e\mu}_{Al,L}&\simeq &-\frac{\alpha_e}{72\pi\lambda_H^2v^2}
{\Big[} m^\dagger_{\nu} \ln \left(\frac{M_\Delta}{\tilde{m}_\a}\right)  m_{\nu} {\Big]}_{\mu e }~~~,
\label{ping2t2}
\eea
where  $\tilde{m}_\tau = m_\tau$ and $\tilde{m}_\a = m_\mu$ for $\a\in\{e,\mu\}$ because the logarithm is cut off by the momentum transfer from the leptons to the nucleus\cite{Raidal:1997hq}, which  is of order $m_\mu$.
Notice that   the penguin  diagram of  Figure \ref{fig:t2match} generates a 2-lepton-2-quark operator at scales above  2 GeV, where quarks are matched to nucleons,  then  it continues to mix into  a 2-lepton-2-proton operator, which is why the  logarithm cuts off at $m_\mu$.

The inverse seesaw  generates vector operators,  via  $Z$ and $\g$ penguins above the weak scale:
\bea
{C}^{e\mu}_{Al,L}&\simeq & v^2\frac{\alpha_e}{4\pi}\bigg(-0.05[Y_\nu M_a^{-2}\left(\frac{11}{6}+\ln\left(\frac{m^2_W}{M_a^2}\right)\right)Y_\nu^\dagger]_{e\mu} +0.09 [Y_{\nu}(Y^\dagger_\nu Y_\nu)_{ab}\frac{1}{M^2_{a}-M^2_{b}}\ln\left(\frac{M^2_{a}}{M^2_{b}}\right)Y_\nu^\dagger]_{e\mu}~~~~\label{CAlinv}\\
&+&\mathcal{O}\left(\frac{\alpha_e}{4\pi}\right)\bigg)\nonumber\\
&\to&\frac{v^2}{M^2}\left(8.6\times  10^{-5} (Y_\nu Y^\dagger_\nu)_{e\mu}+5.6\times 10^{-5}(Y_\nu Y^\dagger_\nu Y_\nu Y^\dagger_\nu)_{e\mu}\right) \label{CAlinvdeg} 
\eea
For the leptoquark, which induces scalar and vector $2l2q$ operators,   we obtain
\bea
{C}^{e\mu}_{Al,L}&\simeq &
{\Big (}0.032  \lambda^{e u}_L \lambda^{\mu u *}_L .
\left(1{ +} \frac{{ 2}\alpha}{\pi} \ln \frac{m_{LQ}}{2{\rm GeV}}\right)
+0.033\left(
\frac{g^2}{64\pi^2} \lambda^{e u}_L \lambda^{\mu u *}_L  \ln\frac{m_{LQ}}{m_W}\right)   \nonumber\\&&
- 0.086\frac{\alpha}{6 \pi}
  \left[\lambda_L \ln \frac{m_{LQ}}{\tilde{m}_Q}    \lambda^\dagger_L \right]^{e\mu}  
   -4.1\times 10^{-4}
\left[ \lambda_LY_u  \left(\ln \frac{m_{LQ}}{\tilde{m}_Q}- \frac{5}{6}\right)  Y_u^\dagger  \lambda^\dagger_L \right]^{e\mu} 
\label{mecTi1} \\
    && -\eta^{a_S}
 \left( 0.16  \lambda^{e u}_L \lambda^{\mu u *}_R 
+ \frac{0.035 m_N}{27m_c}  \lambda^{e c}_L \lambda^{\mu c *}_R 
\right) ( 1+  \frac{13 \alpha}{ 6\pi}  \ln \frac{m_{LQ}}{2{\rm GeV}} )
+  \frac{0.094m_N}{27m_t}   \lambda^{e t}_L \lambda^{\mu t *}_R {\Big)}\times \frac{v^2}{m_{LQ}^2}\ , \nonumber
\eea
where  are included 
the tree vector coefficient on $u$ quarks with its QED running,
the electroweak box  contribution to the $d$ vector,
the QED  then  $Z$ penguin  contributions to the $u$ and $d$ vectors (where we  took  $V_{ud} \simeq 1, \sin^2 \theta_W \simeq 1/4$),  and the scalar $u$, $c$ and $t$ contributions { (where the QCD running of the top contribution is negligible). As in Eq. \ref{CVLLeeLQ},    the $Z$-penguin  contribution  only  includes the diagrams $\propto y_Q^2$, with their ``finite parts''. }

\subsubsection{$\muc$ on heavy targets}
\label{sssec:mucAu}

Changing the target in $\muc$ allows to probe a different combination of operator coefficients \cite{KKO}. This is discussed quantitatively in Appendix \ref{app:ops}, where  Eq.s (\ref{uAl}, \ref{uTi}, \ref{uAu})  give the operators probed by light and heavy targets in the quark basis.
The SINDRUM experiment searched for $ \muc$ on Gold (see Table~\ref{tab:bds}), and   there are plans based on the proposal  of Ref.  \cite{Kuno:2005mm} to build experiments that could  probe $\muc$  on heavy targets (see Table~\ref{tab:bds}). However,  we consider these experiments to be to far in the future  for the purposes of our study,  so we suppose that the  data for  Gold remains the bound of SINDRUM given in Table~\ref{tab:bds}. 
The operator probed by Gold can be decomposed into the operator  probed by light targets, plus the remaining part, approximately given in Eq (\ref{OAheavyperp}).
In this section, we discuss  the coefficient of this  orthogonal part, which can be written  as
\bea
C_{Au\perp,L}& \simeq & 
-  {\Big(} 0.2 C^{e\mu uu}_{SR} +0.1 C^{e\mu dd}_{SR} +0.0075 C^{e\mu ss}_{SR} +  0.001C^{e\mu cc}_{SR}+ 0.0003 C^{e\mu bb }_{SR}
{\Big)}\nonumber\\
&&+ 0.56  C^{e\mu uu}_{VL}+   0.7956   C^{e\mu dd}_{VL} ~~~. 
\eea

The type II seesaw only generated a vector operator on protons, (no scalar operators, and no vector operator on neutrons),  so  once the coefficient is measured on a first target, it  can be predicted  on any other. That is, Gold probes approximately the same four-fermion  operator as  light targets, and since the dipole and proton vector coefficients  are weighted by approximately 1/4 and 1/2 in both the  amplitudes on Gold and Aluminium, the ratio of the rates
is 
$$
\frac{BR(\mu Au \to eAu)}{BR(\mu Al \to eAl)}
\approx\frac{\widetilde{B}_{Au}}{\widetilde{B}_{Al}} \approx 2 \pm {\cal O}(10\%)~~,
$$
where $\widetilde{B}_{Al}$ and $\widetilde{B}_{Au}$ are given after Eq.~(\ref{tildeB}).

In the inverse seesaw, $\muc$ on heavy targets could give  complementary information, because the $Z$ penguin  contribution   generates vector $\me$  operators on  both protons and neutrons:
\bea
{C}^{e\mu}_{Au\perp,L}&\simeq & v^2\frac{\alpha_e}{4\pi}\bigg(-0.5[Y_\nu M_a^{-2}\left(\frac{11}{6}+\ln\left(\frac{m^2_W}{M_a^2}\right)\right)Y_\nu^\dagger]_{e\mu} +0.8[Y_{\nu}(Y^\dagger_\nu Y_\nu)_{ab}\frac{1}{M^2_{a}-M^2_{b}}\ln\left(\frac{M^2_{a}}{M^2_{b}}\right)Y_\nu^\dagger]_{e\mu}~~~~\label{CAuinv}\\
&+&\mathcal{O}\left(\frac{\alpha_e}{4\pi}\right)\bigg)\nonumber\\
&\to&\frac{v^2}{M^2}\left(1.1\times  10^{-4} (Y_\nu Y^\dagger_\nu)_{e\mu}+4.9\times 10^{-4}(Y_\nu Y^\dagger_\nu Y_\nu Y^\dagger_\nu)_{e\mu}\right) \label{CAuinvdeg} 
\eea
The coefficient is nonetheless a combination of the same invariants that feature in the other $\mu\to e$ operators, and so can be predicted with a combination of $\mu\to e$ observations. For instance, in the nearly degenerate limit, it can be written as a linear combination of the light targets coefficient and the dipole.

In the leptoquark  model, we   neglect the scalar coefficients on $d,s$ and $b$  quarks, because the model does not generate scalar operators with down-type quarks at tree level, and the estimates of  Appendix \ref{ssec:LQmatchsum} suggest that the loop-induced coefficients are below experimental sensitivity. So   we obtain
\bea
C_{Au\perp,L} &\simeq &
\left(\eta^{a_S} 
\left( 0.1  \frac{\lambda^{e u}_L \lambda^{\mu u *}_R  }{2}
+ 0.001 \frac{\lambda^{e c}_L \lambda^{\mu c *}_R  }{2} \right) \left(1+ \frac{13 \alpha}{ 6\pi} \ln\frac{m_{LQ}}{2{\rm GeV}}
\right) \right.
\nonumber\\&&
+0.14 \lambda^{e u}_L \lambda^{\mu u *}_L (1+   \frac{2\a}{\pi} \ln\frac{m_{LQ}}{2{\rm GeV}} )
\nonumber\\ &&
+ 0.4 \frac {g^2}{32\pi^2 } [\lambda_L V]^{ed}
[V^\dagger \lambda^\dagger_L]^{d\mu}    
-0.32    \frac{\a }{9\pi  }
 [\lambda_L\log \frac{m_{LQ}}{m_Q} \lambda^\dagger_L ]^{e\mu}  ~~~~~~~~~ 
\nonumber\\ && \left.
- \frac{ N_c   }{16\pi^2 } {\Big (}0.28 (g_L^u+  g_R^u) + 0.4 (g_L^d+  g_R^d)  {\Big )}  [\lambda_L Y_u  \left(\log \frac{m_{LQ}}{m_Q} - \frac{5}{6}\right) Y_u^\dagger \lambda^\dagger_L]^{e\mu} \right)\times \frac{v^2}{m_{LQ}^2}  \nonumber
    \nonumber
\eea
where the terms, in order, are the scalar up  and charm quark contribution with their QED and QCD running, the tree vector up quark contribution with its QED running, the leptoquark-$W$ box matching onto the vector  operator for down quarks (neglecting the QED running),  then the QED penguin  contribution to  the vector coefficients for $u$ and $d$ quarks, and finally  the $Z$ penguin contributions to both $u$ and $d$ vector coefficients.

{We claim that in this leptoquark model,  ${C}_{Au\perp,X} $ is  independent  from $C_{Al,X} $, implying that   $C_{Au\perp,X} $ could be just below the current SINDRUMII bound  (see Table~\ref{tab:2}) even if $\me$ flavour change is not observed in upcoming experiments.
Furthermore, we anticipate that this  will remain true, even if the definition of $ {\cal O}_{Au\perp} $ changes  as theoretical calculations are updated
(the  definition of the  $ {\cal O}_{A} $s in  References \cite{Haxton:2022piv,KKO} appears different.).   The point  is that  cancellations can occur, for instance among vector and scalar coefficients in the coefficient on Aluminium, allowing the coefficient on Gold  to be relatively large. 
}

\section{Phenomenology}
 \label{sec:pheno}

 The type II seesaw model is reputed to be predictive, because the  lepton flavour-changing  couplings of the scalar boson  $\Delta$ are proportional to the observed neutrino mass matrix. However, we observed in \cite{Ardu:2023yyw} that  knowing  the neutrino oscillation parameters  does not predict    the  observable $\me$  coefficients. So  Sections \ref{ssec:mnu} and \ref{ssec:tauLFV} explore the  connections  between   $\me$ processes and  other   lepton flavour- and number-changing observables   in the type II seesaw. The remaining two subsections respectively  discuss how $\meg$ constraints suppress dimension eight operators in the leptoquark model (Section \ref{ssec:tLFV}),  and   the impact  of   allowing  operator coefficients to be complex(Section \ref{ssec:C}).

\subsection{The neutrino mass scale in the type II seesaw}
\label{ssec:mnu}

In this section, we explore how the predictions of the type II seesaw model for $\mu\to e$ observables are influenced by the lightest neutrino mass scale, denoted as $m_{\rm min}.$

The lightest neutrino mass plays a crucial role in determining the relevant coefficients in the model, namely $C^{e\mu}_{D,R}$, $C^{e\mu ee}_{V,LL}$, and $C^{e\mu ee}_{V,LR}\propto C_{Al,L},\ C_{Au\perp,L}$. The tau mass cut-off in the two-loop vector mixing and one-loop photon penguin introduces a term $\propto [m^*_\nu]_{\mu\tau} [m_\nu]_{e\tau}$
(see Eq. \ref{CDt2} and \ref{pingt2}) that gives rise to the dipole and $C^{e\mu ee}_{V,LR}$ dependence on the unknown neutrino parameters. Additionally, the $\mu\to 3e_L$ vector depends on $m_{\rm min}$ from both the tree-level $\propto[m_{\nu}]_{ee}[m^*_{\nu}]_{\mu e}$ and the photon penguin contributions.
 
 The magnitude of these unknown terms increases with the lightest neutrino mass, making them more relevant when $m_{\rm min}$ is large. For example, if we allow $m_{\rm min}\sim 0.2$ eV, the two-loop vector-to-dipole mixing can reach the size of the one-loop matching contribution to the dipole, and certain choices of Majorana phases could allow for a cancellation that suppresses the dipole coefficient. Although this has not been pointed out in the literature before (the dipole cannot vanish in the type II seesaw at the leading order), the cancellation requires a high neutrino mass scale $m_{\rm min}$. Values $m_{\rm min}\sim 0.2$ eV are compatible with the laboratory constraint extracted from the tritium decay end-point, which yields $\sqrt{\sum_i m^2_i |U_{ei}|^2}< 0.8$ eV ($90\%$ CL)~\cite{KATRIN:2021uub}, but are disfavored by the cosmological bounds on the neutrino masses sum $\sum m_i$. Assuming $\Lambda$CDM, CMB data constrain the sum to be $\sum m_i<0.26$ eV ($95\%$ CL), while combined with the BAO measurements the constraint is stricter $\sum m_i<0.12$ eV ($95\%$ CL) \cite{Planck:2018vyg}. 
 \begin{figure}[ht]
  \begin{subfigure}{.4\textwidth}
 	\begin{center}
 		\includegraphics[width=\linewidth]{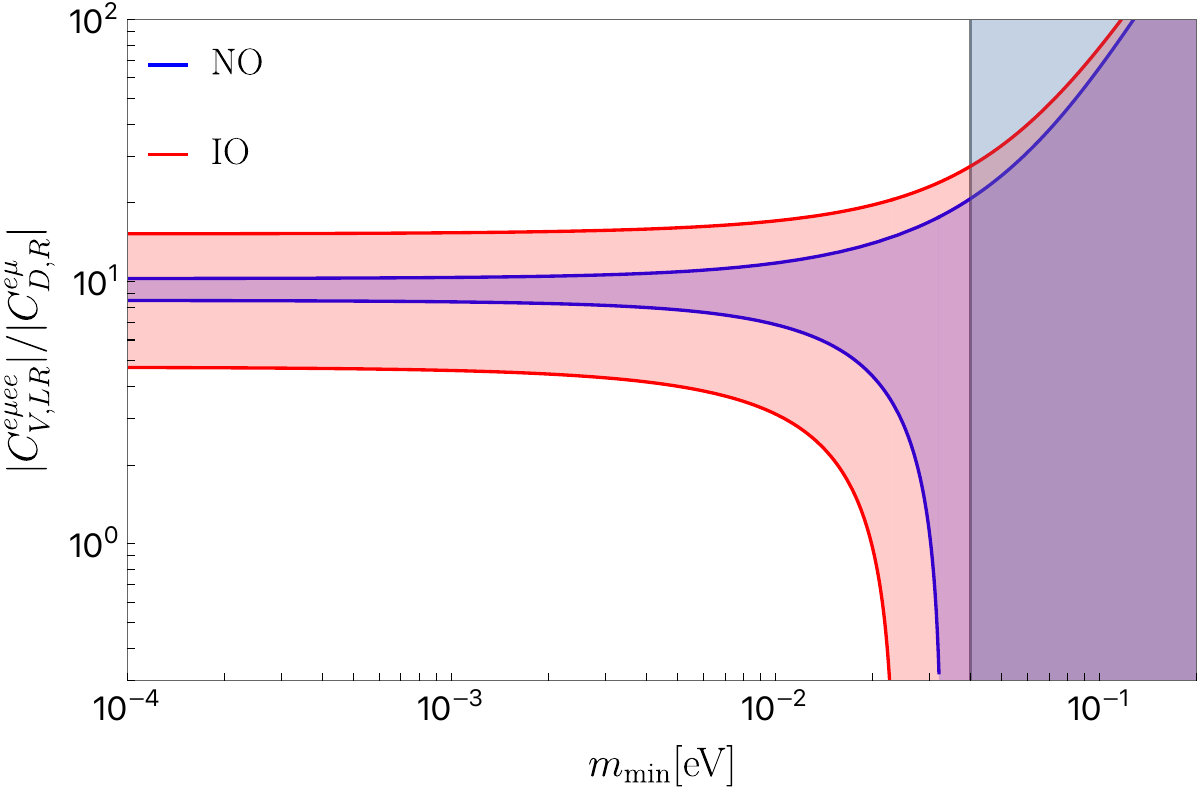}
 	\end{center}
 	\caption{\label{fig:CVLRmmin}} %
 \end{subfigure}
  \begin{subfigure}{.4\textwidth}
 	\begin{center}
 		\includegraphics[width=\linewidth]{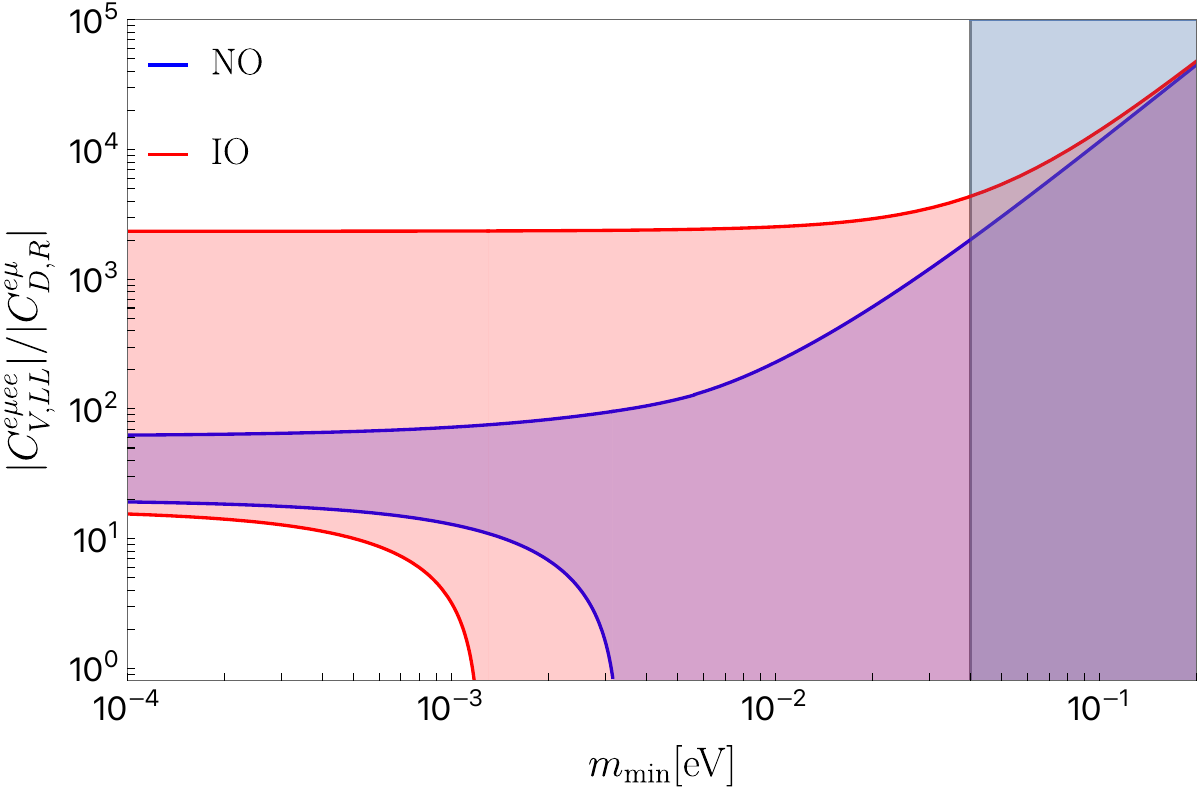}
 	\end{center}
 	\caption{\label{fig:CVLLmmin}} %
 \end{subfigure}
 \caption{$\frac{|C^{e\mu ee}_{V,LR}|}{|C^{e\mu}_{D,R}|}$ and $\frac{|C^{e\mu ee}_{V,LL}|}{|C^{e\mu}_{D,R}|}$ as  functions of $m_{\rm min}$ for normal and inverted ordering in blue and red respectively, for all possible values of the Majorana phases. We consider only the one-loop matching contribution to the dipole since for $m_{\rm min}\lesssim 0.2$ eV the two-loop vector to dipole mixing is negligible. The shaded blue region corresponds to the values of $m_{\rm min}$ above the cosmology preferred upper bound $m_{\rm min}\lesssim 0.04$ eV.  }
  \end{figure}
  
  By imposing the cosmological upper bound $m_{\rm min}\lesssim 0.04$ eV, the $\mu\to e$ coefficients are also constrained. In this case, the dipole is approximately unaffected by the two-loop contribution and is completely determined from the neutrino oscillation parameters, apart from the overall LFV scale. When the dipole coefficient is non-vanishing, the ratios
   \begin{equation}
  	\frac{|C^{e\mu ee}_{V,LL}|}{|C^{e\mu}_{D,R}|},\qquad \frac{|C^{e\mu ee}_{V,LR}|}{|C^{e\mu}_{D,R}|},
  \end{equation} 
  are finite for any value of the Majorana phases. Figures \ref{fig:CVLRmmin} and \ref{fig:CVLLmmin} illustrate that by imposing the upper-bound $m_{\rm min}\lesssim 0.04$ eV and allowing the Majorana phases to vary freely, these ratios are bounded from above
 \begin{equation}
 	m_{\rm min}\leq 0.04\ {\rm eV}\quad\to\quad \frac{|C^{e\mu ee}_{V,LL}|}{|C^{e\mu}_{D,R}|}\leq \begin{cases}
 		  4.3\times 10^{3}\ &{\rm (IO)} \\
 		   2\times 10^{3}\ &{\rm (NO)} \\
 		\end{cases},\qquad \frac{|C^{e\mu ee}_{V,LR}|}{|C^{e\mu}_{D,R}|}\leq \begin{cases}
 		28\ &{\rm (IO)} \\
 		21\ &{\rm (NO)} \\
 		\end{cases}
 \end{equation}
The cosmological bound is generally insufficient to constrain the ratios from below, as there exist $m_{\rm min}$ values for both $\mu\to 3e$ vectors such that the operator coefficient can vanish.  If $m_{\rm min}$ is measured, for instance by observing the neutrinoless double beta decay, additional information can be obtained for the $\mu\to e$ coefficients. The penguin cannot vanish for $m_{\rm min}\lesssim 0.02$ eV regardless of the ordering, while the $\mu\to 3e_L$ vector is also non-vanishing for $m_{\rm min}\lesssim 10^{-3}$ eV:
\begin{equation}
	m_{\rm min}\lesssim 0.02\ {\rm eV}\qquad \to \quad \begin{cases}
		4.7\ &{\rm (IO)} \\
		8.5\ &{\rm (NO)} \\
	\end{cases} \leq  \frac{|C^{e\mu ee}_{V,LR}|}{|C^{e\mu}_{D,R}|}\leq \begin{cases}
		15\ &{\rm (IO)} \\
		10\ &{\rm (NO)} \\
	\end{cases}\nonumber
\end{equation}
\begin{equation}
	m_{\rm min}\lesssim 10^{-3}\ {\rm eV}\qquad \to \quad \begin{cases}
		17\ &{\rm (IO)} \\
		20\ &{\rm (NO)} \\
	\end{cases} \leq  \frac{|C^{e\mu ee}_{V,LL}|}{|C^{e\mu}_{D,R}|}\leq \begin{cases}
		2.3 \times 10^{3}\ &{\rm (IO)} \\
		62\ &{\rm (NO)} \\
	\end{cases}\label{eq:mmin0typeII}
\end{equation}
Therefore, if the ratio of coefficients $\frac{|C^{e\mu ee}_{V,LR}|}{|C^{e\mu}_{D,R}|}$ or $\frac{|C^{e\mu ee}_{V,LL}|}{|C^{e\mu}_{D,R}|}$ were observed outside the ranges of Eq.~(\ref{eq:mmin0typeII}), it would be possible to infer a lower bound on the neutrino mass scale 
{(if} {it originates from the type II seesaw mechanism)}. Since these ranges are significantly narrower in the NO case, measuring the ordering (which is expected to be determined in the upcoming years) would be particularly useful to pinpoint the interplay between the lightest neutrino mass and the $\mu\to e$ predictions in the type II seesaw.
{Furthermore, if the mass ordering is normal, future beta decay experiments might be able to constrain
$m_{\rm min} \lesssim 0.02\, \mbox{eV}$ (the Project 8 experiment~\cite{Project8:2022wqh} aims at a 90\% C. L. sensitivity
of $40\, \mbox{meV}$ on the effective neutrino mass in beta decay, which corresponds to $m_{\rm min} \simeq 0.04\, \mbox{eV}$).
In this case, measuring the ratio of coefficients $\frac{|C^{e\mu ee}_{V,LR}|}{|C^{e\mu}_{D,R}|}$ outside the range
$[8.5, 10]$ would exclude the type II seesaw model.}

\subsection{ $\tau$ LFV in  the type II seesaw}
\label{ssec:tauLFV}

In this section, we explore whether  the  type II seesaw model  can be predictive, when one considers  flavour-changing decays of muons and taus. 
In particular, we focus on   observations of $\tlll$ decays, if one does not see $\me$ decays.   So in practise  in this section, the Majorana phases  are fixed to ensure that the tree contribution  to $\meee$ vanishes, and we  study the $\tlll$ rates as a function of the neutrino mass scale.

There are four $\Delta$LF$=$1   $\tlll$ decays ($\tau \to \mu\bar{\mu} \mu$,
 $ \to e\bar{e} e$,
$ \to e \bar{\mu} \mu$ and
$\to e \bar{e} \mu$) and
two that are $\Delta$LF$=$2
($ \to e\bar{\mu} e$,
  and 
  $\to \mu \bar{e} \mu$), as given in
 \cite{Kitano:2000fg}.   They can all be mediated at tree level by $\Delta$ exchange as illustrated in  the middle diagram of Figure \ref{fig:t2match}.
The branching ratios  for the four $\Delta$LF$=$1  decays  are
analogous to
Eq (\ref{BRmeee}) 
\bea
BR(\tau \to l_L \overline{l} l) 
& =& 0.18 {\Big \{}  2| C^{l\tau l l}_{V,LL}  + 4eC^{l \tau}_{D ,R}|^2  \label{BRtaaa}\\
&&+ (64 \ln\frac{m_\tau}{m_l} -136) 
|eC^{l \tau}_{D ,R}|^2    + |C^{l \tau  l l}_{V,LR}  + 4eC^{l \tau}_{D, R} |^2 {\Big\}} 
 \nonumber\\
BR(\tau \to \tilde{l}_L \overline{l} l) 
& =& 0.18 {\Big \{}  | C^{\tilde{l}\tau l l}_{V,LL}  + 4eC^{\tilde{l} \tau}_{D ,R}|^2  \label{BRtabb}\\
&&+ (64 \ln\frac{m_\tau}{m_l} -136) 
|eC^{\tilde{l} \tau}_{D ,R}|^2    + |C^{\tilde{l} \tau  l l}_{V,LR}  + 4eC^{\tilde{l} \tau}_{D, R} |^2 {\Big\}}  \nonumber\\
BR(\tau \to l_L \overline{\tilde{l}} l) 
& =& 0.18   | C^{l\tau  l \tilde{l} }_{V,LL} |^2  \label{BRtaba}
\eea
where  we restrict to a fixed chirality of the flavour-changing lepton bilinear, $l,\tilde{l} \in\{e,\mu\}, l\neq \tilde{l}$  and the factor 0.18 accounts for the hadronic $\tau$ decays: 
$\Gamma(\tau \to e\bar{\nu} \nu)/\Gamma(\tau \to {\rm all}) \simeq 0.18$. The
final  pair of  branching ratios  (Eq.   \ref{BRtaba})  are $\Delta$LF$=$2;
we assume such coefficients run with QED like  other four-lepton vector operators. 
The current  experimental constraints (see Table~\ref{tab:bds}) are of order 
\bea
BR(\tlll)&\lsim& {\rm few } \times 10^{-8}~{\rm (now)}
\to 10^{-10} ~~{\rm (Belle II)}~~.
\nonumber \eea

 \begin{figure}[h]
\begin{center}
  \includegraphics[width=0.8\linewidth]{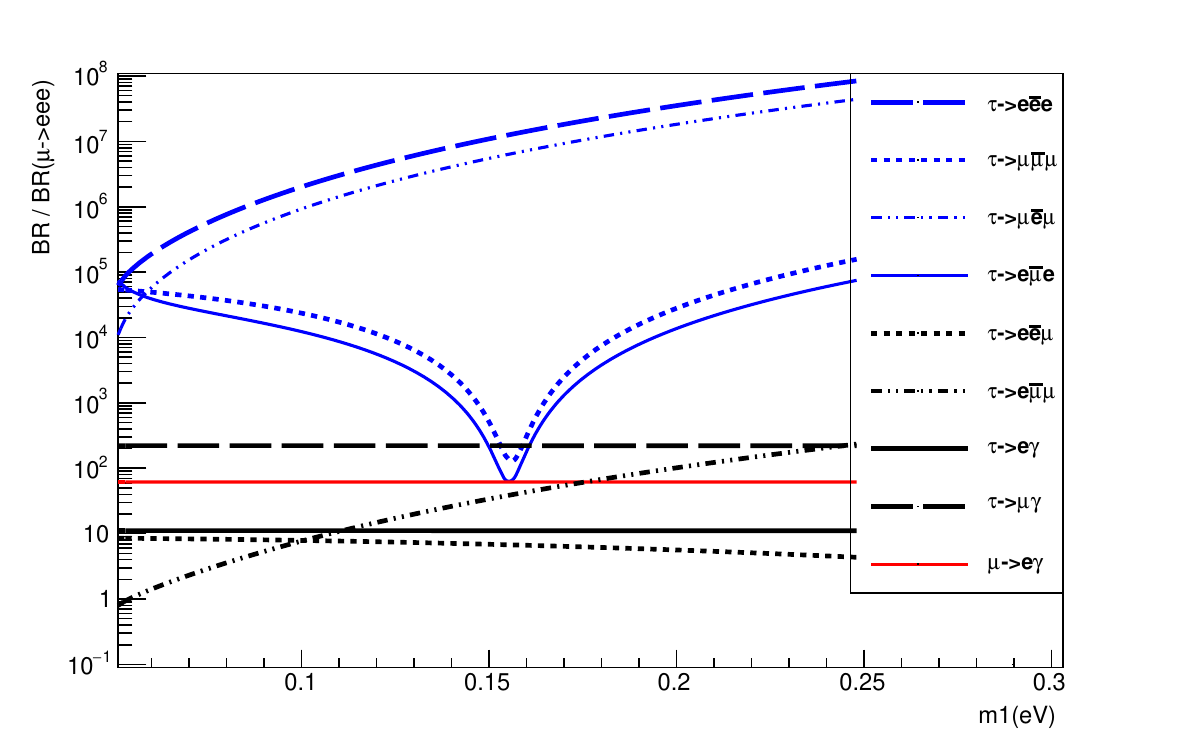}
 \end{center}
\caption{Ratios  BR($\tau\to 3l$)/BR($\meee$) in the type II seesaw model
as a function of $m_1$ in inverted ordering (so $m_1 \geq \sqrt{\Delta^2_{atm}}$)  for vanishing $[m_\nu]_{e \mu}$.
This illustrates that Belle II could  observe  specific $\tau\to 3l$ decays, even if  $\meee$ and $\meg$ are  not observed at upcoming experiments because the tree contribution to $\meee$ vanishes with  $[m_\nu]_{e \mu}$. Some $\tau \to 3l$ processes (in black) also vanish at tree level, and we include $\teg$ and $\tmg$ to illustrate our claim that they are undetectable at Belle II in the Type II seesaw model(the BR$(l_i\to l_j \g)$ are also divided by BR$(\meee)$). 
\label{fig:TauDecmeIH}} 
\end{figure}

 \begin{figure}[ht]
\begin{center}
  \includegraphics[width=0.8\linewidth]{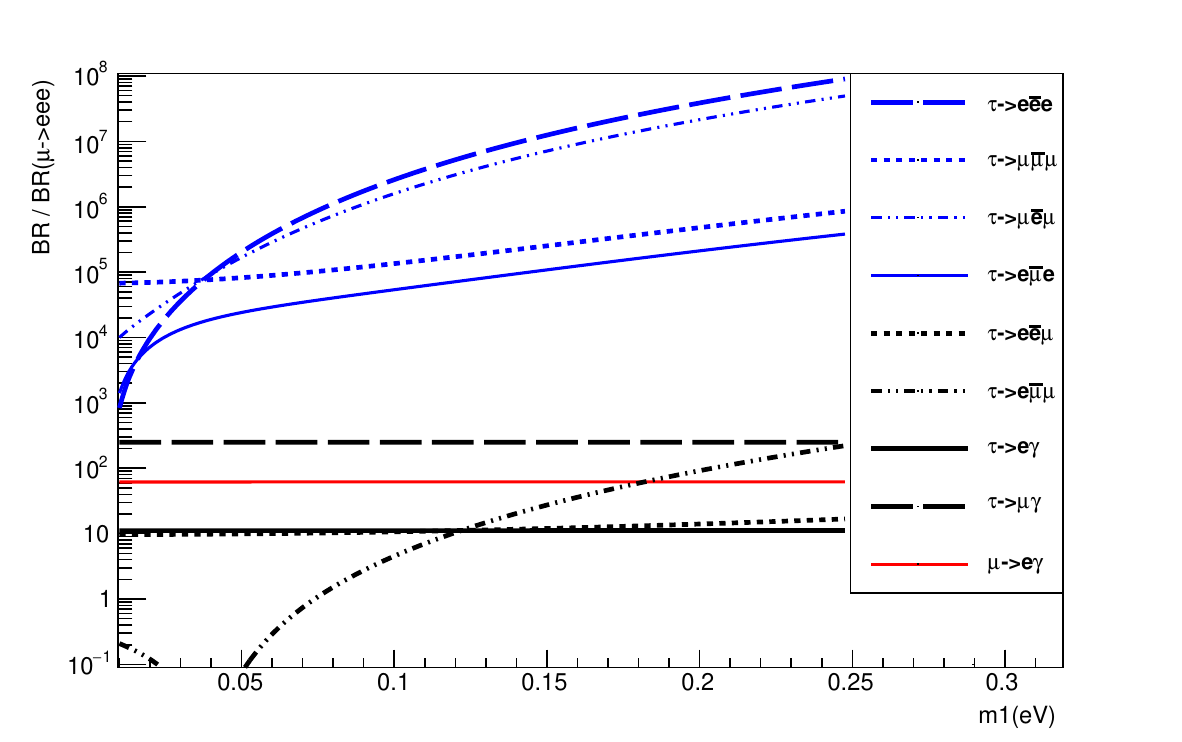}
 \end{center}
\caption{
  Same as Fig.~\ref{fig:TauDecmeIH} for normal ordering {in the case where $\meee$ is suppressed because $[m_\nu]_{e \mu}$ vanishes
    (so $ m_{\rm min} = m_1  \gsim \sqrt{\Delta^2_{sol}}$)}.
\label{fig:TauDecmeNH}} %
\end{figure}

The type II seesaw can predict $\tlll$  within the reach of Belle II, in spite of  the current bound BR$(\meee) \leq 10^{-12}$. For instance, in normal ordering with vanishing lightest neutrino mass $m_1$ 
\bea
 \frac{ BR(\meee)}{BR (\tmmm)}
   &\sim&
 5 \left|\frac{s_{13}  \Delta_{sol} } {\sqrt{2} \Delta_{atm}}\right|^2 
  \sim 10^{-3} \nonumber
 \eea
However, in the coming years,  the Mu3e experiment  will improve the  sensitivity to $\meee$,  so  the  allowed range for the  $\tl$  rates  will  depend on {the results of Mu3e.} To be concrete, we suppose that $\me$ flavour change is not observed at MEG II or Mu3e. So in the following, we suppose that the Majorana phases (and  the lightest neutrino mass  $m_{min}$) are fixed  such that  $C_{VLL}^{e\mu ee}$ vanishes at tree level,  implying  that the  $\meee$  rate is suppressed by ${\cal O}(\a^2)$, because it is   mediated by the dipole and  the penguin-induced $C_{VLR}^{e\mu ee}$.
{Recall that the tree-level contribution to the coefficient $C_{VLL}^{e\mu ee}$,
which is proportional to $[m]^*_{\mu e} [m]_{ee}$,}
can vanish with $[m]_{ee}$ in normal ordering (NO) for $m_{min}\lsim \Delta_{sol}$
(as is well known from neutrinoless double beta decay
\cite{Vergados:2012xy}),   and it can  also vanish with  $[m]_{e\mu}$ both in normal and inverted ordering for $m_{min}\gsim \Delta_{sol}$.

There are  six $\tlll$ decays, whose  tree-level  amplitudes are  proportional 
to products of   only five  neutrino mass matrix elements.  This  suggests  at least one relation among these decays for generic  $m_{min}$ and  Majorana phases ---but  this relation  could be difficult to test in general, because $m_{min}$ and the Majorana phases 
can accidentally suppress  almost any  element of the neutrino mass matrix (as we saw for $[m_\nu]_{e \m}$).   This section considers a scenario where   either $ [m_\nu]_{e\mu} $ or $ [m_\nu]_{ee}$ vanishes in order to suppress $\me$ rates, so testable   predictions  can be expected.

For $m_{ee}\to 0$,   one of the $\Delta LF$ = 2 processes vanishes (at the order we calculate):
 \bea
 BR(\tau \to e  \bar{\mu} e)  \to 0 ~~~,  
 &~~~& {\rm for}~ [m_\nu]_{e e} \to 0~~~~. 
 \eea
 Similarly,  the tree  contribution to $\teee$ vanishes, so that only the dipole and penguin contributions to this decay  remain.
 So  a signature of  the type II seesaw with vanishing $[m_\nu]_{ee}$, is that  these decays  are suppressed  simultaneously with the matrix element for neutrinoless double beta decay.

  In the case where  $\meee$ is suppressed because $[m_\nu]^{e\mu}$ vanishes {(and $m_{min}\gsim \sqrt{\Delta^2_{sol}}$)},
  there are  identities  among the tree-level  coefficients ${ C}_{V,LL}^{l\tau \s\r} \propto m_{l\s}m_{\tau \r}^* $ ( which  by default dominate the rates for the decays   $\tau \to l\overline{\r} \s):$
 \bea
 \frac{m_{e e }} { m_{\mu \mu }}=
 \frac{{ C}_{V,LL}^{e \tau  e e }} { C_{V,LL}^{\mu \tau \mu e}}=
 \frac{{ C}_{V,LL}^{e\tau  e \mu  }} { C_{V,LL}^{\mu\tau \mu \mu}}
~~~,~~~
 \frac{m^*_{\tau  e }} { m^*_{\tau \mu }}=
 \frac{{ C}_{V,LL}^{\mu \tau  \mu e }} { C_{V,LL}^{\mu \tau \mu \mu}}=
 \frac{{ C}_{V,LL}^{e\tau  e e  }} { C_{V,LL}^{e\tau e \mu}}
  =
  \frac{{ C}_{V,LL}^{e\tau  \mu e  }} { C_{V,LL}^{e\tau \mu  \mu}} ~~~.
 \eea
 As a result, for values of the Majorana phases fixed to suppress $\meee$ (and for   compatible   $m_{min}$), ratios of  $\tlll$ decays are predicted. 
 {We plot in Figs.~\ref{fig:TauDecmeNH} (for normal ordering) and~\ref{fig:TauDecmeIH} (for inverted ordering)}  the  rates  of various $\tlll$ decays, normalised to the rate for $\meee$\footnote{In the type II seesaw,  the overall magnitude of LFV $ \sim |f^2v^2/M_\Delta^2|$ is unknown, so we plot ratios of rates. This unknown corresponds to $\lambda_H$ in our parametrisation.}, 
 for the case where $\meee$ is suppressed by  $[m_\nu]_{e\mu} \to 0$.
 This shows  \cite{Chun:2003ej,Akeroyd:2009nu}
that Belle II could see $\tlll$ decays for  BR($\meee) \lsim 10^{-16} $ and BR$(\meg) \lsim 10^{-14}$.
{If several $\tlll$ decays are observed (including $\teee$), we may compare their relative branching ratios with
the predictions of the type II seesaw shown in Figs.~\ref{fig:TauDecmeIH} and~\ref{fig:TauDecmeNH},
and be able to either exclude the model or deduce some constraints on $m_1$ and the mass ordering.}
{Regarding $\muc$, we estimate that}
BR$(\mu Al \to eAl) \sim 10^2$ BR$(\meee)$  in the case of vanishing   $[m_\nu]_{e\mu}$  in  the type II seesaw,  because  $C_{V,LR}^{e\mu ee}$ and $\tilde{C}_{Al,L}$ are  both induced by the photon penguin, see Eqs. (\ref{pingt2},\ref{ping2t2}).

Finally,    the decays $\teg$ and $\tmg$   will not be observed at Belle II if  neutrino masses arise via the  type II seesaw model, because  this requires  $\meg$ or $\meee$ larger than the current constraints
{(this conclusion is fully general and does not assume that the tree-level contribution to $\meee$ vanishes).}
The dipole coefficients $C_{D,R}^{l \tau}$, at the experimental scale $m_\tau$, can be written analogously to Eq. (\ref{megt2}), with  the index replacement $e\to l$ and $\mu\to \tau$, and without the second term, because the RG running ends at $m_\tau$ for all flavours in the loop. At the order we calculate, the $\tau$ dipole coefficients are therefore  given by $[m_\nu m_\nu^\dagger]_{l\tau}$,  so are independent of the neutrino mass scale and Majorana phases (see Eq. \ref{mmdag}), and the ratio
$$
\frac{BR(\tmg)}{BR(\teg)} \simeq \left|\frac{[m_\nu m_\nu^\dagger]_{\mu\tau}}{[m_\nu m_\nu^\dagger]_{e\tau}} \right|^2 \sim \frac{1}{2 s^2_{13}}
$$
is predicted.  However, in order to  be within the reach of Belle II (BR$\gsim 10^{-9}$, see Table~\ref{tab:bds}), these  branching ratios  need to be much larger than BR($\meg)$,  which can be engineered via a cancellation in   $C_{D,R}^{e\mu}$ for  specific Majorana phases at  a large neutrino mass scale. However, when this cancellation arises, the model predicts a larger  branching ratio for $\meee$  than $\tlg$,
because  the coefficient $C_{VLL}^{e\mu ee}  \propto m_1^2$ arises at tree level.

\subsection{The dipole  constraints on  boxes for the leptoquark}
\label{ssec:tLFV}

The leptoquark Lagrangian of  Eq. (\ref{LLQ}) allows the leptoquark to interact with doublet and singlet leptons,  so it can induce lepton flavour-changing dipole, tensor and scalar operators, without any suppression by the lepton Yukawas.  Nonetheless, we neglect four-lepton scalar operators in this model, because the coefficients are suppressed below the  upcoming experimental reach  by the dipole constraint.  This section aims to show that suppression. 

\begin{figure}
  \begin{center}
\hspace{2cm}
\begin{picture}(70,70)(30,10)
\ArrowLine(0,60)(25,60)
\ArrowLine(60,60)(25,60)
\ArrowLine(60,60)(90,60)
\ArrowLine(90,30)(60,30)
\ArrowLine(25,30)(60,30)
\ArrowLine(25,30)(0,30)
\put(-12,60){$\mu_X$} 
\put(-12,30){$l_Y$}
\put(98,60){$e_Y$} 
\put(98,30){$l_X$}
\put(10,40){$S_1$} 
\put(65,40){$S_1$}
\put(50,68){$H$} 
\put(50,18){$H$}
\DashArrowLine(30,30)(30,60){5}
\DashArrowLine(60,60)(60,30){5}
\DashLine(45,60)(45,80){3}
\DashLine(45,30)(45,10){3}
\end{picture}
\hspace{2cm}
%
\begin{picture}(70,70)(30,10)
\ArrowLine(0,60)(30,45)
\ArrowLine(60,45)(90,60)
\ArrowArc(45,45)(15,0,360)
\ArrowLine(90,30)(60,45)
\ArrowLine(25,45)(0,30)
\put(-12,60){$\mu_X$} 
\put(-12,30){$l_Y$}
\put(98,60){$e_Y$} 
\put(98,30){$l_X$}
\GCirc(30,45){5}{.7}
\GCirc(60,45){5}{.7}
\put(50,68){$H$} 
\put(50,18){$H$}
\DashLine(45,60)(45,80){3}
\DashLine(45,30)(45,10){3}
\end{picture}
\hspace{2cm}
\begin{picture}(70,70)(30,10)
\ArrowLine(0,60)(30,45)
\ArrowLine(60,45)(90,60)
\ArrowArc(45,45)(15,0,360)
\ArrowLine(90,30)(60,45)
\ArrowLine(25,45)(0,30)
\put(-12,60){$\mu_X$} 
\put(-12,30){$l_Y$}
\put(98,60){$e_Y$} 
\put(98,30){$l_X$}
\GCirc(30,45){5}{.7}
\GCirc(60,45){5}{.7}
\put(43,58){x}
\put(43,28){x}
\end{picture}
\end{center}
\caption{ A diagram  that generates  the four-lepton scalar operator ${\cal O}_{SXX}^{e \mu ll}$ in the model (on the left), in the RGEs of  SMEFT (centre) and in the RGEs of the QCD$\times$QED invariant EFT (on the right). (There is another diagram with $l_Y\leftrightarrow e_Y$.)
\label{fig:box+fishes}} 
\end{figure}
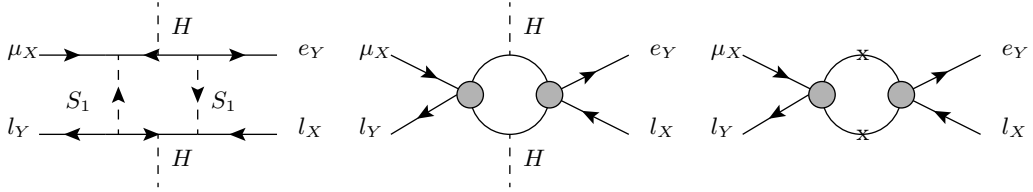

We are interested in scalar four-lepton operators ${\cal O}_{S,XX}$,  such as those in the observable Lagrangian of Eq. (\ref{Lag1}). These operators occur at dimension eight in SMEFT with a pair of Higgs legs, for instance in the form
$(\overline{\ell}_eH \mu_R) (\overline{\ell}_eH e_R)$ and can be generated in the leptoquark model via box diagrams with Higgs legs, as illustrated  on the left in Figure \ref{fig:box+fishes}. Equivalently, these operators are generated in the dimension six$^2 \to$ dimension eight RGEs of SMEFT or the low energy EFT,  respectively by the fish diagrams at the centre or right of Figure  \ref{fig:box+fishes}. It is straightforward to see from the model diagram, that
\beq
C_{SRR}^{e \mu ee} \propto  N_c [\lambda_L Y^*_u \lambda_R^\dagger]^{e\mu} [\lambda_L Y^*_u \lambda_R^\dagger]^{ee} \frac{v^4}{32 \pi^2 m_{LQ}^4} \log
\label{CS4ldim8}
\eeq
so that this coefficient could be $ \sim 3\times 10^{-5}$ for leptoquark couplings  of ${\cal O}(1)$ to the top quark.  This is marginally above the current  experimental bound, which  is  $|C_{SRR}^{e \mu ee}|\leq 2.8 \times 10^{-6}$  from SINDRUM~\cite{SINDRUM:1987nra}.
So now we want to show that  the product
$\lambda_L^{\a t} \lambda_R ^{\b t *}$ is constrained to be $\lsim 0.01$ for almost any combination of lepton flavours $\a, \b$.

In the presence of both the $\lambda_L$ and $\lambda_R$ interactions, the leptoquark  matches onto $2l2q$ tensor  operators involving the up-type quarks (see Eq \ref{CT-Q}). These tensor operators then mix to the dipole (see  the last term of Eq. \ref{megLQ}), generating an ``invariant'' that is not identical to the one appearing  in Eq (\ref{CS4ldim8}):
\beq
\Delta C_{DR}^{ \a\b} \propto \frac{\a}{\pi y_\b}[\lambda_L Y_u^*  \eta_{m_Q}\ln\frac{m_{LQ}}{m_Q} \lambda_R^\dagger]^{\a\b} \sim \frac{\a}{\pi y_\b}{\Big (} {\cal O} {\Big (} \lambda^{\a t}_L  y_t   \lambda_R^{\b t *} {\Big )} +  {\cal O}  {\Big (} \lambda^{\a c}_L  y_c   \lambda_R^{\b c *}  {\Big )}+ ... {\Big)} ~~.
\label{DeltaCDt}
\eeq
However, the   term  $\propto y_t$ in $[\lambda_L Y_u^*  \eta_{m_Q}\ln\frac{m_{LQ}}{m_Q} \lambda_R^\dagger]^{e \mu}$ is of  comparable magnitude  to the term $\propto y_t$  of  $[\lambda_L Y_u^*  \lambda_R^\dagger]^{e \mu}$  from  Eq (\ref{CS4ldim8}).

Now we want to argue that this term  $\lsim y_c,y_\tau$,  because if its larger  it exceeds  the experimental bound on  $\Delta C_{DR}^{ \a\b}$, so must be small enough to  cancel against next biggest term. In this argument,  we
assume that $|\lambda_X^{\s Q}|\lsim 1$, and  use
a different normalisation of the dipole operator:
\bea
 \widetilde{\cal O}_{D}^{\a\b}& =&
  \overline{\ell}_\a H   \sigma^{\r\s}P_{R} e_\b F_{\r\s}
  \label{Drenorm}
  \eea
  where  the lepton Yukawa is removed. This is more convenient for comparing the experimental bounds~\cite{MEG:2016leq,Belletmg,Babartmgteg,Morel:2020dww,Muong-2:2021ojo,Roussy:2022cmp,Muong-2:2008ebm}  on NP contributions to    the  coefficient  of the redefined dipole  operator, expressed as a  matrix in flavour space:
\bea
  |[\widetilde{C}_{D}]^{\a\b}| \leq  \left[\begin{array}{ccc}
 2.3\times 10^{-6}   & 6\times 10^{-12}  
 &7\times 10^{-8}\\
  6\times 10^{-12}& 4\times 10^{-7} 
 &7\times 10^{-8}\\
 7\times 10^{-8} & 7\times 10^{-8}& -
\end{array}
    \right]
  \label{CDbds}
\eea
where  on the diagonal  are the $(g-2)_\b$ constraints\cite{Morel:2020dww,Muong-2:2021ojo}   using 
$|\Delta a_\b| \simeq  2m_\b|C_D^{\b\b} + C_D^{\b\b*}|/(e v)$, and the EDM constraints
\cite{Roussy:2022cmp,Muong-2:2008ebm}
on ${\cal I}m \{C_D^{\b\b}\} \simeq  -  v d_\b/2$ \cite{Pospelov:2005pr} are 
$|C_D^{ee}| \lsim 2.5\times 10^{-14}$, and 
$|C_D^{\m \m}| \lsim 3 \times 10^{-4}$.
The contribution of 
Eq. (\ref{DeltaCDt}), with  $\lambda_L^{\a t} \lambda_R^{\a t *}$ of order 1, is larger than the bounds of 
Eq. (\ref{CDbds})  on all the dipole  coefficients, except possibly
$[\widetilde{C}_{D}]^{\tau\tau}$.
So the top contribution must cancel against the next largest contribution to the dipole,  which   is relatively suppressed  at least by  the charm or $\tau$ Yukawa, because the dipole operator requires a single  Higgs leg. So this implies that
$$
\lambda_X^{\a t} \lambda_Y^{\a t *} \lsim 10^{-2}
$$
 for all flavour combinations $\a\b$ except $\tau\tau$, so the scalar four-lepton coefficients are below upcoming experimental sensiticity and can be neglected.

 We also explored the possibility that constraints on vector four-quark operators\cite{UTfit:2007eik} --- for instance from meson-antimeson oscillations--- allow to set bounds on  the vector four-lepton operators.   Our hope was that  the quark sector could set bounds on the eigenvalues $\{\lambda_i\}$ of the leptoquark coupling matrices $$
[\lambda_L] = V_l^\dagger D_\lambda V_Q 
 $$
 where $ D_\lambda =$ diag$\{\lambda_1, \lambda_2, \lambda_3\}$, and $V_l$ and $V_Q$ are unitary matrices. $K^0\! -\! \overline{K^0}$ and $D^0\! -\! \overline{D^0}$ mixing constrain two independent combinations of   $ V_Q^\dagger D^2_\lambda V_Q $, but in order to set an upper bound on the eigenvalues, at least one more constraint would be required, and we did not  find  useful constraints involving tops.

\subsection{Complex Coefficients---what changes?}
\label{ssec:C}

The coefficients of the observables operator are generally complex numbers, and it is not immediately clear whether experiments can fully determine these coefficients when non-zero phases are present.  When considering upper limits on branching ratios, the only difference is that we have two identical 12-dimensional ellipses, each respectively in the space of the real and imaginary parts of the coefficients. This happens because the branching fractions are functions of the absolute values of operator coefficient combinations, which result in a quadrature sum of the real and imaginary components. Therefore, the two do not mix, and the rate is a combination of the same positive defined quadratic forms, to which the upper limit can separately apply.

The complication arises when we consider the possibility of measuring the complex coefficients from data. Expanding the branching ratios of Eq. (\ref{BRmeg})-(\ref{BRmucAu}) in terms of the (complex) coefficients of Eq. (\ref{Lag1}), we find\footnote{We assume that we can distinguish the processes with electron/positrons of different chiralities via the angular distributions. We do not discuss the scalar operator $C^{e\mu ee}_{S,XX}$ because it is negligible in the models we consider.}
\bea
BR(\mu\to e_X \gamma)&=& 384 \pi^2 |C^{e\mu}_{D,Y}|^2 \\
BR(\mu\to e_X \overline{e}_X e_X) &=& 2 |C^{e\mu ee}_{V,XX}|^2+32e^2\left(\log\left(\frac{m_\mu}{m_e}\right)-1\right)|C^{e\mu}_{D,Y}|^2+16e |C^{e\mu ee}_{V,XX} C^{e\mu }_{D,Y}| \cos(\phi_{V,XX}-\phi_{D,Y} )\\
BR(\mu\to e_X \overline{e}_Y e_Y) &=&  |C^{e\mu ee}_{V,XY}|^2+32e^2\left(\log\left(\frac{m_\mu}{m_e}\right)-\frac{3}{2}\right)|C^{e\mu }_{D,Y}|^2+8e |C^{e\mu ee}_{V,XY} C^{e\mu }_{D,Y}| \cos(\phi_{V,XY}-\phi_{D,Y} )\\
BR(\mu A\to e_X A) &=& B_A\left(d^2_A|C^{e\mu}_{D,Y}|^2+|C^{e\mu}_A|^2+2d_A |C^{e \mu}_{D,Y}C^{e\mu}_A|\cos(\phi_{A}-\phi_{D,Y} )\right)
\eea
where we have defined $C_{\square}=|C_\square|e^{i\phi_\square}$, $X\neq Y$ and $A$ can be Al or Au$\perp$. The observables only depend on relative phases, so for brevity we  relabel $\phi_\square-\phi_{D,Y}\to \phi_\square$.  We thus have 10 branching fractions that can generally depend on 18 parameters: 10 absolute values and 8 relative phases. For example, observing $\mu\to e_X \gamma$ and $\mu\to e_X \overline{e}_X e_X$ ($\mu\to e_X \overline{e}_Y e_Y$) would not be sufficient to measure the real and imaginary parts of the coefficient $C^{e\mu ee}_{V,XX}$ ($C^{e\mu ee}_{V,XY}$). However, some observables may be  directly related to the coefficient relative phases. For instance, it has been shown in \cite{Okadameee} that  the T-odd asymmetry term
\begin{equation}
	A^T_{\mu\to 3e}\propto -{\rm Im}(C_{D,R} (3C^*_{V,LL}-2C^*_{V,LR}) + (L\leftrightarrow R)=3|C^{e\mu ee}_{V,LL} C^{e\mu ee}_{D,R}| \sin\phi_{V,LL}-2|C^{e\mu ee}_{V,LR} C^{e\mu ee}_{D,R}| \sin\phi_{V,LR}+(L\leftrightarrow R)\label{eq:Tasym}
\end{equation}
is accessible via the angular distribution of the outgoing electrons/positrons, and could assist in determining the relative phase of the $\mu\to 3e$ vectors  and dipole. In addition, interpreting data assuming specific models can help in reducing the number of relevant parameters, and measurements may still be used to find regions of parameter space that are incompatible with the model predictions. For the three models we considered here: 

\begin{enumerate}
	\item In the type II seesaw the observable coefficients are related to the neutrino mass matrix, which is complex (given the hints of the CP violating Dirac phase  $\delta\simeq 3/2\pi$ and due the potential presence of non-zero Majorana phases).  In our analysis of the type II seesaw in \cite{Ardu:2023yyw}, we identified the region where the model can sit in the space of the angular variables $\tan\theta=\sqrt{|C_D|^2+|C_{V,LR}|^2}/|C_{V,LL}|$ and $\tan\phi=|C_D/C_{V,LR}|$, which depend on the absolute values of the observable coefficients. One may wonder whether experiments can identify a point in this space despite having complex phases, which make three branching fractions depend on five parameters (three absolute values and two phases). Since the flavour changing interactions with electron singlets are negligible, the $A^T_{\mu\to 3e}$ asymmetry is given by  the relative phases of the $C_{V,LL}, C_{V,LR}$ vectors with the $C^{e\mu}_{D,R}$ dipole. Combined with the measurements of the branching fractions for  $\mu\to e_L \gamma$, $\mu\to e_L \overline{e}_L e_L$ and $\mu\to e_L \overline{e}_R e_R$, one of five physical parameters would still remain undetermined. Taking advantage of the fact that in the type II seesaw $C_{A}\propto C_{V,LR}$, a detection of $\mu A\to e A$ could be used to extract the value of the last unknown, resulting in the complete knowledge of the relevant complex coefficients (modulo an overall phase). This also opens the interesting possibility of taking advantage of $\mu\to e$ observables to determine the unmeasured neutrino parameters, i.e the lightest neutrino mass and the Majorana phases. Since the complex EFT coefficients depends on these three unknown, measuring the coefficients could allow to infer their values. Additionally, since the system is over-constrained, with three complex coefficients being a function of four parameters (including the overall magnitude of LFV), we can use experimental results to check for consistency with the type II seesaw.
	\item The operator coefficients in the inverse seesaw are given by the off-diagonal elements of Hermitian matrices, which can be in general complex. In the case of nearly degenerate sterile neutrinos, we have found that the coefficients satisfy linear relations of the following form \cite{Ardu:2023yyw}
	\begin{equation}
		C^{e\mu ee}_{V,XY}=a_{XY} C^{e\mu }_{A,L}+b_{XY} C^{e\mu ee}_{D,R} \label{eq:inversecorr}
	\end{equation} 
	where $a$ and $b$ are real numbers, and $XY=LL,LR$. These relations hold in general and do not assume real coefficients. However, with non-vanishing phases, two observables are not sufficient to fully determine the $\mu\to e$ predictions of the inverse seesaw.  Observing $\mu\to e_L \gamma$ would give the dipole absolute value, and measuring  $\mu A\to e_L A$ would only yield a combination of $|C^{e\mu}_{2q,L}|$ and dipole-four fermion relative phase. Taking advantage of Equation \ref{eq:inversecorr}, a $\mu \to e_L \overline{e}_R e_R$ signal  could  allow to solve for $C^{e\mu}_{2q,L}$ as a complex number. Then, again by means of Equation \ref{eq:inversecorr} for $XY=LR$, we could predict $BR(\mu\to e_L \overline{e}_R e_R)$ and compare it against experimental results. We conclude that, despite the non-zero phases, the inverse seesaw with degenerate sterile neutrinos is predictive enough that it could be ruled out by a combination of $\mu\to e$ observations
	\item In the leptoquark model the $\mu\to e$ coefficients depend on a number of invariants that is larger than the number of observables. This means that the model could already fill the experimental ellipse (with the exception of the scalar four lepton directions) even in the case of real couplings. Allowing complex couplings would make the model even less constrained, and thus we do not discuss this case further. 
\end{enumerate}

\section{Discussion}
\label{sec:disc}

The purpose of bottom-up EFT is to  take low-energy experimental information to high-scale models. In this section, we   discuss various  aspects of this process, in the light of the three models we considered.

\subsubsection{  What differences among models can be identified by the data?}

We showed in \cite{Ardu:2023yyw} that the data could rule out  the models we consider, because  the models  predict relations between the Wilson coefficients, so are unable to fill the whole  ellipse in coefficient space that is  accessible to experiments.  But it would be more interesting to ask  whether $\me$ observations could identify properties of models.

A  simple question is whether $\me$ data  could distinguish models with LFV couplings to either lepton doublets or singlets, vs models that interact with both. It seems that the answer is yes.
If LFV interactions involve only doublets or singlets, the only lepton-chirality-changing interaction in the theory is the Higgs Yukawa.  So the coefficients of   chirality-changing LFV operators must be proportional to the Yukawa couplnging to an odd power. 
 For instance,    the dipole coefficients would   satisfy
  $$\frac{C^{e\mu}_{DR}}{C^{e\mu}_{DL}} \propto \left(\frac{m_e}{m_\mu}\right)^{\pm 1}
~~~.  $$
If this relation is {\it  not} satisfied,   it suggests that LFV involves doublets and singlets.
  If it is satisfied with exponent $-1$,  then it is probable that LFV involves doublets but not singlets. If in addition, the  coefficients of other  singlet-LFV  operators are negligible, it becomes very probable that   LFV involves  only doublets ---although it could  result from accidental cancellations in a  model with LFV for doublets and singlets.

 It would be interesting and useful if the data   could  also identify other model properties, such as  the loop order at which LFV occurs.  But  our models  suggest this is not  possible,   because   loops
  that occur in matching the model to the EFT are indistinguishable from small couplings, and   because coefficients that arise at tree level can be accidentally small, as  occured in the type II seesaw where $C_{VLL}^{e\mu ee}$ can vanish. 
 For similar reasons, $\me$ observables can not distinguish NP that interacts only with leptons in its renormalisable Lagrangian, from NP that interacts with quarks and leptons.

\subsubsection{What is the role of  the RGEs?}

There are  many reasons to use Renormalisation Group Equations in EFT, as is  well-known in quark flavour physics.
However, the  RGEs may not be motivated in the lepton sector, because  LFV has yet to be discovered, so   precise, scheme-independent predictions are not crucial.  Indeed, the RGEs are often not included in the $\tl$ sector, where   the data separately  constrain  most coefficients.  In  the $\me$ sector where there are fewer  restrictive  bounds, an EFT analysis suggests that the RGEs are   relevant
because  they    mix     difficult-to-probe operators  into   well-constrained processes.  In this section, we explore whether this occurs in our models.

  The inverse seesaw at the TeV-scale is an example of a model where the RGEs  are unneccessary, because  low-energy LFV operators are generated via loop diagrams  in matching, and the RGEs just contribute an ${\cal O}(10\%)$ renormalisation. A few properties of the model contribute to this behaviour:
  first, the new interactions  couple one light SM particle  with a heavy new particle and  a weak boson,   so LFV occurs via  loops  that contribute  in matching. Then, there  is no significant operator mixing via the RGEs,  because
  the  photon penguin   vanishes below the weak scale  at one loop; that is,   LFV operators are vectors(or  $\propto$ a lepton  yukawa coupling) because    only the lepton doublet interacts with new particles,   and vector operators only mix into each other via the penguins.   But in the one-loop penguin diagrams, the gauge boson attaches to the Higgs,   so the penguins are only present above the weak scale, and   generate  $2l2q$ operators in matching.

  The type II seesaw is an example of a model  where the leading contribution to some  observables arises in matching, and via the RGEs for others. This behaviour can be reproduced in EFT, and can  also be  obtained in  model calculations  that  judiciously  include log-enhanced loops \cite{Dinh:2012bp}.
 However,   the correct  lower limits of the logarithms are crucial  to obtain the dependence of LFV observables on the unknown parameters of the neutrino mass matrix in the type II seesaw model.   These lower limits should be implemented automatically in EFT, but to our knowledge were previously missing from the literature.  So it seems that the RGEs are required in this model, in order to  identify the parameter space the model cannot reach.

In the type II seesaw, a careful  one-loop model calculation could include all the terms  of our leading-log EFT, because we  do not resum $(\a \lg)^n$ for all $n$, but rather   work to ${\cal O}(\a\log)$.
However, our EFT also includes the ``leading'' vector to dipole mixing at  ${\cal O}(\a^2 log)$,  which we did not find    in the literature about this model.
This mixing  causes the dipole to depend on the  unknown neutrino  parameters (Majorana phases and $m_{min}$), thereby allowing it to vanish.
It is ``well known'' that the 2-loop electroweak contribution to $(g-2)_\mu$ is comparable to  the  one-loop part, but it seems that the implications of this  may not have been implemented in all model calculations.
However, it is relatively simple to implement in EFT \cite{Crivellin:2017rmk}, illustrating the first reason to do EFT: it is the simpler way to get a more precise  result.

 The leptoquark is our model where the RGEs are most useful, because they allow to simultaneously  include the multitude of  relevant electroweak loops and large QCD effects in an organised  fashion. The RGEs are required to obtain model predictions, because they mix difficult-to-constrain coefficients--- such as tensor operators involving top quarks---into observable coefficients like the dipole, while simultaneously including the QCD running of the operators.  so are required   to obtain model predictions.

\subsubsection{Do cancellations among coefficients occur in models? }

It is common to make  tables  listing the ``sensitivity'' of observables to operator coefficients ({\it e.g.} \cite{Davidson:2020hkf,Crivellin:2017rmk}); these ``one-at-a-time-bounds'' are simple to obtain  by allowing a single operator to have a non-zero coefficient, and computing  the experimental constraint upon it.  It can also be common to take these sensitivities as bounds,  because it is generally considered unlikely that models generate cancellations among operator coefficients, especially  since these coefficients  run with scale. 
However, such  cancellations  can occur,  for instance via the  equations of motion,  so  these ``sensitivities'' are  not true upper bounds (instead, they are the value of the coefficient above which it could be detected).

The   operator  population  in EFTs  is often   reduced  via the equations of motion (as pedagogically discussed in  \cite{EoMEFT,polonais}).  This can sometimes  impose   ``accidental'' but  precise cancellations among operator coefficients.
An example  is discussed   in Appendix \ref{assec:ping}: in a model,  the   $Z$ penguin {\it diagrams}     can be $\propto q^2$ (= the momentum-transfer$^2$ of the $Z$),  or $\propto v^2$.  Both contributions could contribute to the decay $Z\to e^\pm \mu^\mp$, but the part $\propto q^2$   gives  a negligible contribution  to  $\meee$, due to the $q^2\sim m_\mu^2$ suppression. However if the model is matched to SMEFT,    the  equations of motion are used to to transform the $q^2$ part of the diagrams into  four-fermion and  penguin  operators (${\cal O}_{HL3}^{e\mu}$ and ${\cal O}_{HL1}^{e\mu}$), with coefficients  whose sum cancels  in low-energy matrix elements where $q^2 \to 0$.   The ``one at a time bounds'' on penguin and four-fermion operators miss this cancellation,  so would instead suggest that both are strictly constrained by $\meee$.

In the type II seesaw model, the $Z$-penguin diagrams give negligible contribution to $\meee$ because the $\propto v^2$ part is suppressed by lepton Yukawas, and  the $\propto q^2$ part is kinematically  suppressed as discussed above.
So  experimental observations do not exclude a $\bar{e}\Zslash \mu$ interaction just below the sensitivity of the LHC, despite that the ``one-at-a-time-bounds'' suggest that they do.  Nonetheless, the model can not generate such interactions, because they  are controlled by the same model parameters as the photon penguin diagrams, which are constrained by $\meee$.

So in summary,  apparently  accidental cancellations can occur  among coefficients in EFT; whether this affects the constraints on models is model-dependent.

\subsubsection{Does it matter that  coefficients are complex?} 

$\mu\to e$ observables define an experimentally accessible 12-dimensional ellipse in the space of the Wilson coefficients, but these are generally complex numbers. Although this does not complicate the analysis when imposing bounds on the coefficients, because the ellipses for the real and imaginary components are identical, allowing for complex phases could generally hinder the determination of the coefficients from data. If the muon polarization and the electron/positron angular observables could in principle identify 12 real coefficients, unobservable directions  will be present when considering the full parameter space spanned by coefficients with non-zero phases. Measuring for instance the branching fraction for $\mu\to e_X \gamma$ would identify a circle with radius $\sqrt{BR(\mu\to e_X \gamma)/384\pi^2}$ centered at the origin of the complex plane for the dipole coefficient $C_{DX}$, but without further assumptions the real and imaginary parts would remain undetermined. However, if our goal is to use data to exclude models, the model predictions can help in reducing the number of unmeasurable direction. Interpreting data in light of a particular model can lead to the determination of absolute values as well as relative phases for the coefficient. The neutrino mass models we considered in this paper are an example where this determination is possible. 
	
In the type II seesaw, the LFV operators are complex because their coefficients are directly related to the neutrino mass matrix, which contains up to three phases in the case of Majorana neutrinos. We discussed in section \ref{ssec:C} how, taking advantage of the model predictions and making use of observables directly related to the coefficient phases (see Eq.~(\ref{eq:Tasym})), one could interpret $\mu\to e$ data to determine the (complex) coefficients predicted by the type II seesaw. 
	
Similarly, in the inverse seesaw model, the predictions for flavour-changing observables are determined by the magnitude of off-diagonal matrix elements, and the operators can be complex. In section \ref{ssec:C}, we showed that despite the presence of operator phases, we can use the model predictions to identify points in the experimentally accessible ellipse.

\section{Summary}
\label{sec:sum}

 The $\me$ sector is promising for the discovery of LFV, due to the  exceptional  upcoming experimental  sensitivity --- to three processes. So  this project explored what could be learned about  the  New Physics in the lepton sector from  $\me$ observations, by  studying some  ``representative'' TeV-scale models described in Section \ref{sec:models}.
 We take the data to be 12 Wilson coefficients, which can be individually constrained  and distinguished in measurements (with the exception of vector and  scalar four-lepton operators which have indistinguishable angular distributions  in $\meee$, see Section \ref{sec:notn}).

 Bottom-up EFT  is an appropriate formalism to  compare data with models, because data improves slowly while   models can evolve more rapidly. It also gave some relevant effects  (such as the two-loop vector to dipole mixing  in the type II seesaw, see Eq. \ref{megt2}) which we did not find in the literature. Our analysis is at leading order in EFT, meaning that we attempt to  include  the largest contribution  of the model to  all the operators to which the data can be sensitive.  This includes some 2-loop anomalous dimensions and some   operators which are dimension eight in SMEFT.   Our notation  and assumptions are summarised in section \ref{sec:notn}.  Our models are located at  the TeV scale in order    to profit from many complementary observables, but  the ratio $m_W/$TeV is not large, which  implies  that EFT is poorly motivated between $m_W \to $ TeV (see the discussion   in Appendix \ref{assec:quelEFT}), so    in practise we  match our models to the QCD$\times$QED-invariant  EFT that is relevant below $m_W$.

 The  observable operator coefficients are given   in Section  \ref{sec:obs+invar}  in terms of model parameters at the TeV, which usually appear with SM parameters  in  elegant combinations that recall Jarlskog invariants.
  This  unforeseen curiosity  (discussed in Section \ref{ssec:invar}) may be an accident of  our simple leading  order  analysis, which allows analytic  expressions.
  Or possibly it indicates  an  interesting new role for invariants as stepping stones  in the reconstruction of models from EFT coefficients.
  For instance, our models did not fulfill our expectations:  we anticipated that the type II seesaw was predictive because the flavour-changing  couplings $f_{\a\b}$ (Eq.\ref{LII}) are determined by the neutrino mass matrix,  and that the inverse seesaw was unpredictive because  the $Y^{\a a}$ couplings (Eq. \ref{Linverse}) are unknown.
  However,   Section \ref{ssec:CLex} shows that $\me$ flavour change is controlled by
  two  invariants  in the inverse seesaw with degenerate singlets,  whereas three invariants are needed in the type II model. 
In any case, it is interesting that $\me$ flavour change  in these models is controlled by a few  complex numbers of magnitude $\lsim 1$, which could be obtained for a wide variety of flavour structures in Lagrangians.

 We showed in ~\cite{Ardu:2023yyw} that  $\me$ data has the ability to exclude  the models we consider, because they cannot fill the whole ellipse in coefficient space accessible to upcoming experiments.  In  Section \ref{sec:pheno}, we  explored the more interesting question of  whether observations  could  indicate a model--- specifically, the type II seesaw model --- by including    some  complementary observables.
 In Section~\ref{ssec:mnu} we showed that, in the type II seesaw model, some ratios of $\mu\to e$ Wilson coefficients
can be confined to relatively narrow intervals (depending on the mass ordering, see Eq.~\ref{eq:mmin0typeII})
if the lightest neutrino mass is small enough. Therefore, if these ratios where observed outside the ranges quoted
in Section~\ref{ssec:mnu}, a lower bound on the neutrino mass scale could be inferred (assuming that
neutrino masses arise from the type II seesaw mechanism).
In Section~\ref{ssec:tauLFV}, we considered $\tau$-LFV at Belle II --- still in the type II seesaw model---in the case where neither $\meg$ nor $\meee$ are observed  in upcoming experiments (see table \ref{tab:bds}). In this case, the model makes specific predictions for $\tau$-LFV ratios  that could be observed at Belle II, 
as a function of the neutrino mass scale and ordering. As a result, Belle II could contribute to {constraining the neutrino mass scale and ordering} or rule out the type II seesaw model.

   Finally, in the discussion section ~\ref{sec:disc}, we addressed some  questions that  arise  in a bottom-up EFT attempt to reconstruct   New Physics  from (low-energy) data.

\subsubsection*{Acknowledgements}
We thank Ann-Kathrin Perrevoort  for helpful comments abour observables in $\meee$.
{The work of SL is supported in part by the European Union's Horizon 2020 research and innovation programme under the Marie Sklodowska-Curie grant agreement No. 860881-HIDDeN.} The work of MA is supported by  the Spanish AEI-MICINN PID2020-113334GB-I00/AEI/10.13039/501100011033

\appendix

\section{ $\muc$ operators}
\label{app:ops}

 References \cite{KKO,Cirigliano:2009bz} observed that the target-dependence of  the $\muc$ rate could  give information about  the  contributing operators.  
So the aim of this Appendix is to identify the  independent  four-fermion operators
${\cal O}_{Al,X}$ and  ${\cal O}_{Au\perp,X}$, 
probed by  Spin-Independent (SI) $\muc$ on light and heavy targets. We use the formalism of \cite{Davidson:2020ord} and use the nuclear calculation of \cite{KKO}. The
operators ${\cal O}_{Al,X}$ and  ${\cal O}_{Au,X}$ can be constructed in the nucleon basis relevant at  the experimental scale, or in  a quark basis more relevant for comparing to models.  In the following, we first  construct these operators  in the  nucleon basis, where operators, coefficients and parameters wear  ``tildes'', then express them in the quark basis without tildes. Notice that  normalisations change between the bases,  so  there are numerical differences between, {\it eg},  $\widetilde{B}_A$ and  $B_A$.

The results in this Appendix have two peculiarities,  arising in the matching of nucleons  with quarks. The first is that there is a ``loss of information'' in going from nucleons to quarks, because  the scalar $u$ and $d$ content of a nucleon are comparable: $\langle N|\bar{u} u|N\rangle \approx \langle N|\bar{d} d|N\rangle$ (Or in the notation of Eq (\ref{defnG}), $G_S^{N,u}\approx G_S^{N,d}$, which is obtained both with lattice  and  $\chi$PT methods\cite{FlavourLatticeAveragingGroupFLAG:2021npn}).  As a result, the coefficients of scalar $p$ and $n$ operators need to be measured accurately,  in order to distinguish scalar coefficients involving  $u$s  vs $d$s. The second curiosity is that   ${\cal O}_{Au\perp,X}$ is different, when calculated in the quark or nucleon  bases. That is, Ref. \cite{Davidson:2018kud} obtained an orthonormal   basis of nucleon operators for the  two-dimensional space probed by Gold and Titanium. When these   nucleon operators are matching to quarks, they are no longer orthogonal.  In Ref. \cite{Davidson:2022nnl},  the  operators probed by Gold and Titanium were first matching to quarks, then decomposed into orthonormal components. We follow the second approach here, because  the aim is to identify the independent information available about models, and  that is expressed in the quark basis in our bottom-up perspective.

In the nucleon basis at the experimental scale, the
lepton-nucleon operators relevant for Spin-Independent $\muc$  can be added to the Lagrangian as \cite{KKO}
\beq
\d{\cal L} =  2\sqrt{2} G_F \sum_{N\in \{n,p\}}
\sum_{  X\in\{L,R\}}
\left( \widetilde{C}_{S,X}^{(NN)} \widetilde{{\cal O}}_{S,X}^{(NN)}
+ \widetilde{C}_{V,X}^{(NN)} \widetilde{{\cal O}}_{V,X}^{(NN)} + h.c. \right)
\label{Nops}
\eeq
where 
$\widetilde{{\cal O}}_{S,X}^{(NN)} = (\overline{e} P_X \mu) (\overline{N} N)$ and 
$\widetilde{{\cal O}}_{V,X}^{(NN)} =  (\overline{e} \g^\a P_X \mu) (\overline{N}\g_\a  N)$. Notice that the nucleon currents are not chiral: $\widetilde{{\cal O}}_{V,X}^{(NN)} = \widetilde{{\cal O}}_{V,XL}^{(NN)} + \widetilde{{\cal O}}_{V,XR}^{(NN)}$, so
\beq
\widetilde{C}_{V,X}^{(NN)} =\frac{1}{2} {\Big (}\widetilde{C}_{V,XL}^{(NN)} + \widetilde{C}_{V,XR}^{(NN)}{\Big )} ~~~.
\label{factor2}
\eeq
A similar relation holds for the coefficients on quarks used in $\muc$.

The Spin-Independent  conversion rate, normalised to the $\mu$ capture rate\footnote{ Some capture rates are given in 
~\cite{KKO,Suzuki:1987jf}; the capture rates in Aluminium, Titanium and Gold  are respectively
$0.7054$, 2.59, and 13.07 $\times 10^{6}$ sec$^{-1}$}
    $ (\mu + A\to  \nu+ A')$, can be written \cite{KKO}
    \bea
{\rm BR}_{SI}(\mu A \to eA) &=&   \frac{32G_F^2 m_\m^5 }{ \Gamma_{cap}}   
 {\Big ( } \big|     
   \widetilde{C}^{pp}_{V,R} I_{A,V}^{(p)} + \widetilde{C}^{pp}_{S,L}  I_{A, S}^{(p)}
+ \widetilde{C}^{nn}_{V,R} I_{A, V}^{(n)} + \widetilde{C}^{nn}_{S,L}  I_{A,S}^{(n)} 
+  C_{D,L} {\frac{I_{A,D}}{4}}  
 \big|^2   + \{ L \leftrightarrow R \}~ {\Big )} ~~~. 
 \label{BRmecKKO}
 \eea
 The  nucleus($A$) and nucleon($N$) -dependent 
``overlap  integrals''  $ I_{A,V}^{(N)}$, $I_{A,S}^{(N)},I_{A,D}$
correspond to the integral 
over the nucleus  of the lepton wavefunctions
and the appropriate  nucleon  density; we use here
the results of \cite{KKO}.  We neglect smaller 
contributions to the $\mu\to e$ conversion amplitude,
such as the Spin-Dependent part \cite{Cirigliano:2017azj,Davidson:2017nrp,Hoferichter:2022mna} which is not coherently  enhanced by the numerous nucleons, 
and  subdominant Spin-Independent contributions arising 
 from, {\it eg}   $\mu e\g\g$ operators \cite{Davidson:2020ord}, or from  tensor contributions to the scalar coefficient \cite{Cirigliano:2017azj}.

Focusing on an outgoing electron of left helicity, 
  the Branching Ratio can be re-expressed as \cite{Davidson:2018kud,Davidson:2022nnl} ($\widetilde{B}_{A}$ here is written $B_A$ in \cite{Davidson:2018kud})
    \bea
{\rm BR}_{SI}(\mu A \to e_L A) &=&  \widetilde{B}_A |\vec{\tilde{C}}_L\cdot \hat{v}_{A,5} |^2
\label{defn}
\eea
where the coefficients (of operators with outgoing $e_L$)  are lined up in a vector $\vec{\tilde{C}}_L$,
the overlap integrals are lined up in a five-component vector
$\vec{v}_{A5} = (I_{A,S}^{(p)},I_{A,V}^{(p)},I_{A,S}^{(n)},I_{A,V}^{(n)},\frac{1}{4}I_{A,D})$, and the target-dependent constants $ \widetilde{B}_A $  are
\beq
\widetilde{B}_A = \frac{6144\pi^3|\vec{v}_{A5}|^2}{2.197\Gamma_{capt, A}10^{-6} {\rm sec} }
\label{tildeB}
\eeq
with $\widetilde{B}_{Al}=142$,$\widetilde{B}_{Ti}=250$, and $\widetilde{B}_{Au}=300$.

In order to identify the  four-fermion {\it operator} probed by a target $A$ (as opposed to the combination of coefficients),
it is convenient to temporarily neglect the dipole. 
This implies that $\muc$ on the target nucleus $A$
probes the combination of coefficients
$\tilde{C}_{A,L} \equiv \vec{\tilde{C}}_L \cdot \hat{v}_{A4}$
where  $ \vec{v}_{A4} $  contains only the four-fermion overlap integrals
(so neglects the dipole).
So $\tilde{C}_{A,L} $ is the coefficient of the operator
\beq
\widetilde{{\cal O}}_{A,L} \equiv \hat{v}_{A4} \cdot (  \widetilde{{\cal O}}_{S,R}^{(p)},  \widetilde{{\cal O}}_{V,L}^{(p)},
\widetilde{{\cal O}}_{S,R}^{(n)},  \widetilde{{\cal O}}_{V,L}^{(n)})
\label{defOA}
\eeq
because the  Branching Ratio resulting from $\d {\cal L} = 
\tilde{C}_{A,L}  \widetilde{{\cal O}}_{A,L}$ will be
$$
      {\rm BR}_{SI}(\mu A \to e_L A) =  \widetilde{B}_A |\tilde{C}_{A,L}|^2
      \frac{|\vec{v}_{A4}|^2}{|\vec{v}_{A5}|^2} 
$$
      in agreement with eqn (\ref{defn}). In general,
\beq
 {\rm BR}_{SI}(\mu A \to e_L A) =  \widetilde{B}_{Al}\left|\frac{I_A^D}{4|v_{A5}|} C^{e \m}_{DR} + \frac{|v_{A4}|}{|v_{A5}|} \tilde{C}_{Al,L}\right|^2
+ L\leftrightarrow R \label{app:Albd}
\eeq
where  on light targets like Aluminium, $I_A^D/(4|v_{A5}|)= 0.27$ and
$|v_{A4}|/|v_{A5}| \simeq$0.96.

      In order to express the Al and Au operators  in the quark basis, there is one more step.
The nucleon operators  of eqn (\ref{Nops})  can be matched at 2 GeV onto light quark operators
${\cal O}_{S,X}^{qq} = (\overline{e} P_X \mu) (\overline{q} q)
$,
${\cal O}_{V,X}^{qq} = (\overline{e} \g^\a P_X \mu) (\overline{q} \g_\a  q)
$
using
\bea
\langle N(P_f)| \bar{q}(x) \Gamma_O q(x)|N(P_i) \rangle \simeq
G^{N,q}_O \langle N (P_f)| \bar{N}(x) \Gamma_O N(x)|N (P_i)\rangle=G^{N,q}_O\overline{u_N}(P_f) \Gamma_O u_N(P_i) e^{-i(P_f-P_i)x} \label{defnG} ~~~.
\eea
where the  parameters $G^{N,q}_O$ can be determined from sum rules, lattice calculations or experiment, and we use the  values summarised in
\cite{Davidson:2022nnl}.  
Below the heavy quark ($ b$ and $c$)  mass scales, there is also a  two-step contribution  to scalar nucleon operators  from  scalar  quark operators~\cite{SVZ,CKOT}, via   matching first onto the gluon operator $(\bar{e} P_R \mu)GG$, then onto nucleons\footnote{ The scalar operator with top quarks  also   matches at $m_t$ onto  $(\bar{e} P_R \mu)GG$, whose QCD running we suppose is accounted for by the wavefunction  contributions, and  which at 2 GeV 
 matches  onto nucleons like for  the $b$ and $c$:  
  $$G^{NQ}_S \simeq 0.9 \frac{2m_N }{27m_Q(m_Q) }
  \frac{\a_s(m_Q)}{ \a_s(2{\rm GeV}) }~~.$$
However, we  do not write the top contribution because it is absent from 
the EFT below the weak scale,  so  we include it at $m_t$}. 
As a result, the nucleon and quark coefficients are related as
\bea
 \widetilde{{C}}^{N}_{O,X} = \sum_q G_O^{Nq} C_{O,X}^{qq}~~~,\label{text}
\eea
for $q\in \{u,d,s,c,b,t\}$, and $O\in\{S,V\}$.   
This allows to define quark ``overlap integrals'' for target $A$ as 
\bea
I^q_{A,S}& =& G_{S}^{pq} I_{A,S}^p + G_{S}^{nq} I_{A,S}^n \nonumber\\
 I^q_{A,V}& = &G_{V}^{pq} I_{A,V}^p + G_{V}^{nq} I_{A,V}^n
\label{quarkOI}
\eea
which can be assembled into a ``target vector'' $\vec{u}_A$  in the space of  operators describing $\mu\to e$ conversion on quarks: 
\bea
\vec{u}_{A} &=& (I^u_{A,S},I^d_{A,S}, I^s_{A,S}, I^c_{A,S},I^b_{A,S},I^t_{A,S},
I^u_{A,V},I^d_{A,V}) ~~,
\label{uA}\\
\hat{u}_{Al} &=& (0.692,   0.699,   0.0341 ,  0.00400,  0.00121, 0.0000118,   0.125 ,  0.128 ) ~~,~~ |\vec{u}_{Al}| = 0.397 \label{uAl}
\\
\hat{u}_{Ti} &=& (0.690, 
0.699 ,  0.0340 ,  0.00398,   0.00121,  0.0000118,  0.127,   0.134 ) ~~,~~ |\vec{u}_{Ti}| = 0.991 \label{uTi} \\
\hat{u}_{Au} &=& (
0.672 ,  0.689 ,  0.0334 ,  0.00380 ,  0.00118,
 0.0000115 , 0.177 ,  0.202  ) ~~,~~ |\vec{u}_{Au}| =  1.923 \label{uAu}
 \eea
 Analogously to eqn (\ref{defOA}), we  use $\hat{u}_{Al} \approx \hat{u}_{Ti}$ to define ${\cal O}_{Al}$ in eqn
(\ref{OAlight}).  The orthogonal direction probed by Gold, $\hat{u}_{Au \perp}$ is defined as
\bea
\vec{u}_{Au}&=& |\vec{u}_{Au}| (\cos\theta_{\rm A} \hat{u}_{Al} + \sin\theta_{\rm A} 
\hat{u}_{\perp})
\nonumber\\
\Rightarrow \hat{u}_{\perp}& =& \frac{ \hat{u}_{Au} -\cos\theta_{A} \hat{u}_{Al} }{\sin\theta_{A}}
\nonumber\\
~~~~~~~~~~~~~~~~~~~~~~~~
\hat{u}_{\perp}& =& (-0.21, -0.099, -0.0075, -0.001, -0.0003, -0.000003, 0.558,0.7956)
\label{thetaAlAu}
\eea
where the operators are in the order given in Eq. (\ref{uA}),  sin$\theta_{A} \approx 0.093$ and
$\theta_{A}$ is $\approx$ 5.3 degrees\cite{Davidson:2022nnl}.
This gives the definition  of  ${\cal O}_{Au\perp}$ in eqn
(\ref{OAheavyperp}).

Finally, it can be convenient, in the quark  operator basis,  to  write  
\bea
BR(\mucL) &=& B_A {\Big |}\vec{C}_L\cdot \hat{u}_A +
C_{D,R}\frac{ I_D}{4|\vec{u}_A|}{\Big |}^2
\label{dA}
\eea
where $B_A$ is defined as  is eqn (\ref{tildeB}) but with $\vec{v}_{A5} \to \vec{u}_A$,
$B_{Al} =19363$,
$B_{Ti} =32860$,
$B_{Au} =24519$,
and $d_A = I_D/(4|\vec{u}_A|)= 0.0228, 0.0219,0.0246 $ respectively for Al, Ti and Au. 

\section{observable coefficients at $\LNP$}
\label{ssec:CLNP}

The coefficients  of the Lagrangian of Eq (\ref{Lag1}), which are at the experimental scale $m_\mu$, are written  in \cite{Davidson:2020hkf} as linear combinations of coefficients at some higher scale  taken to be $m_W$.  These formulae allows to identify what magnitude of which coefficients needs to be retained in matching to the models. This  Appendix gives  the linear combination of  coefficients probed by $\muc$ on  Aluminium, not given in \cite{Davidson:2020hkf} and also on Gold for completeness.

Using the results of \cite{KKO}, the combination of coefficients at the scale $\Lambda$   which is probed by
$\muc$ on Aluminium, for outgoing $e_L$, is: 
\bea
\sqrt{\frac{BR^{exp}_{Al}}{\widetilde{B}_{Al}}}
&\gsim &  {\Big|} 0.266 C_{D,R} (m_\mu) +
1.454C^{uu}_{V,L} + 1.490C^{dd}_{V,L}
 -0.86 \frac{8\pi m_N}{9\a_sm_t}C_{GG,R} \nonumber\\ &&
  - { \frac{\alpha}{3\pi}}
{\Big (}2( C^{uu}_{V,L} + C^{cc}_{V,L})
-( C^{dd}_{V,L} + C^{ss}_{V,L}  + C^{bb}_{V,L})
-( C^{ee}_{V,L} + C^{\mu \mu}_{V,L}  + C^{\tau \tau}_{V,L}) {\Big )}\widetilde{\ln}
\nonumber\\
&&+\frac{\alpha}{\pi}
{\Big (}(  1.5 C^{dd}_{A,L} -  3C^{uu}_{A,L} ) 
+ \frac{1}{6}
( C^{ee}_{V,L} + C^{\mu \mu}_{V,L}
- 
 C^{ee}_{A,L} -C^{\mu \mu}_{A,L}) {\Big)}\widetilde{\ln}
\nonumber\\ 
&&
+ \eta^{a_S}\left(1+  \frac{13 \alpha}{6 \pi} \widetilde{\ln} \right) (8.06 C^{uu}_{S,R}   + \frac{1.7m_N}{27m_c} C^{cc}_{S,R} )
+ \eta^{a_S}\left(1+  \frac{5 \alpha}{3 \pi} \widetilde{\ln} \right)
( 8.14 C^{dd}_{S,R} + 0.405C^{ss}_{S,R} + \frac{1.7m_N}{27m_b}C^{bb}_{S,R})
\nonumber\\ 
&&
 -\eta^{a_T}f_{TS}  \frac{4\alpha}{ \pi}  
{\Big(} 16.12 C^{uu}_{T,RR}   + \frac{3.4m_N}{27m_c} C^{cc}_{T,RR} 
- 8.14 C^{dd}_{T,RR} - 0.405C^{ss}_{T,RR}
-\frac{1.7m_N}{27m_b}C^{bb}_{T,RR}) {\Big )} \widetilde{\ln}
{\Big|} \label{RGEsAl}
\eea
where the result for Titanium can be approximately obtained  by replacing $
\widetilde{B}_{Al} = 142  \longrightarrow \widetilde{B}_{Ti} = 250$.  The constraint from $\muc$ on  Gold is: 
\bea
4.9 \times 10^{-8}&\gsim&  {\Big|}\left(
0.222 C_{D,R}(m_\mu) + 1.602C^{uu}_{V,L} + 1.830C^{dd}_{V,L} -0.721 f_N^{GG}\frac{8\pi m_N}{9\a_sm_t}C_{GG,R}\right)
\label{bdmecAumW}\\
&&-\frac{\alpha}{2\pi} \widetilde{\ln}
{\Big (} 6.41C^{uu}_{A,L}  { -}  3.66C^{dd}_{A,L}
- 0.305({ C^{ee}_{V,L} } +  C^{\m\m}_{V,L} {- C^{ee }_{A,L}}- C^{\mu \mu}_{A,L}) 
+ 0.228 C_{ping,1}  {\Big )} \nonumber \\
&&{ +\eta^{a_S} {\Big (}
1+  \frac{ \alpha}{ 4\pi}  \frac{26}{3} \widetilde{\ln}  {\Big )} }
 \left(  6.10C^{uu}_{S,R} +\frac{1.30m_N}{27m_c}C^{cc}_{S,R}
  \right)  
{ +\eta^{a_S} {\Big (}
1+  \frac{ \alpha}{ 4\pi}  \frac{20}{3} \widetilde{\ln}  {\Big )} }
\left( 6.26C^{dd}_{S,R} +0.303 C^{ss}_{S,R} 
 +  \frac{1.30m_N}{27m_b}C^{bb}_{S,R}\right)  \nonumber \\
&&
-  f_{TS}  \eta^{a_T} \frac{4 \alpha}{ \pi} \widetilde{\ln}
\left( 2(6.100 C^{uu}_{T,RR}   + \frac{1.30m_N}{27m_c} C^{cc}_{T,RR} )
- 6.258 C^{dd}_{T,RR} - 0.303C^{ss}_{T,RR}  -\frac{1.30m_N}{27m_b}C^{bb}_{T,RR} \right)  {\Big|}
\nonumber
\eea
where
we used the scalar quark densities in the nucleon of Ref. \cite{Hoferichter:2015dsa}, 
the coefficients are at $\Lambda$, $m_f = m_f(m_f)$,
$$
C^{ff}_{V,L}\equiv \frac{1}{2}(C^{e \mu ff}_{V,LR} + C^{e \mu ff}_{V,LL} )~~~,~~~
C^{ff}_{A,L}\equiv \frac{1}{2}(C^{e \mu ff}_{V,LR} - C^{e \mu ff}_{V,LL} ) ~~~,~~~ 
C^{ff}_{S,L}\equiv \frac{1}{2}(C^{e \mu ff}_{S,LR} + C^{e \mu ff}_{S,LL} )
$$
$\eta$ and   $f_{TS}$ are defined  in Eq.(\ref{soln1tG}), 
$\widetilde{\ln} \equiv \ln (\Lambda/m_\psi)$ where
$m_\psi \in\{m_b, 2~{\rm GeV},m_\mu\}$ is a suitably chosen lower cutoff for the logarithm,  and the dipole contribution
is given at the experimental
scale (it can be written in terms of coefficients at $\Lambda$
as given in \cite{Davidson:2020hkf}). A similar bound applies with
$L$ and $R$ interchanged.

.

  \section{Matching  the models with the EFT}
\label{app:matching}

This section summarises the matching of the  models with EFT. We first discuss   which EFT to match onto in Section \ref{assec:quelEFT}, and  how  we define the lepton flavour basis in Section \ref{assec:EH}, then
give  the  matching of   our  three models   onto a QED$\times$QCD-invariant  EFT  in section \ref{assec:matching}. 
In section \ref{assec:ping},  we  discuss some curiosities related to the  EFT description of  penguin  diagrams in the models.

\subsection{Which EFT to match to?}
\label{assec:quelEFT}

The models we study contain new  particles  with masses around the TeV, so we need a recipe to match these models onto the QCD$\times$QED-invariant EFT appropriate below the weak scale.

A formally correct  approach would be to match the models to the EFT at $m_W$. 
  Indeed, this would be similar to  the  weak-scale  matching in the SM, where one removes  the  $t,W,Z$ and $h$ at $m_W$, because there is no  ``large log'' argument for   sequential matching out of $t,h,Z$ then $W$. 
  However,  this   approach has two drawbacks: first, it  implies calculating  many loop diagrams  involving SM particles and  New Physics in the broken SM,
  and  second,    resumming QCD from the TeV to the weak scale   is unobvious because  the loops one calculates are  electroweak. (But the QCD issue may be minor, because   $\a_s\lsim .12$   varies by $\sim 30\%$ from $m_Z\to $TeV.)
  We will follow this approach for  the inverse seesaw,  where the leading contributions arise in  electroweak loops(that induce no field bilinears running under QCD), and  below $m_W$ the  RGEs  of QED only  modify the magnitude of the coefficients. 

If we  match  to an  EFT at $\LNP$,  there could be fewer diagrams to calculate, and QCD can be resummed. 
However,   the scale ratio TeV/$v$ is not large,  so  higher order contributions in the EFT expansions in log-enhanced loops and operator dimension can be relevant.
For instance,   ln(TeV/$m_W)\simeq$2.5 and   ln(TeV/$m_t)\simeq$1.75, so one-loop model diagrams  with  weak-scale particles in the  loop  can have finite  parts comparable to the log-enhanced part --- despite that in EFT power-counting,  these finite parts  should be included at NLL, and not at LL where they can give undesirable dependence on the renormalisation scheme of the operators.
Also,  by power-counting, dimension eight operators are  only suppressed with respect to dimension six by a factor $v^2/{\rm TeV}^2\sim$ 0.03.

The obvious EFT  to use at a TeV  is SMEFT, which can then be matched at  $m_W$  to the  low-energy EFT.
This two-step matching  should allow to  correctly resum QCD at all scales.
Some of the four-fermion operators present in the low-energy EFT, first arise in SMEFT at dimension eight ---  an example is the ${\cal O}_{S,XX}^{e\mu ee}$ operator  that contributes at tree level to $\meee$ (see Eq. \ref{Lag1}).
Matching the models to SMEFT then the low-energy EFT  generates  contributions to these  additional operators  at ${\cal O}(v^2/(\LNP^2 v^2))$ via the ``penguin'' operators (see Eq.s \ref{LLpenguin}-\ref{EEpenguin}) or ${\cal O}_{EH}$, but  in order to recover contributions  at ${\cal O}(v^2/\LNP^4)$, we need to match onto dimension eight operators, and include the dimension six$^2\to$ dimension eight mixing in the RGEs.   Dimension eight operators are legion in SMEFT \cite{Murphy:2020cly,Li:2020gnx}, so this requires  cherry-picking the relevant operators.

We also tried matching out the new particles and the electroweak bosons at $\LNP$,   such that below $\LNP$, we use the RGEs of QCD and QED.
In this approach,  there are four-fermion operators induced by the Higgs and the $Z$ in our EFT, and   including the dimension six$^2 \to$ dimension eight  mixing in the RGEs reproduces  the  ${\cal O}(\alpha \log)$ terms obtained in SMEFT at dimension six, as well as the ${\cal O}(1/\LNP^4)$ terms.
This is simpler than using SMEFT, because   there is no intermediate matching, and its easier to pick out the relevant dimension eight operators.
However, this approximation of treating electroweak particles as contact interactions when they are dynamical,
can in  some  cases
\footnote{  In matching   the leptoquark  to SMEFT at $\LNP$, the  ``higgs-penguin'' diagrams of the model correspond to the mixing of the tree-induced  scalar  ${\cal O}_{LEQU}$ into ${\cal O}_{EH}$.   Similarly, the
  $\propto v^2$-$Z$-penguin diagrams correspond to the mixing of $2l2q$ vector operators into the penguin operators  ${\cal O}_{HL1}$,  ${\cal O}_{HL3}$,  and/or  ${\cal O}_{HE}$. All the operators are of dimension six and involve at most two quarks, so the effect of QCD is as discussed around Eq. (\ref{soln1tG}).
It is straightforward to see that,  for $\eta \simeq 1+\delta$, 
this modification of the electroweak anomalous dimension is ${\cal O}(\d^2)$, so  we neglect this effect.

In the QED$\times $QCD-invariant EFT, the log-enhanced  contribution of box and $Z$-penguin diagrams arises in the dimension six$^2\to $ dimension eight RGEs via  ``fish diagrams'' (see eg Figure \ref{fig:pingZ1}) combining  a  $Z$- or Higgs-induced four-quark operator with a  leptoquark-induced   two-lepton-two-quark   operator. The  QCD running  for these diagrams  is more complicated (four-quark operators mix under QCD), and includes the wrong diagrams.   So using the RGEs of 
QCD and QED between  the weak scale and  $\LNP$,  would  resum the wrong QCD for penguins and boxes. 
It  gives the correct QCD running below $m_W$,  but such fish diagrams have a light ($\neq t$) quark in the loop, so  are negligible due to the  Yukawa suppression (We nonetheless  include the light quarks in the formulae, in order to construct attractive ``invariants'').
} give the wrong    QCD and/or QED  running from $\LNP\to m_W$.
And unfortunately,  numerous loop calculations must be performed in the model to obtain the potentially relevant  finite part of electroweak loops.

In summary, we aim for the first, correct, approach, augmented by resummed one-loop QCD corrections. But sometimes  we neglect finite matching contributions. %
For conciseness, we give the matching results at $\LNP$ (rather than $m_W$), because  the RGEs we implement automatically  will include the log-enhanced QED and QCD loops.
The finite and log$(\Lambda/m_W)$-enhanced  loops involving $h$,$Z$ and $W$, and
the dimension six$^2\to $ dimension eight mixing are included in the matching.  In resumming QCD, we  use five flavour anomalous dimensions  at all scales.

\subsection{The  Yukawa correction ${\cal O}_{EH}$}
\label{assec:EH}

When lepton flavour is {\it not} conserved,
the charged lepton mass eigenstate basis  provides 
a pragmatic and unambiguous definition of lepton flavour.
It is a significant simplification to always work in this basis,
not only in the EFT, but also in models. 
This circumvents a problem that could arise 
in  matching  models onto EFT:
if the model induces  the SMEFT  ``Yukawa correction'' operator
$${\cal O}_{EH} =  H^\dagger H \overline{\ell} H e$$ then the mass eigenstate basis that defines lepton flavour is rotated with respect to the Yukawa eigenbasis, which seems the   obvious definition of flavour in a model.
If the rotation from Yukawa to mass eigenbases is performed  in matching, the resulting corrections are  at dimension 8 in the EFT($\propto C_{EH}\times C_{other}$),  to which  some $\me$ observables can be sensitive~\cite{Ardu:2021koz}.
However, to obtain self-consistent results at dimension 8 requires  including dimension 6 operators in the equations of motion when  reducing the operator basis (this is particularily relevant for the mass correction  ${\cal O}_{EH}$).
In addition, performing this basis rotation appears daunting.
\\

The simplest solution is define the charged lepton  flavours   as the ``mass'' eigenstates also in the model. This means that the Yukawa matrix  $Y_e$ is not quite diagonal,  and that the flavour eigenstates cannot be simply obtained   from the  Lagrangian of the model. Instead, the matrix  $ [C_{EH}]^{\a\b}$  must be calculated,  then one diagonalises
$$
[m_e] \simeq [Y_e] v + [C_{EH}]v ~~~.
$$
However,  since the Lagrangian is  invariant under flavour basis transformations,  an explicit calculation of the basis is rarely required. And current LHC constraints on  flavour-changing Higgs decays impose that  renormalisation group effects of the off-diagonals of $[Y_e]$ are below the sensitivity of upcoming experiments  \cite{Ardu:2021koz}.
Following this approach,  we include the ${\cal O}_{EH}$ operator in the lepton equations of motion in our EFT, so that  we distinguish $[Y_e]$ which appears at Higgs vertices from $[m_e]/v$ which arises from the equations of motion.

\subsection{Matching Results}
\label{assec:matching}

The  matching  results are  given in the usual theory-motivated   operator basis of the EFT, where there are a large number of operators.  However, $\meg$, $\meee$ and $\muc$ can only probe   the coefficients of the 12 operators in the experimentally-observable subspace, which are given in section \ref{ssec:CLNP}. Since  
the 12 observables have  wildly varying  overlaps with many high-scale operators, its important to estimate the magnitude of all coefficients obtained in matching.
So  we  list the coefficients that we include in our results, and also    estimate  subleading corrections to those coefficients, and contributions to coefficients we took to be vanishing.

.

\subsubsection{type II matching-to-QED summary}
\label{ssec:t2matchsum}

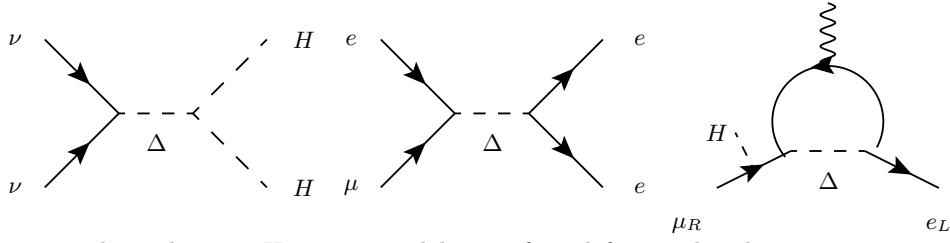
\begin{figure}[ht]
\begin{center}
  \unitlength.5mm
\SetScale{1.4}
\begin{picture}(45,40)(0,0)
\ArrowLine(0,40)(20,20)
\ArrowLine(0,0)(20,20)
\DashLine(40,20)(20,20){3}
\DashLine(60,40)(40,20){5}
\DashLine(40,20)(60,0){5}
\Text(70,40)[c]{$H$}
\Text(33,12)[r]{$\Delta$}
\Text(70,0)[c]{$H$}
\Text(-8,0)[c]{$\nu$}
\Text(-8,40)[c]{$\nu$}
\end{picture}
\hspace{2cm}
\begin{picture}(45,40)(0,0)
\ArrowLine(0,40)(20,20)
\ArrowLine(0,0)(20,20)
\DashLine(40,20)(20,20){3}
\ArrowLine(40,20)(60,40)
\ArrowLine(40,20)(60,0)
\Text(70,40)[c]{$e$}
\Text(33,12)[r]{$\Delta$}
\Text(70,0)[c]{$e$}
\Text(-8,0)[c]{$\mu$}
\Text(-8,40)[c]{$e$}
\end{picture}
\hspace{2cm}
\begin{picture}(45,40)(0,10)
\ArrowLine(0,10)(20,20)
\DashLine(40,20)(20,20){3}
\ArrowLine(40,20)(60,10)
\Text(33,12)[r]{$\Delta$}
\Text(60,0)[c]{$e_L$}
\Text(-8,0)[c]{$\mu_R$}
\Text(0,25)[c]{$H$}
\DashLine(5,25)(10,15){3}
\ArrowArc(30,28)(15,-28,220)
\Photon(30,43)(30,60){2}{4}
\end{picture}
 \end{center}
\caption{
Diagrams matching   the type II seesaw model onto, from left to right, the neutrino mass operator, the four-doublet-lepton operator, and the dipole. 
\label{fig:t2match}} 
\end{figure}

The type II seesaw model and its Lagrangian are  discussed in Section (\ref{sec:models}). At tree level in the model, one obtains the neutrino mass matrix 
\bea
\frac{ [m_\nu]_{\r\s}}{v}  &\simeq&
\frac{ [f^*]_{\r\s} \lambda_H v}{M_\Delta}  ~~~.
\label{mnutypeIIb}
\eea
Above the weak scale,  the lepton number changing neutrino mass matrix operator runs,  and  mixes with itself to generate LFV operators \cite{dim52} with  coefficients 
$\propto m_\nu^2/v^2$. We neglect  the running because TeV $\to m_W$ is not far, and
and the dimension five$^2 \to$ dimension six mixing because the effect is tiny.

There are also  four-doublet-lepton {\bf vector} operators with coefficient
\bea
C^{\a\b ll }_{VLL}  &\simeq &
\frac{[m^*_\nu]_{\beta l} [m_\nu]_{ \a l} }{ (1+ \d_{\a l} + \d_{\b l})|\lambda_H|^2 v^2 }
\label{CVLLt2v2}
\eea
where
$ \a,\b,l\in \{ e,\mu,\tau\}$ and  $\a\neq\b$.
Then at one-loop in the model, there is a contribution to the {\bf  dipole} coefficients:
\bea
(C^{e\mu  }_{DL}, C^{e\mu }_{DR}) &\simeq &\left(
 \frac{m_e}{m_\mu} \frac{3 e   }{ 128 \pi^2},
   \frac{3 e   }{ 128 \pi^2} \right) \times
 \frac{ [m_\nu m^*_\nu]_{ e \mu} }{|\lambda_H|^2  v^2 }
 \label{CD}
 \eea
where we recall that our dipole operator  includes a factor $m_\mu$.

In practise, we neglect $C^{e\mu  }_{DL} \propto m_e$, and the matching contributions to all the other operators;  we  estimate  below  the  largest  matching contributions we neglected  for each coefficient.

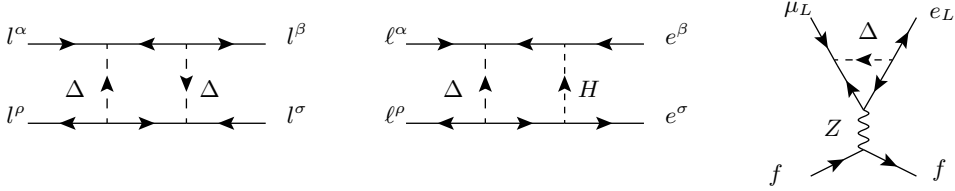
\begin{figure}[htb]
\begin{center}
\begin{picture}(70,70)(30,10)
\ArrowLine(0,60)(30,60)
\ArrowLine(60,60)(30,60)
\ArrowLine(60,60)(90,60)
\ArrowLine(90,30)(60,30)
\ArrowLine(30,30)(60,30)
\ArrowLine(30,30)(0,30)
\put(-8,60){$l^\a$} 
\put(-8,30){$l^\r$}
\put(98,60){$l^\b$} 
\put(98,30){$l^\s$}
\put(14,40){$\Delta$} 
\put(65,40){$\Delta$}
\DashArrowLine(30,30)(30,60){5}
\DashArrowLine(60,60)(60,30){5}
\end{picture}
\hspace{2cm}
\begin{picture}(70,70)(20,10)
\ArrowLine(0,60)(30,60)
\ArrowLine(60,60)(30,60)
\ArrowLine(90,60)(60,60)
\ArrowLine(60,30)(90,30)
\ArrowLine(30,30)(60,30)
\ArrowLine(30,30)(0,30)
\put(-8,60){$\ell^\a$} 
\put(-8,30){$\ell^\r$}
\put(98,60){$e^\b$} 
\put(98,30){$e^\s$}
\put(14,40){$\Delta$} 
\put(65,40){$H$}
\DashArrowLine(30,30)(30,60){5}
\DashArrowLine(60,30)(60,60){3}
\end{picture}
\hspace{2cm}
\begin{picture}(70,70)(0,-10)
\ArrowLine(0,50)(10,32)
\ArrowLine(31,34)(20,15)  
\ArrowLine(31,34)(40,50)
\ArrowLine(20,15)(10,32)
\put(-10,52){$\mu_L$} 
\put(45,50){$e_L$}
\put(18,42){$\Delta$} 
\ArrowLine(0,-10)(20,0) 
\ArrowLine(20,0)(40,-10)
\DashArrowLine(31,34)(9,34){3}
\put(-16,-12){$f$} 
\put(46,-10){$f$}
\put(5,5){$Z$} 
\Photon(20,15)(20,0){2}{3} 
\end{picture}
%
%
\end{center}
\caption{
  Representative diagrams  of matching contributions that we neglect, because they are subdominant or  below upcoming experimental sensitivity: from left to right, a box with two   triplets $\Delta$ that contributes to vector four-lepton operators, a box with a $\Delta$ and a Higgs,  and   the $Z$ penguin. 
  $\ell$ are doublet leptons,  and $e$ are singlets.
\label{fig:p+bt2}} 
\end{figure}

\ben
\item 
{\bf Vector four-fermion operators} ${\cal O}_{VLX}$ will be generated  with log-enhanced coefficients by the photon penguin  in renormalisation group running.
We list here some other one-loop  matching contributions to four lepton operators (illustrated in the first three diagrams of  Figure \ref{fig:p+bt2}):
there are boxes  exchanging   two $\Delta$s (first line of the expression  below), or a Higgs and a $\Delta$ (second line where $m_\d = {\rm max}\{m_\a,m_\b\}$)  and finally  the  $Z$-penguins  which are discussed in Appendix \ref{assec:ping}: 
\bea
(C^{\a\b ll }_{VLL}, C^{\a\b  ll}_{VLR}, C^{\a\b  ll }_{VRR}, 
C^{e\a\b  ll }_{VRL}) 
& \sim& -\left(\frac{5( [m_\nu m_\nu ^\dagger]^{ \a\b}  [f^* f]^{ll} +  [m_\nu m_\nu^*]^{\a l}  [f^* f ]^{ l \b}) }{64\pi^2 |\lambda_H|^2 v^2},   
0,0,0
\right) \nonumber
\\
&  + & \left(
0,
 [Y_e^\dagger m^*_\nu]^{ l \b} [Y^T_e m_\nu]^{l \a}   \log\frac{M_\Delta}{m_l} ,0,  [m^*_\nu Y_e]^{l \b} [m_\nu Y_e^*]^{l \a} \log\frac{M_\Delta}{m_\d} 
\right) \times \frac{3   }{32\pi^2 |\lambda_H|^2v^2}  \nonumber
\\
&+ &  {\Big (}2 g^e_L ,
2   g^e_R  , 0,0
{\Big )}\times 
\frac{1 }{ (16 \pi^2) |\lambda_H|^2 v^4 } {\Big [}m_\nu m^*_e \log \frac{M_\Delta}{m_e}  m_e^T m^*_\nu{\Big]}_{ \a \b}\nonumber
\eea
where recall  that $[m_\nu m_\nu^*]^{\a\b}$  that appears in the triplet-triplet boxes is determined by the neutrino oscillation parameters so  is independent of  the neutrino mass scale or Majorana phases.
In the $Z$-penguin diagrams, there are mass insertions on the internal lepton lines, for which we use a Feynman rule $-i[m_e]_{\a\a}/v$ rather than $-i [Y_e]_{\a\a}$, as discussed in Section \ref{assec:EH}. {The $g^f_X$ couplings here and in the following equations refer to the interactions of the $Z$ boson with a fermion $f$ vector current of chirality $X$. For instance, the Feynman rule for the $Z$ couplings to leptons reads $-i g/(2 c_W)\gamma^\alpha (g^e_L P_L+g^e_R P_R)$ }
\item {\bf vector operators with a quark bilinear} can also be generated by 
the $Z$-penguin: %
\bea
 (C^{e\mu QQ }_{VLL}, C^{e\mu QQ}_{VLR} , C^{e\mu QQ }_{VRR}, 
C^{e\mu QQ }_{VRL}) &\sim  &
  {\Big (}2 g^Q_L ,
2   g^Q_R  , 0,0
{\Big )}\times 
\frac{1 }{ (16 \pi^2) |\lambda_H|^2 v^4 } {\Big [}m_\nu m^*_e \log \frac{M_\Delta}{m_e}  m_e^T m^*_\nu{\Big]}_{ e \m} \label{CV-Qt2}
 \eea

\item {\bf Scalar four-lepton  operators} involving a $\tau$ bilinear can be generated by a  box where the Higgs and the triplet scalar $\Delta$ are exchanged, as illustrated in the middle diagram of Figure \ref{fig:p+bt2}. 
  This  gives ${\cal O}^{e \tau \tau \mu}_{V XY}$ operators,  which can be   Fiertz-transformed to $LR$ and $RL$ scalars, given on the first line below:
\bea
(C^{e\mu \tau\tau }_{SLL}, C^{e\mu \tau\tau }_{SRR},
 C^{e\mu \tau\tau }_{SLR}, C^{e\mu \tau\tau }_{SRL})
&\sim &
\label{CS-tau}
 {\Big (}
0,0,  [ Y^T_e m_\nu^*]^{\tau \m} [m_\nu Y^*_e]^{\tau e} ,
  [m_\nu^* Y_e]^{\tau \mu} [Y_e^\dagger  m_\nu]^{\tau e} {\Big )} \times \frac{3  }{32\pi^2  |\lambda_H|^2v^2} \log\frac{m_\Delta}{m_\d} 
  \nonumber \\
  && -   y_\tau
{\Big (} [ m_\nu m_\nu ^*  ]_{\mu e} y_e,   [  m_\nu m_\nu^*  ]_{e \mu} y_\mu,
 [  m_\nu m_\nu^*  ]_{\mu e}y_e ,   [  m_\nu m_\nu^* ]_{e \mu} y_\mu {\Big )}
\frac{3 \lambda_4 }{32\pi^2  |\lambda_H|^2 v^2}
\nonumber
\eea
where on  the first line $m_\d$ that provides the lower cutoff of the logarithm is the heaviest internal  lepton.  The scalar operators on the second line are  mediated by ``Higgs penguins'', or Higgs  exchange with  a   flavour-changing vertex corresponding  to the loop-induced  SMEFT coefficient
$C_{EH}^{e\mu} \simeq -3 \lambda_4 [m_\nu m_\nu^*Y_e]^{e\mu}/(64\pi^2|\lambda_H|^2 v^2)$, and we used $m_h^2/v^2 \simeq 1/2$ and a Higgs-triplet interaction $\delta {\cal L}\supset \lambda_4 \Delta^\dagger \Delta H^\dagger H$.
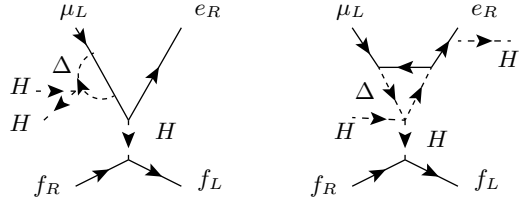
\begin{figure}[htb]
\begin{center}
\begin{picture}(70,70)(0,-10)
  \ArrowLine(0,50)(5,42)
  \Line(5,42)(20,15)
\ArrowLine(20,15)(40,50)
\DashArrowArcn(10,32)(9,300,110){2}
\put(-6,55){$\mu_L$} 
\put(45,55){$e_R$} 
\ArrowLine(0,-10)(20,0) 
\ArrowLine(20,0)(40,-10)
\DashArrowLine(-15,26)(0,26){4}
\DashArrowLine(0,26)(-12,15){4}
\put(-9,33){$\Delta$} 
\put(-16,-12){$f_R$} 
\put(46,-10){$f_L$}
\put(-25,25){$H$}
\put(-25,11){$H$}
\put(30,8){$H$} 
\DashArrowLine(20,15)(20,0){4} 
\end{picture}
\hspace{1cm} 
\begin{picture}(70,70)(0,-10)
\ArrowLine(0,50)(10,35)
\DashArrowLine(10,35)(20,15){2}
\DashArrowLine(20,15)(30,35){2}
\ArrowLine(30,35)(40,50)
\ArrowLine(30,35)(10,35)
\put(-6,55){$\mu_L$} 
\put(45,55){$e_R$} 
\ArrowLine(0,-10)(20,0) 
\ArrowLine(20,0)(40,-10)
\DashArrowLine(0,16)(20,15){4}
\DashArrowLine(40,45)(60,45){4}
\put(1,23){$\Delta$} 
\put(-16,-12){$f_R$} 
\put(46,-10){$f_L$}
\put(55,35){$H$}
\put(-7,8){$H$}
\put(28,5){$H$} 
\DashArrowLine(20,15)(20,0){5} 
\end{picture}
\end{center}
\caption{
  Representative diagrams  generating scalar four-fermion operators  via Higgs exchange. If the model is matched to SMEFT, the  external fermion current $\bar{f}f$ should not be included and the diagrams match onto the operator
  ${\cal O}^\dagger_{EH}= H^\dagger H \bar{e} H^\dagger \ell$, which generates LFV Higgs couplings. Or the diagrams can be matched directly to a QED$\times$QCD invariant four-fermion operator. \label{fig:sp+bt2}} 
\end{figure}

\item
 Higgs ``penguin diagrams'', as illustrated in Figure \ref{fig:sp+bt2} and which generated the second line of the previous equation,  can also   generate {\bf  scalar operators} involving quarks  $Q \in \{d,u, s,c,b\}$, and other flavours of lepton  $l\in \{e,\mu \}$.   They are all  suppressed by two Yukawa couplings and a loop:
\bea
( C^{e\mu ll}_{SLL} , C^{e\mu ll }_{SRR} )
\label{CS-l}
&\sim & -  y_l
{\Big (} [ m_\nu m_\nu ^*  ]_{\mu e} y_e,   [  m_\nu m_\nu^*  ]_{e \mu} y_\mu{\Big )}
\frac{3 \lambda_4 }{32\pi^2  |\lambda_H|^2 v^2}
\nonumber\\
(C^{e\mu QQ }_{SLL}, C^{e\mu QQ}_{SRR},
 C^{e\mu QQ }_{SLR}, C^{e\mu QQ }_{SRL})
& & -   y_Q
{\Big (} [ m_\nu m_\nu ^*  ]_{\mu e} y_e,   [  m_\nu m_\nu^*  ]_{e \mu} y_\mu,
 [  m_\nu m_\nu^*  ]_{\mu e}y_e ,   [  m_\nu m_\nu^* ]_{e \mu} y_\mu {\Big )}
\frac{3 \lambda_4 }{32\pi^2  |\lambda_H|^2 v^2} ~~~. 
\nonumber
\eea
\item
Finally,  we neglect the {\bf tensor operators }in this model. The $\tau$-tensors can be induced at two loop,  suppressed by two lepton Yukawas,  via Higgs exchange with $\tau \to \{e,\mu\}$ flavour-changing couplings at both vertices~\cite{Dorsner:2015mja} ({\it e.g.} as illustrated in Figure \ref{fig:sp+bt2}).

\een

\subsubsection{Inverse seesaw matching-to-QED summary}\label{app:invmatch}
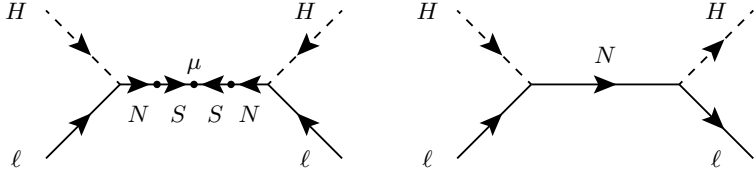
\begin{figure}[ht]
	\begin{center}
		\unitlength.5mm
		\SetScale{1.4}
		\begin{picture}(45,40)(20,0)
			\DashArrowLine(0,40)(20,20){3}
			\ArrowLine(0,0)(20,20)
			\ArrowLine(20,20)(30,20)
			\Vertex(30,20){1}
			\ArrowLine(30,20)(40,20)
			\Vertex(40,20){1}
			\ArrowLine(50,20)(40,20)
			\Vertex(50,20){1}
			\ArrowLine(60,20)(50,20)
			\DashArrowLine(80,40)(60,20){3}
			\ArrowLine(80,0)(60,20)
			\Text(40,25)[c]{$\mu $}
			\Text(70,40)[c]{$H $}
			\Text(28,12)[r]{$N$}
			\Text(38,12)[r]{$S$}
			\Text(48,12)[r]{$S$}
			\Text(58,12)[r]{$N$}
			\Text(70,0)[c]{$\ell$}
			\Text(-8,0)[c]{$\ell$}
			\Text(-8,40)[c]{$H$}
		\end{picture}
		\hspace{3 cm}
		\begin{picture}(45,40)(20,0)
	\DashArrowLine(0,40)(20,20){3}
	\ArrowLine(0,0)(20,20)
	\ArrowLine(20,20)(60,20)
	\DashArrowLine(60,20)(80,40){3}
	\ArrowLine(60,20)(80,0)
	\Text(40,28)[c]{$N$}
	\Text(70,40)[c]{$H $}
	\Text(70,0)[c]{$\ell$}
	\Text(-8,0)[c]{$\ell$}
	\Text(-8,40)[c]{$H$}
\end{picture}
	\end{center}
	\caption{
		Tree level matching diagrams in the inverse seesaw. The left diagram contributes to the Weinberg operator and gives neutrino masses, while the diagram on the right matches onto the penguin operators $\mathcal{O}_{HL1,3}$.
		\label{fig:seesawtree}} 
\end{figure}
The inverse seesaw is introduced in Section \ref{sec:models} and the Lagrangian is written in Eq.~(\ref{Linverse}). At tree-level, the left diagram of Fig. \ref{fig:seesawtree} gives the following neutrino mass matrix 
\begin{equation}
   [m_\nu]^{\a\b} \simeq  [Y_\nu M^{-1} \mu M^{-1} Y_\nu^T]^{\a\b} v^2.
\end{equation}
In the following we neglect the matching contributions that are $\propto m_\nu$, thus suppressed by the small neutrino masses,  and we only consider the insertion of lepton number conserving interactions. The right diagram in Fig.~(\ref{fig:seesawtree}) gives tree-level matching contribution to the SMEFT penguins $C_{HL1,3}$, contributing to the difference $C_{HL1}-C_{HL3}$ but without modifying the fermion $Z$ couplings $\propto( C_{HL1}+C_{HL3})$.
\begin{figure}[ht]
	\begin{center}
		\unitlength.5mm
		\SetScale{1.3}
		\hspace{3 cm}
		\begin{picture}(65,60)(20,0)
			\DashArrowLine(40,40)(20,20){3}
			\ArrowLine(0,0)(20,20)
			\ArrowLine(20,20)(40,20)
			\Vertex(40,20){1}
			\ArrowLine(40,20)(60,20)
			\Photon(60,20)(40,40){2}{4}
			\ArrowLine(60,20)(80,0)
			\DashArrowLine(20,60)(40,40){3}
			\DashArrowLine(40,20)(40,0){3}
			\Photon(60,60)(40,40){2}{4}
			\Text(70,0)[c]{$\ell$}
			\Text(30,14)[c]{$N$}
			\Text(50,14)[c]{$\ell$}
			\Text(-8,0)[c]{$\ell$}
		\end{picture}
		\hspace{3 cm}
		\begin{picture}(65,60)(20,0)
			\ArrowLine(0,0)(20,20)
			\ArrowLine(20,20)(60,20)
			\DashArrowArc(40,20)(20,0,180){3}
			\ArrowLine(60,20)(80,0)
			\DashArrowLine(30,20)(30,0){3}
			\DashArrowLine(50,0)(50,20){3}
			\Photon(40,40)(40,60){2}{4}
			\Text(70,0)[c]{$\ell$}
			\Text(40,25)[c]{$\ell$}
			\Text(-8,0)[c]{$\ell$}
			\Text(25,25)[c]{$N$}
			\Text(55,25)[c]{$N$}
		\end{picture}
	\end{center}
	\caption{
		Representative one-loop diagrams contributing to the $Z$ fermion couplings in the inverse seesaw. Additional topologies are possible.
		\label{fig:seesawpengloop}} 
\end{figure}
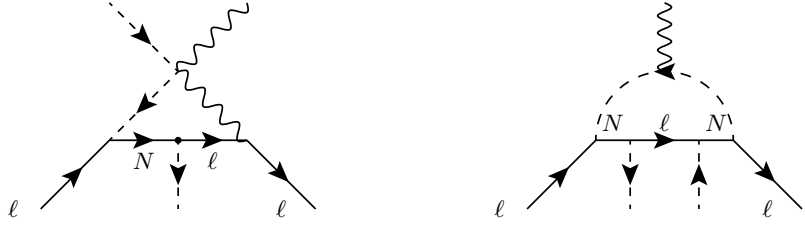
Including the one-loop corrections arising from the (representative) diagrams in Fig.~\ref{fig:seesawpengloop}, we find
\begin{eqnarray}
	(C_{HL1}-C_{HL3})_{\alpha\beta}&=&\frac{v^2}{2}(Y_\nu M_a^{-2} Y^\dagger_\nu)_{\a\b}+{\rm one-loop}
	\\
	(C_{HL1}+C_{HL3})_{\alpha\beta}&=&\frac{v^2}{384\pi^2}(g^{\prime2}+17 g^2)(Y_\nu M_a^{-2} Y^\dagger_\nu)_{\a\b}\left(\log\left(\frac{m_W}{M_a}\right)+\frac{11}{6}\right)\nonumber
	\\
	&-&\frac{v^2}{32\pi^2}\left(Y_{\nu}(Y^\dagger_\nu Y_\nu)_{ab}\frac{1}{M^2_{a}-M^2_{b}}
	\ln\left(\frac{M^2_{a}}{M^2_{b}}\right)Y_\nu^\dagger\right)_{\a\b}
\end{eqnarray}
where $g', g$ are respectively the hypercharge and SU(2) gauge couplings.
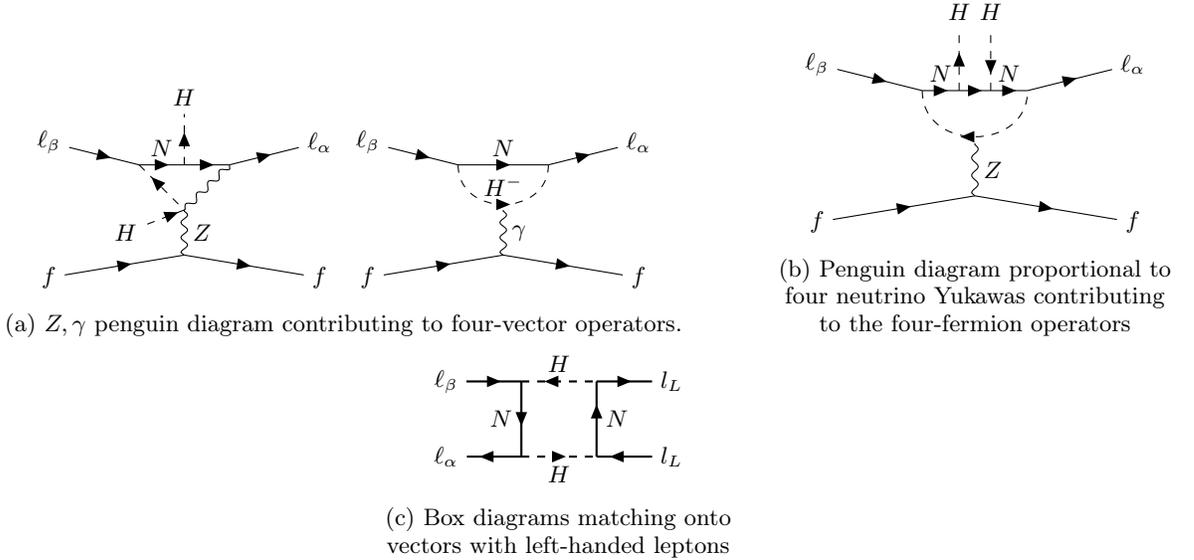
\begin{figure}[ht]
	\centering
	\begin{subfigure}{0.6\textwidth}
		\begin{tikzpicture}[scale=1.2]
			\begin{feynman}
				\vertex (a) at (0, 2);
				\vertex (lin) at (-1, 2.25) {\(\ell_\beta\)};
				\vertex (b) at (1, 2);
				\vertex (Hlin) at (0.5, 2);
				\vertex (H) at (0.5, 2.75) {\(H\)};
				\vertex (H1) at (-0.15, 1.25)  {\(H\)};
				\vertex (gb1) at (0.5, 1.5);
				\vertex (gb2) at (0.5, 1);
				\vertex (f1) at (-1, 0.75) {\(f\)};
				\vertex (f2) at (2, 0.75) {\(f\)};
				\vertex (lout) at (2, 2.25) {\(\ell_\alpha\)};
				
				\diagram* [small]{
					(lin) -- [fermion] (a) -- [fermion, edge label=\(N\)] (Hlin) --  [fermion] (b) -- [fermion]  (lout),
					(Hlin) -- [charged scalar] (H),
					(a) -- [anti charged scalar]  (gb1),
					(gb1) -- [photon] (b),
					(gb1) -- [anti charged scalar] (H1),
					(gb1) -- [photon, edge label=\(Z\)] (gb2), 
					(f1) -- [fermion] (gb2) -- [fermion] (f2),
				};
			\end{feynman}
		\end{tikzpicture}
		\begin{tikzpicture}[scale=1.2]
			\begin{feynman}
				\vertex (a) at (0, 2);
				\vertex (lin) at (-1, 2.25) {\(\ell_\beta\)};
				\vertex (b) at (1, 2);
				\vertex (Hline1) at (0.35, 2);
				\vertex (Hline2) at (0.65, 2);
				\vertex (gb1) at (0.5, 1.5);
				\vertex (gb2) at (0.5, 1);
				\vertex (f1) at (-1, 0.75) {\(f\)};
				\vertex (f2) at (2, 0.75) {\(f\)};
				\vertex (lout) at (2, 2.25) {\(\ell_\alpha\)};
				
				\diagram* [small]{
					(lin) -- [fermion] (a) --[fermion, edge label=\(N\)] (b) -- [fermion] (lout),
					(a) -- [charged scalar, half right,  edge label=\(H^-\)]  (b),
					(gb1) -- [photon, edge label=\(\gamma\)]  (gb2), 
					(f1) -- [fermion] (gb2) -- [fermion] (f2),
				};
			\end{feynman}
		\end{tikzpicture}
		\caption{ $Z,\gamma$ penguin diagram contributing to four-vector operators.}
		\label{fig:penguinYsq}
	\end{subfigure}\quad
	\begin{subfigure}{0.3\textwidth}
		\begin{tikzpicture}[scale=1.4]
			\begin{feynman}
				\vertex (a) at (0, 2);
				\vertex (lin) at (-1, 2.25) {\(\ell_\b\)};
				\vertex (b) at (1, 2);
				\vertex (Hline1) at (0.35, 2);
				\vertex (H1) at (0.35, 2.75) {\(H\)};
				\vertex (Hline2) at (0.65, 2);
				\vertex (H2) at (0.65, 2.75) {\(H\)};
				\vertex (gb1) at (0.5, 1.5);
				\vertex (gb2) at (0.5, 1);
				\vertex (f1) at (-1, 0.75) {\(f\)};
				\vertex (f2) at (2, 0.75) {\(f\)};
				\vertex (lout) at (2, 2.25) {\(\ell_\a\)};
				
				\diagram* [small]{
					(lin) -- [fermion] (a) --[fermion, edge label=\(N\)] (Hline1) -- [fermion] (Hline2) -- [fermion, edge label=\(N\)] (b) -- [fermion] (lout),
					(a) -- [anti charged scalar, half right]  (b),
					(gb1) -- [photon, edge label=\(Z\)]  (gb2), 
					(f1) -- [fermion] (gb2) -- [fermion] (f2),
					(Hline1) -- [charged scalar] (H1),
					(Hline2) -- [anti charged scalar] (H2),
				};
			\end{feynman}
		\end{tikzpicture}
		\caption{Penguin diagram proportional to four  neutrino Yukawas contributing to the four-fermion operators}
		\label{fig:penguinfourY}
	\end{subfigure}
	\quad
	\begin{subfigure}{0.3\textwidth}
		\begin{tikzpicture}
			\begin{feynman}[large]
				\diagram* [inline=(c.base), horizontal=a to b] {
					i1 [particle=\(\ell_\beta\)]
					-- [fermion] a
					-- [anti charged scalar, edge label=\(H\)] b
					-- [fermion] f1 [particle=\(l_L\)],
					i2 [particle=\(\ell_\alpha\)]
					-- [anti fermion] c
					-- [charged scalar, edge label'=\(H\)] d
					-- [anti fermion] f2 [particle=\(l_L\)],
					{ [same layer] a -- [fermion, edge label'=\(N\)] c },
					{ [same layer] b -- [anti fermion, edge label=\(N\)] d},
				};
			\end{feynman}
		\end{tikzpicture}
		\caption{Box diagrams matching onto vectors with left-handed leptons}
		\label{fig:boxll}
	\end{subfigure}
	\caption{Matching contributions to flavour changing vector operators in the inverse seesaw. The diagrams are illustrative and we do not draw all possible topologies~\label{fig:diagramstypeI}}
\end{figure}
The vector four  lepton operators get contribution from the $\gamma$ and $Z$ penguins (Figs \ref{fig:penguinYsq} and \ref{fig:penguinfourY}), as well as from box diagrams (Fig \ref{fig:boxll}) when the external leptons are left-handed
\bea
(C^{\a\b ll }_{VLL}, C^{\a\b  ll}_{VLR}, C^{\a\b  ll }_{VRR}, 
C^{\a\b  ll }_{VRL})&=& \frac{1}{192\pi^2}\left(\frac{g^{\prime 2}+g^2}{(1+\delta_{\alpha l}+\delta_{\beta l})}, g'^2, 0, 0\right)\times (Y_\nu M_a^{-2} Y^\dagger_\nu)_{\a\b}\left(\log\left(\frac{m_W}{M_a}\right)+\frac{11}{6}\right)\nonumber\\
&+&(C_{HL1}+C_{HL3})_{\alpha\beta}\left(g^e_L, g^e_R, 0, 0\right)\nonumber\\
+\left(1, 0, 0, 0\right)\times \frac{1}{64 \pi^2}	\bigg[ Y^{\alpha a}_\nu Y^{*\beta a}_\nu Y^{l b}_\nu Y^{*l b}_\nu &+& (1-\delta_{\a l})(1-\delta_{\beta l})Y^{l a}_\nu Y^{*\beta a}_\nu Y^{\a b}_\nu Y^{*l b}_\nu\bigg]\frac{1}{M^2_{a}-M^2_{b}}\ln\left(\frac{M^2_{a}}{M^2_{b}}\right)
\eea
where $\alpha,\beta,l\in \{e,\mu,\tau\}$ and $\alpha\neq \beta$. Similarly, the two-lepton two-quark vectors matching conditions are
\bea
(C^{\a\b QQ }_{VLL}, C^{\a\b QQ}_{VLR} , C^{\a\b QQ }_{VRR}, 
C^{\a\b QQ }_{VRL})&=& \frac{1}{384\pi^2}\left(-\frac{g^{\prime 2}}{3}+\eta_Q g^2, -2q_Qg'^2, 0, 0\right)\times (Y_\nu M_a^{-2} Y^\dagger_\nu)_{\a\b}\left(\log\left(\frac{m_W}{M_a}\right)+\frac{11}{6}\right)\nonumber\\
&+&(C_{HL1}+C_{HL3})_{\alpha\beta}\left(g^Q_L, g^Q_R, 0, 0\right)
\eea
where $q_Q$ is the electric charge of the $Q=u,d$ quark and $\eta_{u}=-\eta_d=1$.  
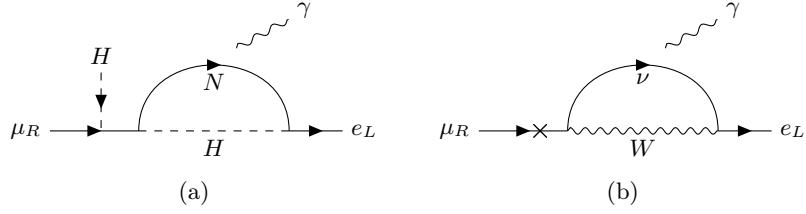
\begin{figure}
		\begin{subfigure}{0.3\textwidth}
		\begin{tikzpicture}
			\begin{feynman}[small]
				\vertex (mu) at (-1.5,0) {\(\mu_R\)};
				\vertex (e) at (3,0) {\(e_L\)};
				\vertex (H1) at (-0.5,1) {\(H\)};
				\vertex (H2) at (-0.5,0);
				\vertex (a)  at (0,0) ;
				\vertex (b) at (2,0) ;
				\vertex (boson1) at (1.5, 1);
				\vertex (boson2) at (2, 2);
				\vertex (c) at (1.3,1.1) ;
				\vertex (H3) at (2.2,1.6) {\(\gamma\)};
				\diagram* [inline=(a.base)]{
					(H1) -- [charged scalar] (H2),
					(H3) -- [photon] (c),
					(mu) -- [fermion] (a),
					(b) -- [fermion] (e),
					(a) -- [fermion, half left, edge label'=\(N\)] (b)  -- [scalar, edge label=\(H\)] (a),
				};
			\end{feynman}
		\end{tikzpicture}
		\caption{\label{fig:dipoleinverseN}}
	\end{subfigure}\quad
			\begin{subfigure}{0.3\textwidth}
		\begin{tikzpicture}
			\begin{feynman}[small]
				\vertex (mu) at (-1.5,0) {\(\mu_R\)};
				\vertex (e) at (3,0) {\(e_L\)};
				\vertex (H2) at (-0.5,0);
				\vertex (a)  at (0,0) ;
				\vertex (b) at (2,0) ;
				\vertex (boson1) at (1.5, 1);
				\vertex (boson2) at (2, 2);
				\vertex (c) at (1.3,1.1) ;
				\vertex (H3) at (2.2,1.6) {\(\gamma\)};
				\diagram* [inline=(a.base)]{
					(H3) -- [photon] (c),
					(mu) -- [fermion, insertion=0.69] (a),
					(b) -- [fermion] (e),
					(a) -- [fermion, half left, edge label'=\(\nu\)] (b)  -- [photon, edge label=\(W\)] (a),
				};
			\end{feynman}
		\end{tikzpicture}
		\caption{\label{fig:dipoleinversenu}}
	\end{subfigure}
	\caption{One-loop diagrams matching onto the $\mu\to e$ dipole in the inverse seesaw. In the right diagram a flavour changing W coupling arising from the $\mathcal{O}_{HL3}$ operator is inserted between the charged lepton and an active neutrino}
\end{figure}
We neglect all vectors with right-handed flavour changing currents because they are suppressed by two insertions of the lepton Yukawas, and arise at dimension eight or at higher-loops
\bea
(C^{\a\b QQ }_{VRR}, 
C^{\a\b QQ }_{VRL})\sim ( C^{\a\b QQ }_{VLR}, 
C^{\a\b QQ }_{VLL})\times y_\alpha y_\beta\left\{\frac{v^2}{M^2}, \frac{1}{16\pi^2}\right\}\nonumber\\
(C^{\a\b ll }_{VRR}, 
C^{\a\b ll}_{VRL})\sim ( C^{\a\b ll }_{VLR}, 
C^{\a\b ll }_{VLL})\times y_\alpha y_\beta \left\{\frac{v^2}{M^2}, \frac{1}{16\pi^2}\right\}\nonumber
\eea
giving contributions below the future experimental sensitivities. Similar Yukawa suppressions are expected for the scalar operators, which we neglect.

 The dipole receive matching contributions from diagrams involving virtual sterile neutrinos (Fig \ref{fig:dipoleinverseN}), as well as from the exchange of a virtual active neutrino between flavour conserving and flavour changing ($\propto C_{HL3}$) $W$ couplings (Fig \ref{fig:dipoleinversenu}). The resulting dipole coefficients are
\bea
(C^{e\mu  }_{DL}, C^{e\mu }_{DR}) &= &-\left(
\frac{m_e}{m_\mu} \frac{e   }{ 32 \pi^2},
\frac{e   }{ 32 \pi^2} \right) \times v^2(Y_\nu M_a^{-2} Y^\dagger_\nu)_{e\mu }
\label{CDinv}
\eea 
\begin{figure}
	\SetScale{1.25}
   \begin{picture}(65,60)(20,0)
	\ArrowLine(0,0)(20,20)
	\DashArrowLine(0,20)(10,10){3}
	\ArrowLine(20,20)(60,20)
	\DashArrowArc(40,20)(20,0,180){3}
	\ArrowLine(60,20)(80,0)
	\DashArrowLine(30,20)(30,0){3}
	\DashArrowLine(50,0)(50,20){3}
	\Photon(40,40)(40,60){2}{4}
	\Text(70,0)[c]{$e_L$}
	\Text(40,25)[c]{$\ell$}
	\Text(-8,0)[c]{$\mu_R$}
	\Text(25,25)[c]{$N$}
	\Text(55,25)[c]{$N$}
    \end{picture}\quad
    \hspace{2 cm}
	\begin{picture}(65,60)(20,0)
		\ArrowLine(0,0)(20,20)
		\DashArrowLine(0,20)(10,10){3}
		\ArrowLine(20,20)(60,20)
		\DashArrowArc(40,20)(20,0,180){3}
		\ArrowLine(60,20)(80,0)
		\DashArrowArc(40,20)(10,180,360){3}
		\Photon(40,40)(40,60){2}{4}
		\Text(70,0)[c]{$e_L$}
		\Text(40,25)[c]{$\ell$}
		\Text(-8,0)[c]{$\mu_R$}
		\Text(25,25)[c]{$N$}
		\Text(55,25)[c]{$N$}
	\end{picture}
	\caption{Diagrams featuring four insertion of the neutrino Yukawas matching on the $\mu\to e$ dipole. At the leading order, these arise at dimension eight (left) or at the two-loop level (right)\label{fig:dipole4ynu}}
\end{figure}
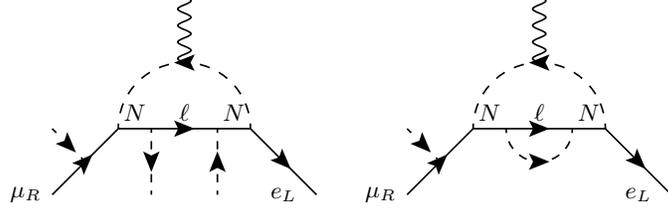
As depicted in Fig \ref{fig:dipole4ynu}, matching contributions featuring four neutrino Yukawa matrices can emerge either at dimension eight or via higher-loops. To estimate their magnitude we consider the dimension-six results and account for the appropriate suppression factors, leading to
\bea
(C^{e\mu  }_{DL}, C^{e\mu }_{DR})\bigg|_{Y^4_\nu} &\sim &\left(
\frac{m_e}{m_\mu} \frac{e   }{ 32 \pi^2},
\frac{e   }{ 32 \pi^2} \right) \times v^2\left(Y_{\nu}(Y^\dagger_\nu Y_\nu)_{ab}\frac{1}{M^2_{a}-M^2_{b}}
\ln\left(\frac{M^2_{a}}{M^2_{b}}\right)Y_\nu^\dagger\right)_{\a\b}\times  \left\{\frac{v^2}{M^2}, \frac{1}{16\pi^2}\right\}\nonumber.
\eea
Although the dimension-eight and two-loop suppressions are insufficient to push these contributions beyond the reach of experiments with no further assumption on the neutrino Yukawas, they nonetheless have a negligible impact on the correlations between $\mu\to e$ observables that the model can predict.

\subsubsection{Leptoquark matching-to-QED summary}
\label{ssec:LQmatchsum}

{
The leptoquark model is introduced in Section \ref{sec:models}, and the Lagrangian is given in Eq. (\ref{LLQ}).
The $S_1$ leptoquark alone  does not  induce  neutrino masses; for instance, adding another leptoquark \cite{Deppisch:2016qqd}, such as an SU(2)-doublet leptoquark of hypercharge Y=1/6,   could  generate neutrino masses at one loop via diagrams familiar from $R$-parity-violating Supersymmetry \cite{Dreiner:1997uz,Davidson:2000ne}.  Or, in the presence of   a colour-octet Majorana fermion, the $S_1$ leptoquark  can generate neutrino masses at two-loop \cite{Cai:2017wry}.}

At tree level in the model,  the leptoquark  induces  vector, scalar and tensor  flavour-changing operators involving $u$-type  quarks and  charged leptons of both chiralities. Then at one-loop, there are various penguin and box diagrams, which  match onto   dimension six operators of SMEFT and dimension eight   four-fermion operators  in SMEFT.

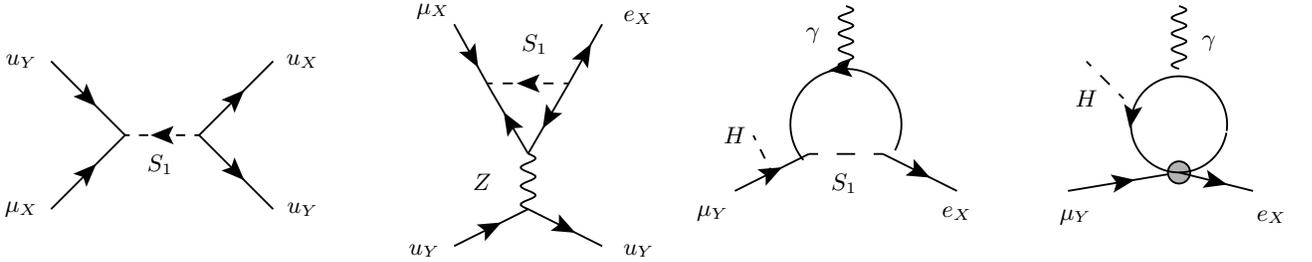
\begin{figure}[ht]
\begin{center}
  \unitlength.5mm
\SetScale{1.4}
\begin{picture}(45,40)(60,-10)
\ArrowLine(0,40)(20,20)
\ArrowLine(0,0)(20,20)
\DashArrowLine(40,20)(20,20){3}
\ArrowLine(40,20)(60,40)
\ArrowLine(40,20)(60,0)
\Text(68,40)[c]{$u_X$}
\Text(33,12)[r]{$S_1$}
\Text(68,0)[c]{$u_Y$}
\Text(-8,0)[c]{$\mu_X$}
\Text(-8,40)[c]{$u_Y$}
\end{picture}
%
%
\begin{picture}(70,70)(0,-10)
\ArrowLine(0,50)(10,32)
\ArrowLine(31,34)(20,15)  
\ArrowLine(31,34)(40,50)
\ArrowLine(20,15)(10,32)
\put(-10,52){$\mu_X$} 
\put(45,50){$e_X$}
\put(18,42){$S_1$} 
\ArrowLine(0,-10)(20,0) 
\ArrowLine(20,0)(40,-10)
\DashArrowLine(31,34)(9,34){3}
\put(-12,-12){$u_Y$} 
\put(45,-12){$u_Y$}
\put(5,5){$Z$} 
\Photon(20,15)(20,0){2}{3} 
\end{picture}
\hspace{1cm}
\begin{picture}(45,40)(0,-5)
\ArrowLine(-20,10)(0,20)
\DashLine(20,20)(0,20){5}
\DashLine(-10,15)(-15,25){3}
\Text(3,52)[r]{$\g$}
\ArrowLine(20,20)(40,10)
\Text(13,12)[r]{$S_1$}
\Text(40,4)[c]{$e_X$}
\Text(-26,4)[c]{$\mu_Y$}
\Text(-20,25)[c]{$H$}
\ArrowArc(10,28)(15,-28,220)
\Photon(10,43)(10,60){2}{4}
\ArrowLine(70,10)(100,15)
\GCirc(100,15){3}{.7}
\ArrowLine(100,15)(120,10)
\Text(125,3)[c]{$e_X$}
\Text(72,3)[c]{$\mu_Y$}
\DashLine(85,35)(75,45){3}
\Text(110,50)[r]{$\g$}
\Text(75,35)[c]{$H$}
\ArrowArc(100,28)(13,-8,350)
\Photon(100,43)(100,60){2}{4}
\end{picture}
 \end{center}
\caption{
  Diagrams matching   the $S_1$ leptoquark onto, from left to right,
  two-lepton two-quark  operators (vector for $X=Y$, scalar and tensor for $X\neq Y$, where $X,Y \in \{L,R\}$),  a $Z$-penguin contribution to the $2l2u$ vector operators,  and two  diagrams that contribute to the  dipole: a finite matching part $\propto \lambda_X \lambda_X^\dagger y_\mu $, and the tensor$\to$ dipole mixing   $\propto \lambda_Y y _Q \lambda_X^\dagger  $. 
\label{fig:LQmatch1}} 
\end{figure}

  At tree level in the model,
 {\bf tensor and scalar  coefficients involving  leptons and up-type quarks}  arise as illustrated in  Figure \ref{fig:LQmatch1}:
 \bea
 (C^{e\mu QQ }_{TLL}, C^{e\mu QQ }_{TRR})
&\simeq&\left(\frac{  \lambda^{e Q}_R \lambda^{\mu Q *}_L  }{8},
\frac{ \lambda^{e Q}_L \lambda^{\mu Q *}_R  }{8}\right)
\times\frac{v^2}{m_{LQ}^2}~~~,~~~ Q\in \{u,c,t\}
~~\label{CT-Q} \\
(C^{e\mu QQ }_{SLL}, C^{e\mu QQ}_{SRR},
C^{e\mu QQ }_{SLR}, C^{e\mu QQ }_{SRL})
&\simeq&- \left(\frac{  \lambda^{e Q}_R \lambda^{\mu Q *}_L  }{2},
\frac{ \lambda^{e Q}_L \lambda^{\mu Q *}_R  }{2},
0,0 \right)
\times\frac{v^2}{m_{LQ}^2}~~~,~~~ Q\in \{u,c,t\}
\label{CS-Q}
\eea

The {\bf vector  $2l2u$ operators}, for  $Q\in \{u,c,t\}$,   are
\bea
(C^{e\mu QQ }_{VLL}, C^{e\mu QQ }_{VRR}, 
 C^{e\mu QQ }_{VRL}, C^{e\mu QQ}_{VLR}) &\simeq &\left(
  \frac{ \lambda^{eQ}_{L} \lambda^{\mu Q*}_{L}}{2}  ,  \frac{ \lambda^{eQ}_{R} \lambda^{\mu Q*}_{R}}{2}  , 0, 0 \right)
 \times \frac{v^2}{m_{LQ}^2} ~~~
 \label{CV-Q}\\
 && + {\Big (}
 g_L^u (C_{HL1}^{e\mu}+ C_{HL3}^{e\mu}) , g_R^u C_{HE}^{e\mu}~ ,
 g_R^u (C_{HL1}^{e\mu}+ C_{HL3}^{e\mu})~ , g_L^u C_{HE}^{e\mu} {\Big )}
 ~~,~~\label{CV-Qping}
 \eea
 where  the second line gives   the contribution of the  $Z$-penguin  for $Q\in \{u,c\}$(illustrated  in Figure \ref{fig:LQmatch1}, and discussed further in Section
 \ref{assec:ping}), and 
\bea
\Delta C^{e\mu}_{HL1} = \Delta C^{e\mu}_{HL3} &\simeq&  { -} \frac{ N_c v^2  }{32\pi^2 m_{LQ}^2} \left( \ [\lambda_L Y_u \ln \frac{m_{LQ}}{m_Q} Y_u^\dagger \lambda^\dagger_L]^{e\mu} - \frac{5}{6} [\lambda_L Y_uY_u^\dagger \lambda^\dagger_L]^{e\mu} \right)
\label{LQHL13}\\
\Delta C^{e\mu}_{HE} &\simeq&   { -} \frac{ N_c v^2  }{16\pi^2 m_{LQ}^2} \left(  [\lambda_R Y^\dagger_u \ln \frac{m_{LQ}}{m_Q} Y_u \lambda_R]^{e\mu} - \frac{5}{6} [\lambda_R Y_u^\dagger Y_u \lambda^\dagger_R]^{e\mu} \right)~~~.
\label{LQHE}
\eea

There is a finite matching contribution to  the {\bf dipole} coefficients, illustrated in  Figure \ref{fig:LQmatch1}: 
\bea
 C^{e\mu  }_{DL} &\simeq &\left( 
 \frac{ e   [\lambda_R\lambda^\dagger_R]_{e\mu}  }{128\pi^2}
 + \frac{ 3 e N_c Q_u  }{64\pi^2 y_\mu}
 {\Big [}\lambda_R Y_u(m_Q)  \lambda^{\dagger}_L
  {\Big ] }_{ e\mu} 
 \right) \times\frac{v^2}{m_{LQ}^2}
 ~~~~~~
 \label{CDLLQ}\\
 C^{e\mu }_{DR} &\simeq &\left(
  \frac{ e  [\lambda_L\lambda^\dagger_L]_{e\mu} }{128\pi^2}
+ \frac{3  e N_c Q_u  }{64\pi^2 y_\mu}
{\Big [} \lambda_L Y_u(m_Q) 
  \lambda^{\dagger}_R {\Big ]}_{e\mu} 
 \right) \times\frac{v^2}{m_{LQ}^2}~~~~~~
 \label{CDRLQ}
 \eea
 where   the second term is the finite part of the tensor to dipole mixing (represented in the  last diagram of   Figure \ref{fig:LQmatch1}),  which will be  included via  the QED RGEs evolving down from  $m_{LQ}$.

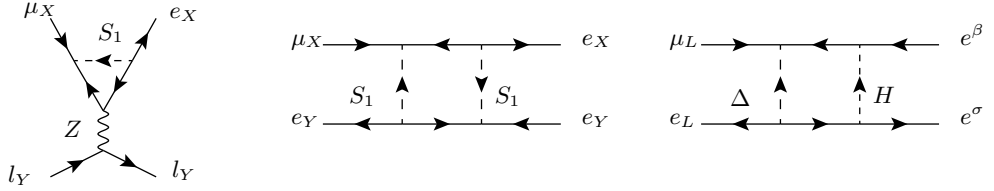
\begin{figure}[htb]
\begin{center}
\hspace{2cm}
\begin{picture}(70,70)(0,-10)
\ArrowLine(0,50)(10,32)
\ArrowLine(31,34)(20,15)  
\ArrowLine(31,34)(40,50)
\ArrowLine(20,15)(10,32)
\put(-10,52){$\mu_X$} 
\put(45,50){$e_X$}
\put(18,42){$S_1$} 
\ArrowLine(0,-10)(20,0) 
\ArrowLine(20,0)(40,-10)
\DashArrowLine(31,34)(9,34){3}
\put(-16,-12){$l_Y$} 
\put(46,-10){$l_Y$}
\put(5,5){$Z$} 
\Photon(20,15)(20,0){2}{3} 
\end{picture}
\hspace{2cm}
\begin{picture}(70,70)(30,10)
\ArrowLine(0,60)(30,60)
\ArrowLine(60,60)(30,60)
\ArrowLine(60,60)(90,60)
\ArrowLine(90,30)(60,30)
\ArrowLine(30,30)(60,30)
\ArrowLine(30,30)(0,30)
\put(-12,60){$\mu_X$} 
\put(-12,30){$e_Y$}
\put(98,60){$e_X$} 
\put(98,30){$e_Y$}
\put(10,38){$S_1$} 
\put(65,38){$S_1$}
\DashArrowLine(30,30)(30,60){5}
\DashArrowLine(60,60)(60,30){5}
\end{picture}
\hspace{2cm}
\begin{picture}(70,70)(20,10)
\ArrowLine(0,60)(30,60)
\ArrowLine(60,60)(30,60)
\ArrowLine(90,60)(60,60)
\ArrowLine(60,30)(90,30)
\ArrowLine(30,30)(60,30)
\ArrowLine(30,30)(0,30)
\put(-12,60){$\mu_L$} 
\put(-12,30){$e_L$}
\put(98,60){$e^\b$} 
\put(98,30){$e^\s$}
\put(11,37){$\Delta$} 
\put(65,38){$H$}
\DashArrowLine(30,30)(30,60){5}
\DashArrowLine(60,30)(60,60){3}
\end{picture}
%
%
\end{center}
\caption{
  Representative diagrams illustrating the $Z$-penguin and box contribution to vector four-lepton operators, and the leptoquark-Higgs box that can generate $2l2d$ operators. 
\label{fig:LQmatch2}} 
\end{figure}

{\bf  Vector four-lepton operators} ${\cal O}_{V,XY}$ can arise  via $Z$-penguins and boxes,
 as illustrated in Figure  \ref{fig:LQmatch2},
 and give
\bea %
(C^{e\mu ll }_{VLL}, C^{e\mu ll }_{VRR}, 
C^{e\mu ll }_{VRL}, C^{e\mu }_{VLR}) &\simeq &{\Big (}
 g_L^e (C_{HL1}^{e\mu}+ C_{HL3}^{e\mu}) , g_R^e C_{HE}^{e\mu}~ ,
 g_R^e (C_{HL1}^{e\mu}+ C_{HL3}^{e\mu})~ , g_L^e C_{HE}^{e\mu} {\Big )}
  ~~~~
 ~~~~~~~~~~~~~~~~~~
 \label{CV-l}\\
~~~~~-\frac{N_cv^2 }{64\pi^2m_{LQ}^2 }\times &&
{\Big (}
  [\lambda_L \lambda_L^\dagger]_{e \mu} [\lambda_L \lambda_L^\dagger]_{ll}
 ,
   [\lambda_R \lambda_R^\dagger]_{e \mu} [\lambda_R \lambda_R^\dagger]_{ll}
 , [\lambda_R \lambda_R^\dagger]_{e \mu} [\lambda_L \lambda_L^\dagger]_{ll}
 , [\lambda_L \lambda_L^\dagger]_{e \mu} [\lambda_R \lambda_R^\dagger]_{ll}
    {\Big )}
 \label{CV-lbox}
 \eea
 where $l\in \{ e,\mu,\tau\}$, the first line is  the $Z$-penguin, and  the second is the boxes.

 Finally, {\bf vector operators involving leptons and $d$-type quarks} can arise  via $Z$-penguins and boxes with a leptoquark and an electroweak boson, which give 
 \bea
(C^{e\mu FF }_{VLL}, C^{e\mu FF }_{VRR}, 
 C^{e\mu FF }_{VRL}, C^{e\mu FF}_{VLR}) &\simeq &\left(
  \frac{ g^2 [ \lambda_L V]^{e F} [\lambda_{L}V]^{ \mu F*}}{32\pi^2}  , 0 , 0, 0 \right)
  \times \frac{v^2}{m_{LQ}^2} \ln  \frac{m_{LQ}}{m_W} ~~~~~~~  
  \label{CV-F}\\
  &&+ {\Big (}
 g_L^d (C_{HL1}^{e\mu}+ C_{HL3}^{e\mu}) , g_R^d C_{HE}^{e\mu}~ ,
 g_R^d (C_{HL1}^{e\mu}+ C_{HL3}^{e\mu})~ , g_L^d C_{HE}^{e\mu} {\Big )}
  ~~,~~  \label{CV-Fping}
 \eea
 where $F\in \{d,s,b\}$ and the  box has no finite part at dimension six.

\begin{figure}[htb]
\begin{center}
\begin{picture}(70,70)
\ArrowLine(0,60)(8,45)
\ArrowLine(33,45)(40,60)
\ArrowLine(20,20)(8,45)
\ArrowLine(33,45)(20,20)
\put(-6,63){$\mu_X$} 
\put(45,63){$e_Y$}
\put(20,52){$S_1$} 
\ArrowLine(0,-10)(20,0) 
\ArrowLine(20,0)(40,-10)
\DashLine(7,45)(33,45){3}
\put(-16,-12){$u_Y$} 
\put(46,-10){$u_X$}
\put(23,10){$H$} 
\DashLine(20,20)(20,0){3}
\DashLine(0,5)(20,10){3}
\DashLine(0,15)(20,10){3} 
\end{picture}
\hspace{2cm}
\begin{picture}(70,70)(30,10)
\ArrowLine(0,60)(25,60)
\ArrowLine(60,60)(25,60)
\ArrowLine(60,60)(90,60)
\ArrowLine(90,30)(60,30)
\ArrowLine(25,30)(60,30)
\ArrowLine(25,30)(0,30)
\put(-12,60){$\mu_X$} 
\put(-12,30){$e_X$}
\put(98,60){$l_Y$} 
\put(98,30){$l_Y$}
\put(10,40){$S_1$} 
\put(65,40){$S_1$}
\put(50,68){$H$} 
\put(50,18){$H$}
\DashArrowLine(30,30)(30,60){5}
\DashArrowLine(60,60)(60,30){5}
\DashLine(45,60)(45,80){3}
\DashLine(45,30)(45,10){3}
\end{picture}
%
%
\hspace{2cm}

\end{center}
\caption{
  Representative diagrams  of matching contributions that we neglect, because they are subdominant or  below upcoming experimental sensitivity: from left to right, a Higgs penguin for $u$ quarks (no Higgs self-interaction is required on the Higgs line for $d$ quarks), and a scalar four-lepton box. 
 \label{fig:p+bLQ}} 
\end{figure}
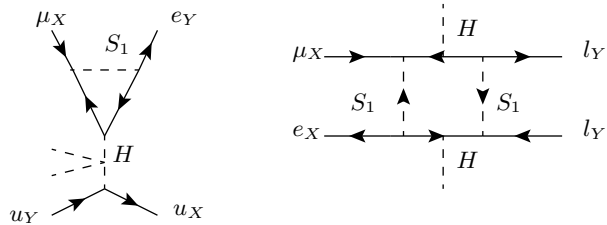

{

{\bf Scalar   $\bm{2l2d}$  operators}  can be generated by ``Higgs-penguin'' diagrams, as illustrated for instance in  the first diagram of Figure \ref{fig:p+bLQ}.   There could also be  boxes  with a leptoquark and a Higgs, but these would be suppressed by  a small quark Yukawa squared,  so appear relatively suppressed with respect to the Higgs penguins. 
  Such diagrams can also  contribute to  scalar $2l2u$ operators, which already  arise   at tree level as given  in  Eq. (\ref{CS-Q}):
\bea
(C^{e\mu QQ }_{SLL}, C^{e\mu QQ}_{SRR},
(C^{e\mu QQ }_{SLR}, C^{e\mu QQ }_{SRL})
&\simeq &  -2 y_Q\left(
 C_{EH}^{\mu e*} ,  C_{EH}^{e\mu},
 C_{EH}^{\mu e*} ,  C_{EH}^{e\mu}
\right)
~~~,~~~ Q\in \{u,c\}
~~~~~~~~~~~~~~~~~~
 \nonumber
\\
(C^{e\mu FF }_{SLL}, C^{e\mu FF }_{SRR},
C^{e\mu FF }_{SLR}, C^{e\mu FF}_{SRL})
&&\sim -2y_F
\left( C_{EH}^{\mu e*} ,  C_{EH}^{e\mu} ,
 C_{EH}^{\mu e*} ,  C_{EH}^{e\mu}\right)  
\nonumber
\eea
where
$ F \in \{d,s,b\}$, and   the  lepton-flavour-changing vertex  of the Higgs is $\propto C_{EH}^{e \mu}$. 
This  SMEFT coefficient can be generated at one-loop in the model via  various diagrams,   and is of order: 
\bea
C_{EH}^{e \mu} &\simeq&\frac{ N_c v^2 }{64\pi^2 m_{LQ}^2} \left(
 \lambda_4 [ \lambda_L \lambda^\dagger_L Y_e ]_{e \mu} 
+4 \lambda_4 [ \lambda_L Y_u \lambda_R^\dagger ]_{e \mu} 
+   [\lambda_L Y_uY_u^\dagger \lambda^\dagger_L  Y_e]_{e \mu} 
 +4 [\lambda_L Y_uY_u^\dagger  Y_u\lambda^\dagger_R]_{e \mu}   
\left[1 -2\ln\frac{m_{LQ}}{m_Q}\right]
\right)~~~~~~~~~~
\label{EHLQfin}
\eea
where the final log-enhanced term   could apparently be large for top quarks in the loop, but $\meg$ imposes $\lambda_L^{et}\lambda_R^{\mu t} \lsim 10^{-2}$, as
discussed in section  \ref{ssec:tLFV}. As a result the Higgs penguin contributions to the scalar $2l2q$ operators are negligible.

{\bf Scalar four-lepton operators} also receive contributions from Higgs penguins,  which  are simple to extrapolate from the quark results, but not listed because they are negligible due to the lepton Yukawa coupling.
For instance,  in the case of $C^{e\mu ee}_{S,XX}$ --- the  scalar four-lepton operator  appearing in the Lagrangian  of ``observables'' (\ref{Lag1}) ---  the Higgs penguin contribution   is  $\propto y_e$ so suppressed below the reach of upcoming experiments.  

 At ${\cal O}(1/m_{LQ}^4)$, there are log-enhanced box diagrams that generate  scalar four-lepton operators (see the right diagram of  Figure \ref{fig:p+bLQ}), with no  suppression  by  small Yukawa couplings when the internal quarks are tops.
 However, the resulting coefficients  are below the sensitivity of upcoming experiments due to dipole constraints,  as discussed in section \ref{ssec:tLFV}.  Finally,
  there could  be two-loop
penguin$^2$ contributions to $XY$ scalars with a $\tau$-bilinear, which are not listed because they  seem suppressed with respect to the ``Higgs penguin'' contribution in this leptoquark model.

We neglect {\bf tensor operators involving  a   pair of $\tau$s, or down-type quarks}, because they seem at least as suppressed as the scalar  operators.

}

\subsection{Penguins}
\label{assec:ping}

``Penguin'' is a widely used word in physics.  In this manuscript, ``penguin diagrams'' are broadly  understood to have the shape
illustrated in figure \ref{fig:ping}:  there is a 4-particle interaction  mediated by a contact interaction or heavy particle exchange(illustrated as a grey ellipse),  then  two of the four legs are closed to a loop. One  or several  boson propagator(s)  attach to the loop (drawn as an  ellipse surrounding a dashed line), and  may  connect to an external particle line.   This definition includes  LFV ``Higgs penguins'', which could contribute  to the ${\cal O}_{EH}$ SMEFT operator which  gives flavour-changing Higgs couplings.

\begin{figure}[ht]
\begin{center}
\includegraphics[width=0.45\linewidth]{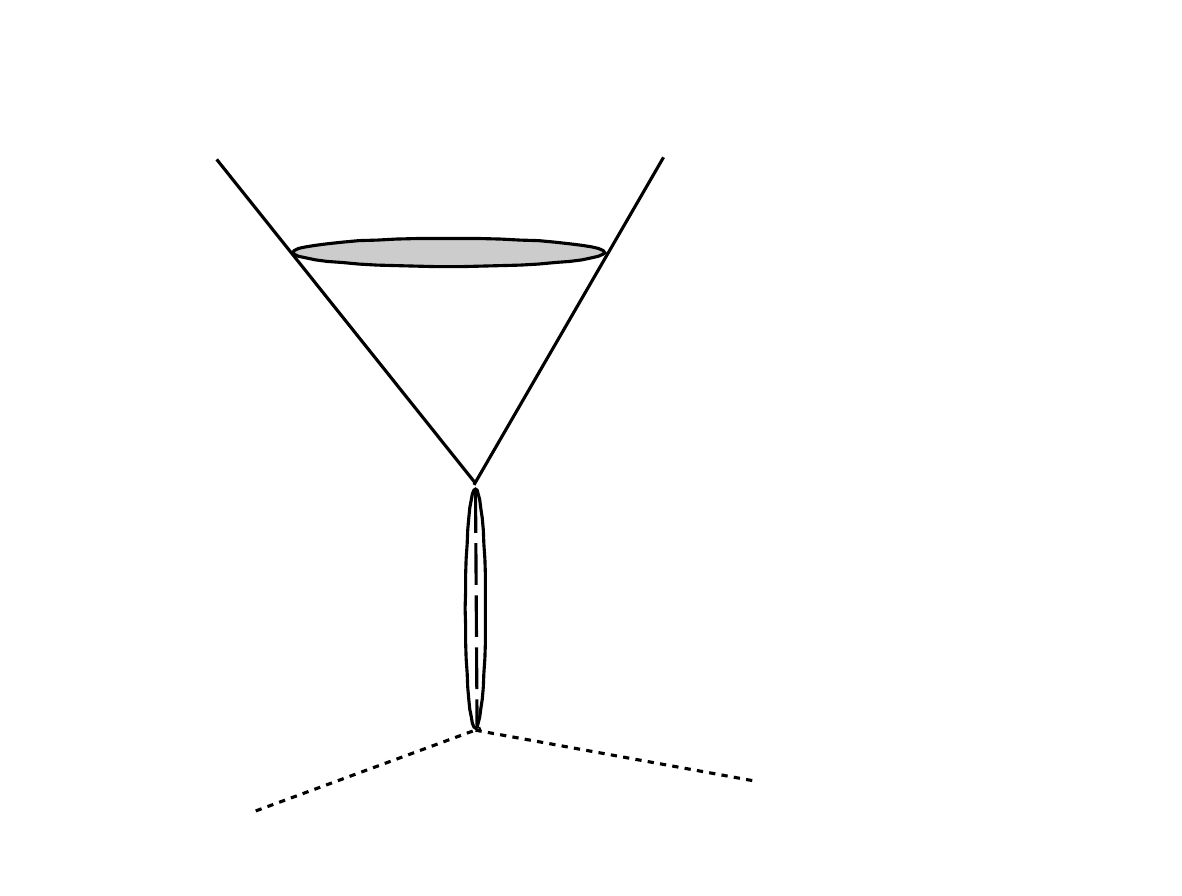}
\end{center}
\caption{
 A schematic representation of a ``penguin'' diagram: the grey ellipse is a heavy particle exchange, the  solid lines are light bosons or fermions,the ellipse containing a dashed line is a SM boson ($\g$,$Z$,$h$), and the dashed lines are possibly-present light particles.
\label{fig:ping}} 
\end{figure}

``Penguin'' diagrams are often  contributions   to  $Z$ or $\g$ vertices, in which case the external  currents must be vector.
In the case of { photon penguins}, the loop is $\propto q^2$ = the four-momentum$^2$ of the  off-shell photon, which has    a vector current bilinear at the other  end of its propagator, so  the diagram contributes to a vector four-particle operator.
For { Z penguins}, there are two types of diagrams: those $\propto v^2$, where $v = \langle H_0 \rangle$ is the Higs vev,  and those $\propto q^2$.

The curiosity we  discuss  here   arises in SMEFT, where the equations of motion of the $Z$ boson are used to  match the $\propto q^2$ part of the  penguin  diagrams onto  momentum-independent operators; the  coefficients of these operators  must  then cancel precisely in low-energy processes to mimic this kinematic suppression as $q^2 \to 0$.
This occurs  in  the type II seesaw  model, 
  where  the $q^2$-$Z$-penguin  diagrams give the dominant   contribution to flavour-changing  $Z$ decays at the electroweak scale ($eg$  $Z\to e^\pm \mu^\mp$),
   but a negligible  contribution  to low-energy observables.
  This  is generically expected in models  where the $Z$ penguin diagrams  are $\propto q^2$, and  motivates LHC searches for   $Z\to e^\pm \mu^\mp$ \cite{LHCZ}.

 We  aim to calculate  in EFT the $Z$-penguin  contribution to  the  low-energy matrix element ${\cal M} (\mu_L u_R \to e_L u_R)$,
  that could arise  in the leptoquark model or the type II seesaw and contributes to $\muc$.   We  calculate in three ways:  in the model,  by matching directly onto the low-energy QCD$\times$QED-invariant EFT at $\LNP$, and by  matching to SMEFT at $\LNP$ then to the low-energy EFT at the weak scale.
The  diagrams are illustrated in 
 Figures \ref{fig:pingZ1} and \ref{fig:pingZ2}.

\begin{figure}[thb]
\begin{center}
\begin{picture}(70,70)
\ArrowLine(0,60)(20,15)
\ArrowLine(20,15)(40,60)
\put(-6,60){$\mu$} 
\put(45,60){$e$} 
\ArrowLine(0,-10)(20,0) 
\ArrowLine(20,0)(40,-10)
\DashLine(7,45)(33,45){3}
\put(39,31){$f$} 
\put(-16,-12){$u_R$} 
\put(46,-10){$u_R$} 
\Photon(20,15)(20,0){2}{3} 
\end{picture}
\hspace{2cm}
\begin{picture}(70,70)
\ArrowLine(0,60)(20,50)
\ArrowLine(20,50)(40,60)
\put(-6,60){$\mu$} 
\put(45,60){$e$} 
\ArrowLine(0,-10)(20,10) 
\ArrowLine(20,10)(40,-10)
\ArrowArc(20,30)(15,0,360)
\GCirc(19.5,47){5}{.7}
\GCirc(19.5,13){5}{.7}
\put(39,31){$f$} 
\put(-16,-12){$u_R$} 
\put(46,-10){$u_R$} 
\end{picture}
\end{center}
\caption{penguin diagrams in the model  and QCD$\times$QED invariant EFT
\label{fig:pingZ1}}
\end{figure}
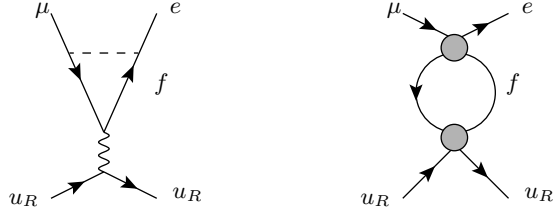

 In the models,   there are in principle  four  diagrams contributing to the ``$Z$-penguin'':  the $Z$  attached to either external fermion, to   the internal scalar or  to the internal fermion(as illustrated on the left in  Figure \ref{fig:pingZ1}).   They give a log-enhanced contribution   to the coefficient of
  $ (\bar{e} \g P_L \mu) (\bar{u}\g P_R u) $:
\bea
Z\!\!\!-\!\!Penguin &\sim&
\sum_f \frac{N_c C_{V XY}^{e\mu ff}}{ 8 \pi^2 } [m_f^2 + {\cal O}(g^f_Y\frac{q^2}{2})]
\log \frac{\LNP }{m_f }
\frac{  g^2  g^u_R  }{4\cos^2\theta_W (q^2 -m_Z^2)} 
\label{ZP1}
\eea
where  $N_c=3$ for the leptoquark and $N_c=1$ for type II, the mass and couplings of the heavy boson are represented via the  vector  four-fermion coefficient induced at tree-level $ C_{V XY}^{e\mu ff}$, and  $q^2 $  is the momentum transfer on the $Z$ line. 
A better approximation for the  lower cutoff of the logarithm  would be  max$\{m_f^2, q^2_{min}\}$, where $q^2_{min}$ for $\muc$ is  $m_\mu^2$ \footnote{The lower cutoff for the diagram of  Figure \ref{fig:pingZ1}) is neccessarily 2 GeV, because the quarks are matching to hadrons at that scale. However, below 2 GeV, the $u$ quark can be replaced  in the diagram by a proton, and this is equivalent to retaining the $u$ quark, given that the matching of quark onto nucleon vector currents in the nucleus  respects charge conservation.}. However, putting $m_f$ gives a nicer invariant, and the difference is numerically irrelevant  for  current experimental sensitivities. 
The  $\propto q^2$ part of the matrix element  is  negligible   for low-energy muon processes  where  $q^2 \sim m_\mu^2$, because  $m_\mu^2/(16 \pi^2 v^2) \lsim 10^{-8}$  suppresses the coefficients beneath the reach of upcoming experiments. 

 In  matching the models  directly to  the  QCD$\times$QED invariant EFT, four-fermion   operators can be generated by the heavy New Physics, and also  by  electroweak bosons of the SM ($Z$ and $h$).  The (log-enhanced part of the) model  amplitude   arises in the  [dimension 6]$^2 \to$ dimension eight running of the EFT is  represented in the  fish  diagram to the right in figure \ref{fig:pingZ1}.  Calculating the fish at zero-external-momentum (because $q^2$ is negligible)  gives \cite{Ardu:2022pzk}:
 \bea
\Delta C^{e \mu uu}_{VXR} &\sim&
\frac{N_c C_{V XY}^{e\mu ff} m_f^2 g^u_R}{ 8 \pi^2 m_{LQ}^2 } 
\log \frac{\LNP }{m_f }
\label{ZP2}
\eea
where there is a contribution with  an $m_f$ insertion on both lines connecting the grey bubble to a Higgs,  as well as contributions with two mass insertions on one of the $f$ lines,  which combine $\propto |g^f_R - g^f_L| = 1$.

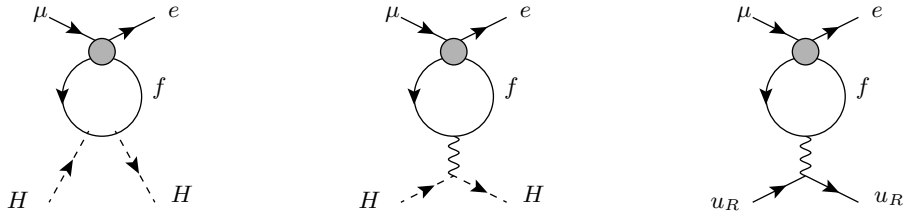
\begin{figure}[thb]
\begin{center}
\begin{picture}(70,70)
\ArrowLine(0,60)(20,50)
\ArrowLine(20,50)(40,60)
\put(-6,60){$\mu$} 
\put(45,60){$e$} 
\DashArrowLine(0,-10)(15,17){3} 
\DashArrowLine(25,17)(40,-10){3}
\ArrowArc(20,30)(15,0,360)
\GCirc(20,47){5}{.7}
\put(39,31){$f$} 
\put(-16,-12){$H$} 
\put(46,-10){$H$} 
\end{picture}
\hspace{2cm}
\begin{picture}(70,70)
\ArrowLine(0,60)(20,50)
\ArrowLine(20,50)(40,60)
\put(-6,60){$\mu$} 
\put(45,60){$e$} 
\DashArrowLine(0,-10)(20,0){3} 
\DashArrowLine(20,0)(40,-10){3}
\ArrowArc(20,30)(15,0,360)
\GCirc(20,47){5}{.7}
\put(39,31){$f$} 
\put(-16,-12){$H$} 
\put(46,-10){$H$} 
\Photon(20,15)(20,0){2}{3} 
\end{picture}
\hspace{2cm}
\begin{picture}(70,70)
\ArrowLine(0,60)(20,50)
\ArrowLine(20,50)(40,60)
\put(-6,60){$\mu$} 
\put(45,60){$e$} 
\ArrowLine(0,-10)(20,0) 
\ArrowLine(20,0)(40,-10)
\ArrowArc(20,30)(15,0,360)
\GCirc(20,47){5}{.7}
\put(39,31){$f$} 
\put(-16,-12){$u_R$} 
\put(46,-10){$u_R$} 
\Photon(20,15)(20,0){2}{3} 
\end{picture}
\end{center}
\caption{SMEFT penguin diagrams.
\label{fig:pingZ2}}
\end{figure}

 Alternatively, one could match the model to SMEFT at $\LNP$, run to $m_W$, then match to the  QED$\times$QCD-invariant EFT. The penguin diagrams arise in the SMEFT RGEs as illustrated in Figure \ref{fig:pingZ2}:  
  the grey blob is a New-Physics induced  vector  four-fermion operator 
  which mixes  via the first two diagrams  into the ``penguin operators'', defined as
\bea
    {\cal O}^{ e\m}_{HL1} &=& i[(H^\dagger   D_\m H) - ( D_\m H)^\dagger H](\overline{\ell}_e \gamma^\mu \ell_\m )
    \label{LLpenguin} \\
   & \to & \sqrt{g^2 + g^{'2}} v^2 Z_\mu  (\overline{\ell}_e \gamma^\mu \ell_\m )
\nonumber\\
      {\cal O}^{ e\m}_{HL3 } &=&  i[(H^\dagger  \tau^a D_\m H) - ( D_\m H)^\dagger \tau^a H ] (\overline{\ell}_e \gamma^\mu \tau^{a}\ell_\m )
 \label{LL3penguin} \\      &\to& -\sqrt{g^2 + g^{'2}} v^2 Z_\mu (\overline{\ell}_e \gamma^\mu \tau^{3}\ell_\m ) \nonumber\\
      {\cal O}^{ e\m}_{HE} &=& i[(H^\dagger  D_\m H) - ( D_\m H)^\dagger H ](\overline{e}_e \gamma^\mu e_\m )
\label{EEpenguin} \\     & \to&   \sqrt{g^2 + g^{'2}} v^2 Z_\mu (\overline{e}_e \gamma^\mu e_\m )\nonumber
\eea
where after the arrows are the 
flavour-changing $Z$ vertices to
which the  operators reduce when the Higgs gets a vev
(so $a=3$ in the triplet case, where  $\langle H^\dagger \tau_3 H \rangle = - v^2$), and   we used    $Z^\mu = -s_W B^\mu + c_W W_3^\m$,
where $s_W =\sin \theta_W = g'/\sqrt{g^2 + g^{'2}}$ and
$2m_Z^2 = (g^2 + g^{'2}) v^2$.
 In the SMEFT RGEs,    the grey blob also
mixes via      $B$ and $W_0$ exchange (which can be written as $\g$ and $Z$ exchange by a basis rotation) into other four-fermion operators, as illustrated by the last diagram of  figure \ref{fig:pingZ2}.
Then at the weak scale,  the penguin and four-fermion operators of SMEFT
match onto  four-fermion operators in the low-energy theory.
The component $\propto m_f^2$   of the penguin matrix element  of  Eqn (\ref{ZP1})   corresponds to the first diagram of Figure \ref{fig:pingZ2},    induces the SMEFT penguin operators,  and therefore  a  four-fermion operator at low-energy. However, the $\propto q^2$ component of  matrix element in the model  generates both the penguin operators and  four-fermion operators. These coefficients will cancel in low energy four-fermion processes, as one can see  by using the equations of motion for the $Z$ :$ (q^2 - m_Z^2) Z^\mu = -\frac{g}{2c_W} \sum_f g^f_X  (\overline{f_X}\g_\mu f_X)$ (which apply in ``on-shell'' bases such as SMEFT) in calculating the contribution of a  $Z$ vertex $\propto q^2$ to  the $S$-matrix element $\langle e f \bar{f} | S |\mu\rangle$: 
$$
(\bar{e}\g_\mu \mu) q^2   \frac{-i}{q^2 - m_Z^2}  \frac{-i g_X^f g}{2c_W} (\overline{f_X}\g^\mu f_X)  = (\bar{e}\g_\mu \mu)  \left\{ m_Z^2 \frac{-i}{q^2 - m_Z^2}  \frac{-ig_X^f g}{2c_W}   -  \frac{g_X^f g}{2c_W}   \right\}  (\overline{f_X}\g^\mu f_X) 
$$
where $f_X$ is a  chiral SM  fermion.

So all the calculations give the same result, but  in SMEFT,
    the  kinematics of the matrix element is modified by using the  equation of motion to reduce the operator basis.  That is,  the $\propto q^2$ part of the $Z$ penguin diagrams  vanishes as $q^2\to 0$ via an increasingly precise cancellation between the penguin and four-fermion operators. This illustrates that models can  ``naturally'' engineer very precise cancellations among operator coefficients.

\end{document}